\documentclass[twocolumn]{aastex62}

\usepackage{graphics,graphicx,xspace,natbib,amssymb}
\usepackage[caption=false]{subfig}
\usepackage{amsmath}
\usepackage{threeparttable}
\usepackage{makecell}

\usepackage{multirow}
\usepackage{fontawesome5}
\usepackage[encapsulated]{CJK}

\newcommand{\comment}[1]{}

\let\itAA\AA
\renewcommand{\AA}{\mathrm{\itAA}}

\usepackage{lineno}

\shorttitle{MINERVA}

\begin{document}

\title{MINERVA: A NIRCam Medium Band and MIRI Imaging Survey to Unlock the Hidden Gems of the Distant Universe} 
\shortauthors{}

\author[0000-0002-9330-9108]{Adam Muzzin}
\altaffiliation{Equal contribution}
\affiliation{Department of Physics and Astronomy, York University, 4700 Keele St., Toronto, Ontario, M3J 1P3, Canada}

\author[0000-0002-1714-1905]{Katherine A. Suess}
\altaffiliation{Equal contribution}
\affiliation{Department for Astrophysical \& Planetary Science, University of Colorado, Boulder, CO 80309, USA}

\author[0000-0001-9002-3502]{Danilo Marchesini}
\altaffiliation{Equal contribution}
\affiliation{Department of Physics and Astronomy, Tufts University, 574 Boston Avenue, Suite 304, Medford, MA 02155, USA}

\author[0000-0002-6265-2675]{Luke Robbins}
\affiliation{Department of Physics and Astronomy, Tufts University, 574 Boston Avenue, Suite 304, Medford, MA 02155, USA}

\author[0000-0002-4201-7367]{Chris J. Willott}
\affiliation{NRC Herzberg, 5071 West Saanich Rd, Victoria, BC V9E 2E7, Canada}

\author[0000-0002-8909-8782]{Stacey Alberts}
\affiliation{AURA for the European Space Agency (ESA), Space Telescope Science Institute, 3700 San Martin Dr., Baltimore, MD 21218, USA}

\author[0000-0002-0243-6575]{Jacqueline Antwi-Danso}
\altaffiliation{Banting Postdoctoral Fellow}
\affiliation{David A. Dunlap Department of Astronomy \& Astrophysics, University of Toronto 50 St George Street, Toronto, ON M5S 3H4, Canada}

\author[0000-0003-3983-5438]{Yoshihisa Asada}
\affiliation{Department of Astronomy and Physics and Institute for Computational Astrophysics, Saint Mary's University, 923 Robie Street, Halifax, Nova Scotia B3H 3C3, Canada}
\affiliation{Department of Astronomy, Kyoto University, Sakyo-ku, Kyoto 606-8502, Japan}
\affiliation{Waseda Research Institute for Science and Engineering, Faculty of Science and Engineering, Waseda University, 3-4-1 Okubo, Shinjuku, Tokyo 169-8555, Japan}

\author[0000-0003-2680-005X]{Gabriel Brammer}
\affiliation{Cosmic Dawn Center (DAWN), Copenhagen, Denmark}
\affiliation{Niels Bohr Institute, University of Copenhagen, Jagtvej 128, DK-2200 Copenhagen N, Denmark}

\author[0000-0002-7031-2865]{Sam E. Cutler}
\affiliation{Department of Astronomy, University of Massachusetts, Amherst, MA 01003, USA}

\author[0000-0001-9298-3523]{Kartheik G. Iyer}
\altaffiliation{NHFP Hubble Fellow}
\affiliation{Columbia Astrophysics Laboratory, Columbia University, 550 West 120th Street, New York, NY 10027, USA}

\author[0000-0002-2057-5376]{Ivo Labb\'e}
\affiliation{Centre for Astrophysics and Supercomputing, Swinburne University of Technology, Melbourne, VIC 3122, Australia}

\author[0000-0003-3243-9969]{Nicholas S. Martis}
\affiliation{University of Ljubljana, Faculty of Mathematics and Physics, Jadranska ulica 19, SI-1000 Ljubljana, Slovenia}

\author[0000-0001-8367-6265]{Tim B. Miller}
\affiliation{Center for Interdisciplinary Exploration and Research in Astrophysics (CIERA), Northwestern University, 1800 Sherman Ave, Evanston IL 60201, USA}

\author[0000-0001-7300-9450]{Ikki Mitsuhashi}
\affiliation{Department for Astrophysical \& Planetary Science, University of Colorado, Boulder, CO 80309, USA}

\author[0000-0001-8592-2706]{Alexandra Pope}
\affiliation{Department of Astronomy, University of Massachusetts, Amherst, MA 01003, USA}

\author[0000-0002-1917-1200]{Anna Sajina}
\affiliation{Department of Physics and Astronomy, Tufts University, 574 Boston Avenue, Suite 304, Medford, MA 02155, USA}

\author[0000-0001-8830-2166]{Ghassan T. E. Sarrouh}
\affiliation{Department of Physics and Astronomy, York University, 4700 Keele St., Toronto, Ontario, M3J 1P3, Canada}

\author[0000-0003-1078-9706]{Monu Sharma}
\affiliation{Departament d'Astronomia i Astrofisica, Universitat de Valencia, C. Dr. Moliner 50, E-46100 Burjassot, Valencia, Spain}

\author[0000-0001-7768-5309]{Mauro Stefanon}
\affiliation{Departament d'Astronomia i Astrofisica, Universitat de Valencia, C. Dr. Moliner 50, E-46100 Burjassot, Valencia, Spain}
\affiliation{Unidad Asociada CSIC ``Grupo de Astrofisica Extragalactica y Cosmologi'' (Instituto de Fisica de Cantabria - Universitat de Valencia)}

\author[0000-0001-7160-3632]{Katherine E. Whitaker}
\affiliation{Department of Astronomy, University of Massachusetts, Amherst, MA 01003, USA}
\affiliation{Cosmic Dawn Center (DAWN), Copenhagen, Denmark}

\author[0000-0002-4542-921X]{Roberto Abraham}
\affiliation{David A. Dunlap Department of Astronomy \& Astrophysics, University of Toronto 50 St George Street, Toronto, ON M5S 3H4, Canada}

\author[0000-0002-7570-0824]{Hakim Atek}
\affiliation{Institut d’Astrophysique de Paris, CNRS, Sorbonne Universit\'e, 98bis Boulevard Arago, 75014, Paris, France}

\author[0000-0001-5984-0395]{Maru\v{s}a Brada{\v c}}
\affiliation{University of Ljubljana, Faculty of Mathematics and Physics, Jadranska ulica 19, SI-1000 Ljubljana, Slovenia}
\affiliation{Department of Physics and Astronomy, University of California Davis, 1 Shields Avenue, Davis, CA 95616, USA}

\author[0000-0001-7549-5560]{Samantha Berek}
\affiliation{David A. Dunlap Department of Astronomy \& Astrophysics, University of Toronto 50 St George Street, Toronto, ON M5S 3H4, Canada}
\affiliation{Dunlap Institute for Astronomy \& Astrophysics, University of Toronto, 50 St. George Street, Toronto, ON M5S 3H4, Canada}
\affiliation{Data Sciences Institute, University of Toronto, 17th Floor, Ontario Power Building, 700 University Avenue, Toronto, ON M5G 1Z5, Canada}

\author[0000-0001-5063-8254]{Rachel Bezanson}
\affiliation{Department of Physics and Astronomy and PITT PACC, University of Pittsburgh, Pittsburgh, PA 15260, USA}

\author[0000-0002-6741-078X]{Westley Brown}
\affiliation{Department of Physics and Astronomy, York University, 4700 Keele St., Toronto, Ontario, M3J 1P3, Canada}

\author[0000-0002-6523-9536]{Adam J.\ Burgasser}
\affiliation{Department of Astronomy \& Astrophysics, UC San Diego, La Jolla, CA 92093}

\author[0009-0005-1143-495X]{Nathalie Chicoine}
\affiliation{Department of Physics and Astronomy and PITT PACC, University of Pittsburgh, Pittsburgh, PA 15260, USA}

\author[0000-0001-9978-2601]{Aidan P. Cloonan}
\affiliation{Department of Astronomy, University of Massachusetts, Amherst, MA 01003, USA}

\author[0000-0003-3881-1397]{Olivia R. Cooper}\altaffiliation{NSF Astronomy and Astrophysics Postdoctoral Fellow}
\affiliation{Department for Astrophysical \& Planetary Science, University of Colorado, Boulder, CO 80309, USA}

\author[0000-0001-8460-1564]{Pratika Dayal}
\affiliation{Kapteyn Astronomical Institute, University of Groningen, 9700 AV Groningen, The Netherlands}

\author[0000-0002-2380-9801]{Anna de Graaff}
\affiliation{Max-Planck-Institut f\"ur Astronomie, K\"onigstuhl 17, D-69117, Heidelberg, Germany}

\author[0000-0001-8325-1742]{Guillaume Desprez}
\affiliation{Kapteyn Astronomical Institute, University of Groningen, P.O. Box 800, 9700AV Groningen, The Netherlands}

\author[0000-0002-1109-1919]{Robert Feldmann}
\affiliation{Department of Astrophysics, Universit\"at Z\"urich, Zurich, CH-8057, Switzerland}

\author[0000-0001-6003-0541]{Ben Forrest}
\affiliation{Department of Physics and Astronomy, University of California Davis, One Shields Avenue, Davis, CA, 95616, USA}

\author[0000-0002-8871-3026]{Marijn Franx}
\affiliation{Leiden Observatory, Leiden University, P.O.Box 9513, NL-2300 AA Leiden, The Netherlands}

\author[0000-0001-7440-8832]{Yoshinobu Fudamoto} 
\affiliation{Center for Frontier Science, Chiba University, 1-33 Yayoi-cho, Inage-ku, Chiba 263-8522, Japan}

\author[0000-0001-7201-5066]{Seiji Fujimoto}\altaffiliation{NHFP Hubble Fellow}
\affiliation{Department of Astronomy, The University of Texas at Austin, Austin, TX 78712, USA}

\author[0000-0001-6278-032X]{Lukas J. Furtak}
\affiliation{Department of Physics, Ben-Gurion University of the Negev, P.O. Box 653, Be’er-Sheva 84105, Israel}

\author[0000-0002-3254-9044]{Karl Glazebrook}
\affiliation{Centre for Astrophysics and Supercomputing, Swinburne University of Technology, Melbourne, VIC 3122, Australia}

\author[0009-0007-8470-5946]{Ilias Goovaerts}
\affiliation{Space Telescope Science Institute, 3700 San Martin Drive, Baltimore, Maryland 21218, USA}

\author[0000-0002-5612-3427]{Jenny E. Greene}
\affiliation{Department of Astrophysical Sciences, 4 Ivy Lane, Princeton University, Princeton, NJ 08544, USA}

\author[0009-0009-9848-3074]{Naadiyah Jagga}
\affiliation{Department of Physics and Astronomy, York University, 4700 Keele St., Toronto, Ontario, M3J 1P3, Canada}

\author{William W.H. Jarvis}
\affiliation{Department of Astronomy, University of Massachusetts, Amherst, MA 01003, USA}

\author[0000-0002-7613-9872]{Mariska Kriek}
\affiliation{Leiden Observatory, Leiden University, P.O.Box 9513, NL-2300 AA Leiden, The Netherlands}

\author[0000-0002-3475-7648]{Gourav Khullar}
\affiliation{Department of Physics and Astronomy and PITT PACC, University of Pittsburgh, Pittsburgh, PA 15260, USA}

\author[0009-0002-5758-6025]{Valentina La Torre}
\affiliation{Department of Physics and Astronomy, Tufts University, 574 Boston Avenue, Suite 304, Medford, MA 02155, USA}

\author[0000-0001-6755-1315]{Joel Leja}
\affiliation{Department of Astronomy \& Astrophysics, The Pennsylvania State University, University Park, PA 16802, USA}
\affiliation{Institute for Computational \& Data Sciences, The Pennsylvania State University, University Park, PA 16802, USA}
\affiliation{Institute for Gravitation and the Cosmos, The Pennsylvania State University, University Park, PA 16802, USA}

\author[0000-0002-3101-8348]{Jamie Lin}
\affiliation{Department of Physics and Astronomy, Tufts University, 574 Boston Avenue, Suite 304, Medford, MA 02155, USA}

\author[0000-0002-5337-5856]{Brian Lorenz}
\affiliation{Department of Astronomy, University of California, Berkeley, CA 94720, USA}

\author[0000-0002-7847-1107]{Daniel Lyon}
\affiliation{Centre for Astrophysics and Supercomputing, Swinburne University of Technology, Melbourne, VIC 3122, Australia}

\author[0000-0002-5694-6124]{Vladan Markov}
\affiliation{University of Ljubljana, Faculty of Mathematics and Physics, Jadranska ulica 19, SI-1000 Ljubljana, Slovenia}

\author[0000-0003-0695-4414]{Michael V. Maseda}
\affiliation{Department of Astronomy, University of Wisconsin-Madison, 475 N. Charter St., Madison, WI 53706, USA}

\author[0000-0002-2446-8770]{Ian McConachie}
\affiliation{Department of Astronomy, University of Wisconsin-Madison, 475 N. Charter St., Madison, WI 53706, USA}

\author[0009-0000-5385-8674]{Maya Merchant}
\affiliation{Niels Bohr Institute, University of Copenhagen, Jagtvej 128, DK-2200 Copenhagen N, Denmark}

\author[0000-0001-8115-5845]{Rosa M. M\'erida}
\affiliation{Department of Astronomy and Physics and Institute for Computational Astrophysics, Saint Mary's University, 923 Robie Street, Halifax, Nova Scotia B3H 3C3, Canada}

\author[0000-0002-8530-9765]{Lamiya Mowla}
\affiliation{Whitin Observatory, Department of Physics and Astronomy, Wellesley College, 106 Central Street, Wellesley, MA 02481, USA}

\author[0009-0009-2307-2350]{Katherine Myers}
\affiliation{Department of Physics and Astronomy, York University, 4700 Keele St., Toronto, Ontario, M3J 1P3, Canada}

\author[0000-0003-2895-6218]{Rohan P.\ Naidu}
\altaffiliation{NHFP Hubble Fellow}
\affiliation{MIT Kavli Institute for Astrophysics and Space Research, 77 Massachusetts Ave., Cambridge, MA 02139, USA}

\author[0000-0003-2804-0648 ]{Themiya Nanayakkara}
\affiliation{Centre for Astrophysics and Supercomputing, Swinburne University of Technology, Melbourne, VIC 3122, Australia}

\author[0000-0002-7524-374X]{Erica J. Nelson}
\affiliation{Department for Astrophysical \& Planetary Science, University of Colorado, Boulder, CO 80309, USA}

\author{Gaël Noirot}
\affiliation{Space Telescope Science Institute, 3700 San Martin Drive, Baltimore, Maryland 21218, USA}

\author[0000-0001-5851-6649]{Pascal A. Oesch}
\affiliation{Department of Astronomy, University of Geneva, Chemin Pegasi 51, 1290 Versoix, Switzerland}
\affiliation{Cosmic Dawn Center (DAWN), Copenhagen, Denmark}
\affiliation{Niels Bohr Institute, University of Copenhagen, Jagtvej 128, DK-2200 Copenhagen N, Denmark}

\author[0000-0002-8432-6870]{Kiyoaki C. Omori}
\affiliation{Department of Astronomy and Physics and Institute for Computational Astrophysics, Saint Mary's University, 923 Robie Street, Halifax, Nova Scotia B3H 3C3, Canada}

\author[0000-0002-9651-5716]{Richard Pan}
\affiliation{Department of Physics and Astronomy, Tufts University, 574 Boston Avenue, Suite 304, Medford, MA 02155, USA}

\author[0009-0001-0715-7209]{Natalia Porraz Barrera}
\affiliation{Department for Astrophysical \& Planetary Science, University of Colorado, Boulder, CO 80309, USA}

\author[0000-0002-0108-4176]{Sedona H. Price}
\affiliation{Space Telescope Science Institute, 3700 San Martin Drive, Baltimore, Maryland 21218, USA}

\author[0000-0002-5269-6527]{Swara Ravindranath}
\affiliation{Astrophysics Science Division, NASA Goddard Space Flight Center, 8800 Greenbelt Road, Greenbelt, MD 20771, USA} \affiliation{Center for Research and Exploration in Space Science and Technology II, Department of Physics, Catholic University of America, 620 Michigan Avenue N.E., Washington, DC 20064, USA}

\author[0000-0002-7712-7857]{Marcin Sawicki}
\affiliation{Department of Astronomy and Physics and Institute for Computational Astrophysics, Saint Mary's University, 923 Robie Street, Halifax, Nova Scotia B3H 3C3, Canada}

\author[0000-0003-4075-7393]{David J. Setton}\thanks{Brinson Prize Fellow}
\affiliation{Department of Astrophysical Sciences, 4 Ivy Lane, Princeton University, Princeton, NJ 08544, USA}

\author[0000-0001-8034-7802]{Renske Smit}
\affiliation{ Astrophysics Research Institute, Liverpool John Moores University, 146 Brownlow Hill, Liverpool L3 5RF, UK}

\author[0000-0003-0780-9526]{Visal Sok}
\affiliation{Department of Physics and Astronomy, York University, 4700 Keele St., Toronto, Ontario, M3J 1P3, Canada}

\author[0000-0003-2573-9832]{Joshua S. Speagle(\begin{CJK*}{UTF8}{gbsn}沈佳士\ignorespacesafterend\end{CJK*})}
\affiliation{Department of Statistical Sciences, University of Toronto, 9th Floor, Ontario Power Building, 700 University Ave, Toronto, ON M5G 1Z5, Canada}
\affiliation{David A. Dunlap Department of Astronomy \& Astrophysics, University of Toronto 50 St George Street, Toronto, ON M5S 3H4, Canada}
\affiliation{Dunlap Institute for Astronomy \& Astrophysics, University of Toronto, 50 St. George Street, Toronto, ON M5S 3H4, Canada}
\affiliation{Data Sciences Institute, University of Toronto, 17th Floor, Ontario Power Building, 700 University Avenue, Toronto, ON M5G 1Z5, Canada}

\author[0000-0002-5522-9107]{Edward N.\ Taylor}
\affiliation{Centre for Astrophysics and Supercomputing, Swinburne University of Technology, Melbourne, VIC 3122, Australia}

\author[0000-0002-3503-8899]{Vivian Yun Yan Tan}
\affiliation{Department of Physics and Astronomy, York University, 4700 Keele St., Toronto, Ontario, M3J 1P3, Canada}

\author[0000-0002-9909-3491]{Roberta Tripodi}
\affiliation{INAF - Observatory of Rome, Via Frascati 33, 00044, Rome, Italy}

\author[0000-0002-5027-0135]{Arjen van der Wel}
\affiliation{Sterrenkundig Observatorium, Universiteit Gent, Krijgslaan 281 S9, 9000 Gent, Belgium}

\author[0009-0002-2209-4813]{Edgar Perez Vidal}
\affiliation{Department of Physics and Astronomy, Tufts University, 574 Boston Avenue, Suite 304, Medford, MA 02155, USA}

\author[0000-0001-9269-5046]{Bingjie Wang (\begin{CJK*}{UTF8}{gbsn}王冰洁\ignorespacesafterend\end{CJK*})}
\affiliation{Department of Astronomy \& Astrophysics, The Pennsylvania State University, University Park, PA 16802, USA}
\affiliation{Institute for Computational \& Data Sciences, The Pennsylvania State University, University Park, PA 16802, USA}
\affiliation{Institute for Gravitation and the Cosmos, The Pennsylvania State University, University Park, PA 16802, USA}

\author[0000-0003-1614-196X]{John R. Weaver}\thanks{Brinson Prize Fellow}
\affiliation{MIT Kavli Institute for Astrophysics and Space Research, 77 Massachusetts Ave., Cambridge, MA 02139, USA}

\author[0000-0003-2919-7495]{Christina C.\ Williams}
\affiliation{NSF National Optical-Infrared Astronomy Research Laboratory, 950 North Cherry Avenue, Tucson, AZ 85719, USA}

\author[0009-0000-8716-7695]{Sunna Withers}
\affiliation{Department of Physics and Astronomy, York University, 4700 Keele St., Toronto, Ontario, M3J 1P3, Canada}

\author[0000-0002-1163-7790]{Kumail Zaidi}
\affiliation{Department of Physics and Astronomy, Tufts University, 574 Boston Avenue, Suite 304, Medford, MA 02155, USA}
\affiliation{HEP Division, Argonne National Laboratory, 9700 South Cass Avenue, Lemont, IL 60439, USA}



\begin{abstract}
We present an overview of the MINERVA survey, a 259.8 hour (prime) and 127 hour (parallel) Cycle 4 treasury program on the James Webb Space Telescope (JWST).  MINERVA is obtaining 8 filter NIRCam medium band imaging (F140M, F162M, F182M, F210M, F250M, F300M, F360M, F460M) and 2 filter MIRI imaging (F1280W, F1500W) in four of the five CANDELS Extragalactic fields: UDS, COSMOS, AEGIS and GOODS-N.  These fields were previously observed in Cycle 1 with 7 - 9 NIRCam filters by the PRIMER, CEERS and JADES programs.  MINERVA reaches a 5$\sigma$ depth of 28.1 mag in F300M and covers $\sim$ 542 arcmin$^2$, increasing the area of existing JWST medium-band coverage in at least 8 bands by $\sim$ 7$\times$.  The MIRI imaging reaches a 5$\sigma$ depth of 23.9 mag in  F1280W and covers $\sim$ 275 arcmin$^2$ in at least 2 MIRI filters.  When combined with existing imaging, these data will provide a photometric catalog with 20-26 JWST filters (depending on field) and 26-35 filters total, including HST.  This paper presents a detailed breakdown of the filter coverage, exposure times, and field layout relative to previous observations, as well as an overview of the primary science goals of the project.  These include uncovering the physics of enigmatic sources hiding in current broadband catalogs, improving systematics on stellar mass functions and number densities by factors of $\gtrsim$ 3, and resolved mapping of stellar mass and star formation at 1 $< z <$ 6.  When complete, MINERVA will become an integral part of the treasury deep field imaging datasets, significantly improving population studies with well-understood completeness, robust photometric redshifts, stellar masses, and sizes, and facilitating spectroscopic follow up for decades to come.

\end{abstract}

\keywords{Galaxy evolution (594) --- Galaxy formation (595) --- Galaxy structure (622) --- High-redshift galaxies (608)}

\section{Introduction}

High-quality multi-wavelength imaging has been essential in nearly all major breakthroughs in the modern study of galaxy formation. Multi-color data facilitate complete demographic studies of galaxies via photometric redshifts, and are necessary for computing key parameters for classifying galaxies such as stellar masses and rest-frame colors. While spectroscopy remains the quintessential tool for more detailed studies of galaxies, most spectroscopic studies are pre-selected from photometric catalogs. Developing high-quality multi-wavelength photometric catalogs is therefore essential to move the study of galaxy formation forward.  

While multi-wavelength imaging catalogs from the ground have been compiled for many decades using both broadbands, e.g., CFHTLS \citep{ilbert06, arnouts07}, FIRES \citep{forster04,wuyts08}, MUSYC \citep{gawiser06,taylor09,cardamone10}, UKIDSS \citep{lawrence07}, UltraVISTA \citep{mccracken12}, as well as medium bands\footnote{Medium bands are generally defined as filters with R $\gtrsim$ 10, whereas broadbands are typically filters with R $\sim$ 5}, e.g., COMBO-17 \citep{wolf03,bell04}, COSMOS \citep{scoville07,taniguchi07,mccracken12},  NMBS \citep{brammer09,whitaker11}, zFOURGE \citep{tomczack14,straatman16}, FENIKS \citep{zaidi24}, true multi-wavelength imaging covering a wide wavelength range using only space-based instruments (which provide substantial gains in both depth and spatial resolution) has progressed slowly, and has been relatively rare until the last decade.  This is primarily because of the limited Field of View (FoV) of space-based imagers, both optical and NIR, requiring substantial investments in time to cover representative areas.   

The beginning of true multi-wavelength space-based imaging began with the Hubble Deep Field \citep{williams96} which was followed by the GOODS survey \citep{giavalisco04}, and ultimately the Hubble Ultra Deep Field \citep{beckwith96}. Despite these impressive datasets, without NIR observations it was challenging to do reliable demographic studies of galaxies at $z >$ 2, as ACS covers only the rest-frame UV wavelength range there.  Multi-wavelength spaced-based surveys were substantially augmented by the HST/WFC3 CANDELS treasury program \citep{grogin11, koekemoer11} which added several NIR filters out to $1.6\mu$m, as well as additional ACS data taken in parallel, and expanded into 5 fields: COSMOS, UDS, AEGIS, GOODS-N and GOODS-S.  These fields were also followed up by Spitzer/IRAC out to 8$\mu$m, leading to extensive space-based multi-wavelength fields covering $\sim$ 0.25 deg$^2$ \citep[e.g.,][]{guo13,galametz13,skelton14,stefanon17}. The CANDELS catalogs have been essential for thousands of studies of high-redshift galaxies, including wide-field spectroscopic surveys such as 3D-HST \citep{brammer12,momcheva16}, MOSDEF \citep{kriek15} and KMOS3D \citep{wisnioski15,wisnioski19}.  Hubble's deep multi-wavelength efforts culminated in the Hubble Frontier Fields \citep{lotz17, merlin16, shipley18} which served as a stepping stone into the James Webb Space Telescope (JWST) era.

The next era of space-based multi-wavelength imaging is well underway with JWST, which provides high-resolution Near Infrared (NIR) imaging out to 4.8$\mu$m with the NIRCam \citep{rieke23} and NIRISS \citep{doyon23} instruments, and Mid Infrared (MIR) imaging to 25$\mu$m with the MIRI instrument \citep{argyriou23}. 
In its first three years of operations, JWST has accumulated a significant amount of multi-wavelength imaging, adding to the legacy of HST and Spitzer. The largest deep extragalactic imaging surveys have been COSMOS-Web \citep{casey23}, PRIMER \citep[e.g.,][]{donnan24}, JADES \citep{eisenstein23a,rieke23_jades}, and CEERS \citep{finkelstein25}; these surveys combined effectively cover all five of the CANDELS fields.  

The majority of these surveys have adopted the same strategy: observing 4-9 NIRCam broadbands and a handful of MIRI bands in a subset of the survey area. This strategy balances photometric redshift ($z_{phot}$) accuracy and resources by allowing deep observations across a wide wavelength range in a minimum amount of time. 
While this was an effective initial use of NIRCam and MIRI imaging time, most of the NIRCam/MIRI broadbands have spectral resolutions of R $\sim$ 4-5, effectively the same resolution as Spitzer/IRAC. This significantly limits their accuracy in defining galaxy SEDs, particularly in the cases of (a) strong emission lines that are common at high-redshift and (b) confusion between the Balmer and Lyman breaks. As with Spitzer before JWST \citep[e.g.,][]{labbe13,stark13,smit14}, targeted studies are now showing that these degeneracies prevent robust measurements of photometric redshifts and stellar masses \citep{robertsborsani21,sarrouh24,harvey25,asada25}. Indeed, line emission has already caused several spurious results, including false conclusions about ``$\Lambda$CDM breaking galaxies" \citep[e.g.,][]{fujimoto2023a,kocevski2023,desprez24,Franco2024} as well as the persistent confusion of intermediate-redshift dusty galaxies for $z >$ 12 galaxies \citep{naidu22,arrabal-haro23, fujimoto2023b,furlanetto2023,jin24}.

There is a clear way forward to improve the spectral resolution issue for these large imaging fields as was shown by previous medium-band surveys \citep[e.g.,][]{wolf03,whitaker11,straatman16,zaidi24}.  NIRCam has an impressive suite of 12 medium band (MB) filters covering 1.4 - 4.8$\micron$, which are already being used for a wide range of science in the JADES Origins Field \citep{eisenstein23b}, JEMS \citep{williams23}, CANUCS/Technicolor \citep{sarrouh25}, and UNCOVER/Megascience \citep{suess24} programs.  These studies are showcasing the value of the NIRCam medium bands, demonstrating their effectiveness in highly accurate photometric redshifts and stellar masses \citep{sarrouh24,suess24} as well as the robust identification of ultra-high redshift galaxies \citep{asada25}, strong line emitters \citep[e.g.,][Porraz in prep]{withers23,rinaldi2023,williams23,wold25}, and dusty galaxies \citep{martis25,lorenz25} as well as Balmer breaks in distant galaxies \citep[e.g.,][Robbins in prep, Antwi-Danso in prep, Khullar in prep]{trussler25,mintz25}.  While exceptionally powerful and versatile for science, the existing JWST MB surveys with coverage in most medium bands comprise just a handful of pointings totaling $\sim$ 70 arcmin$^2$ down to limiting magnitudes of $\sim$ 28 - 29 AB, hence are extremely limited in both volume and cosmic variance.  Given the limitations of broadband imaging and the strengths of medium band imaging, it is an opportune time to add these data to our multi-wavelength extragalactic imaging fields.

In this paper we present an overview of the “Medium-band Imaging with NIRCam to Explore ReVolutionary Astrophysics” (MINERVA) survey, an approved cycle-4 JWST treasury survey (PID: 7814, PI: Muzzin, Co-PIs: Marchesini, Suess) that will observe four of the CANDELS fields: PRIMER-COSMOS, PRIMER-UDS, CEERS-AEGIS, and JADES-GOODS-N in 8 NIRCam medium bands spanning 1-5 $\micron$. In parallel MINERVA will obtain imaging in two MIRI bands, F1280W and F1500W. The survey will increase the area of existing medium band coverage $\sim$ 7$\times$, to $\sim$ 542 arcmin$^2$, and when combined with existing data will boast a total of 20-26 JWST filters (depending on fields) and 26-35 space-based filters total including HST. 

\begin{deluxetable*}{lcccccll}
    \tablecaption{MINERVA Survey Fields
        \label{tab:field_info}}
    \tablewidth{0pt}
    \tablehead{
    \colhead{Field} & \colhead{R.A.} & \colhead{Dec.} & \colhead{MIRI\tablenotemark{a}}& \colhead{NIRCam}  & \colhead{N Filters} & \colhead{JWST Survey\tablenotemark{b}} & \colhead{HST Survey}\\ 
    \colhead{} & \colhead{(J2000)} & \colhead{(J2000)} & \colhead{Area} & \colhead{Area} & \colhead{JWST+HST} & \colhead{Coverage} & \colhead{Coverage}
    }
    \startdata
    UDS         & 02:17:30 & -05:12:00 & 125(54) & 234 & 26 & PRIMER & CANDELS\\
        & & & & & &  &  3D-HST \\
    \hline
    COSMOS & 10:00:30 & +02:20:00 & 111(48) & 144 & 30 & PRIMER  & CANDELS \\ 
     & & & & & & COSMOS-Web & UVCANDELS \\
    & & & & & & COSMOS-3D & 3D-HST \\
            & & & & & &  &  COSMOS \\
    \hline
    AEGIS & 14:19:40 & +52:52:00 & 23(10)\tablenotemark{b} & 96 & 34 & CEERS & CANDELS\\
    & & & & &  & SPAM & UVCANDELS \\
     & & & & & & MEGA & EGS  \\
    \hline
    GOODS-N & 12:36:50 & +62:13:00 & 16(7)\tablenotemark{c} & 68 & 35 & JADES & CANDELS \\
    & & & & & & FRESCO & UVCANDELS \\
         & & & & & & MEOW &   \\
    \hline
    \hline
    Total & & & 275(119) & 542 &  & &
    \enddata
    \tablenotetext{a}{Total MIRI area in arcmin$^2$ is quoted with number of pointings in parentheses}
    \tablenotetext{b}{The MEGA Survey contains a unique set of 25 MIRI pointings in 4 bands (F770W, F1000W, F1500W, F2100W) with high overlap with the MINERVA NIRCam observations, bringing the total MIRI coverage in AEGIS to 35 pointings covering 80 arcmin$^{2}$} 
    \tablenotetext{c}{The MEOW Survey contains a unique set of 30 MIRI pointings in 2 bands (F1000W, F2100W) with high overlap with the MINERVA NIRCam observations, bringing the total MIRI coverage in GOODS-N to 37 pointings covering 85 arcmin$^{2}$} 
\end{deluxetable*}

\section{Observational Design}
\label{sec:observations}
\subsection{The MINERVA Fields}
The MINERVA fields are four of the five well-studied CANDELS fields: UDS, COSMOS, AEGIS, and GOODS-N \citep{grogin11,koekemoer11}.  The fifth CANDELS field, GOODS-S, was not included as part of MINERVA as it already has a rich layout of JWST NIRCam/MIRI observations from the JADES survey, including additional medium band coverage from  FRESCO \citep{oesch23}, JEMS \citep{williams23} and the JADES Origins Field \citep{eisenstein23b}.  The CANDELS fields are within some of the best-studied extragalactic fields with the wealth of photometric coverage spanning from the X-rays to the radio. They have also been the the focus of extensive observing campaigns with both HST and JWST. 
Relevant details of each of the MINERVA fields is presented in Table \ref{tab:field_info}.  In the next section we discuss key aspects of the existing data in each field and the geometric layout of each.

\subsubsection{The CANDELS/PRIMER UDS Field}
\begin{figure*}
    \centering
\includegraphics[width=184mm]{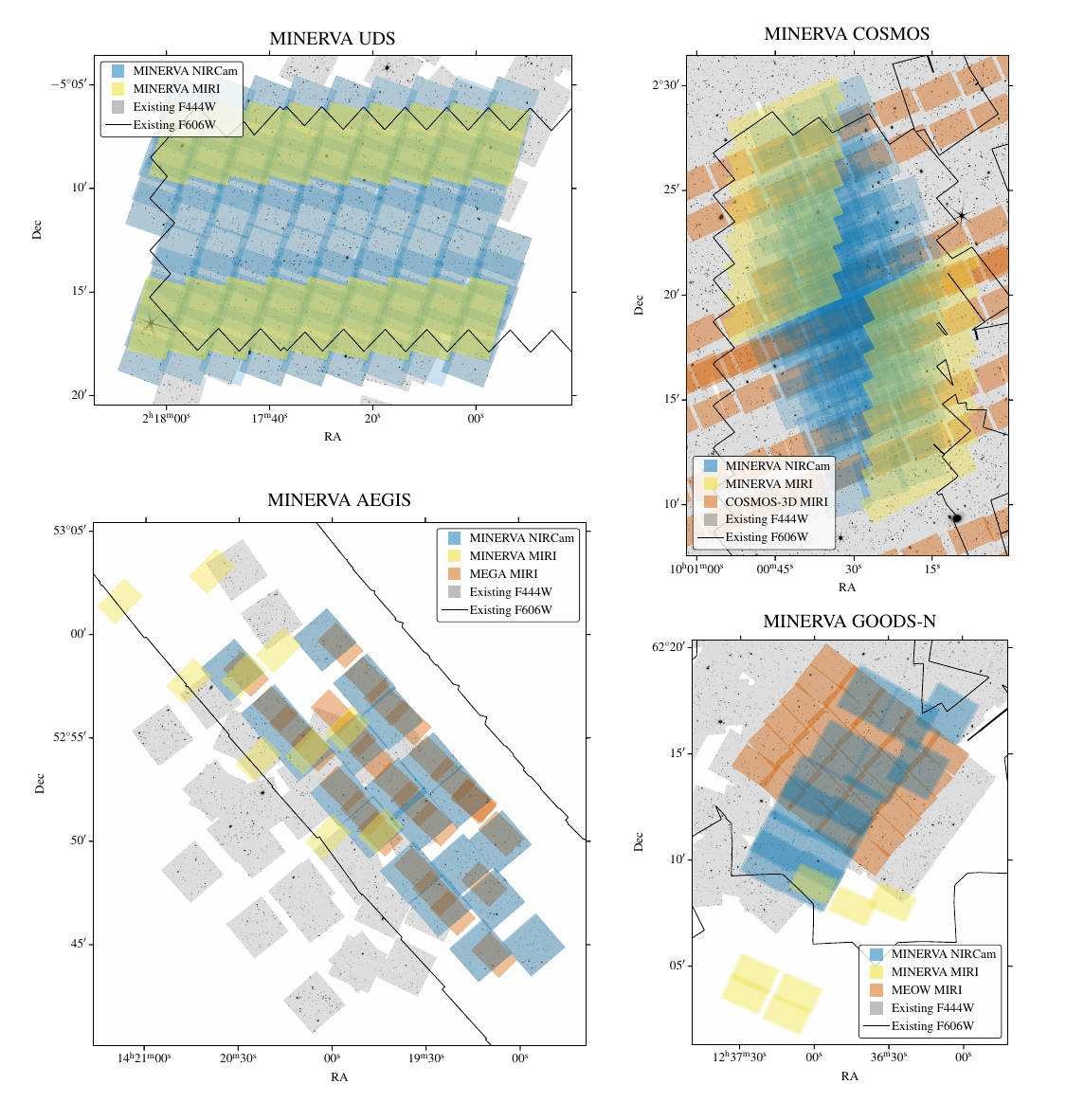}
\caption{The layout of the MINERVA coverage of the UDS, COSMOS, AEGIS and GOODS-N fields.  The background greyscale images are the existing F444W imaging from PRIMER (UDS and COSMOS), CEERS (AEGIS), and JADES (GOODS-N) and includes  additional parallel observations.  The wider-field F444W COSMOS-Web data are also shown for the COSMOS field.  Blue squares are the MINERVA NIRCam coverage in 8 medium bands, and yellow squares are the MINERVA MIRI coverage in two MIRI filters.  Existing F606W HST imaging, primarily from CANDELS, is shown as the solid line.  For COSMOS the F1000W/F2100W coverage of the field from the COSMOS-3D survey is shown in orange.  For AEGIS the footprint of the F770W, F1000W, F1500W, and F2100W MIRI imaging from the MEGA survey is shown in orange.  For GOODS-N the footprint of the F1000W and F2100W MIRI imaging from the MEOW survey is shown in orange.}
    \label{fig:all_coverage}
\end{figure*}

The UDS field is the largest field by area observed as part of the PRIMER and MINERVA surveys.  As Table \ref{tab:field_info} shows, it comprises 54 total MIRI pointings in prime with corresponding NIRCam pointings in parallel, totaling 125 and 234 arcmin$^2$, respectively. Figure \ref{fig:all_coverage} shows the layout of observations on the field. The greyscale is the PRIMER F444W imaging with a few additional archival pointings from pure parallel surveys.  Overlaid in blue is the location of MINERVA medium band imaging and in yellow is the MINERVA MIRI coverage.  Figure \ref{fig:all_coverage} demonstrates that the mosaic is built by tiling MIRI observations with nearly zero overlap in two parts along the major axis of the field. Given the major axis of the field runs east-west, this MIRI configuration creates a NIRCam parallel mosaic with modest overlaps and a large area. In contrast, the major axis of the PRIMER-COSMOS field runs 90 degrees perpendicular, in the north-south direction, and therefore the same  minimally-overlapping MIRI mosaic leads to larger overlaps in the NIRCam mosaic, and thus greater depth and smaller area (see Figure \ref{fig:all_coverage}). Therefore, although the UDS NIRCam mosaic has just a few more pointings than COSMOS, it has nearly double the NIRCam area -- with the tradeoff of having about half the effective exposure time per pixel and therefore being $\sim$ 0.4 mag shallower in some regions. 

The UDS field has substantial existing space-based imaging coverage prior to the MINERVA medium band observations.  Details of the various large imaging campaigns are listed in Table \ref{tab:UDS_info}.  The field was originally defined as part of the UKIDSS survey \citep{lawrence07} and then a smaller subset of the field was used by the CANDELS survey \citep{grogin11,koekemoer11}.  The CANDELS footprint is what is used by MINERVA and has coverage in the WFC3 F125W and F160W filters as part of the CANDELS program.  It has additional WFC3/IR coverage in the F140W filter taken as part of the 3D-HST survey \citep{brammer12,momcheva16}.  The CANDELS survey also obtained F606W and F814W imaging in parallel that has good overlap with much of the field and this is shown in Figure \ref{fig:all_coverage} as the solid black line.  UDS was also followed up in the F435W filter with similar coverage as in CANDELS as part of PID: 16872 (PI: Grogin) and therefore has good coverage in the three primary HST optical bands.  

The UDS field was chosen as one of the PRIMER fields \citep[e.g.,][]{donnan24} in JWST Cycle-1 and has considerable coverage in both NIRCam and MIRI from that program.  As mentioned previously, the PRIMER observations consisted of 54 slightly overlapping MIRI observations in prime in the F770W and F1800W filters, with corresponding NIRCam parallels in F090W, F115W, F150W, F200W, F277W, F356W, F410M, and F444W.  The exposure times in MIRI are $\sim$ 30 minutes per pixel in both filters, which gives NIRCam parallel filters (taken two per MIRI filter at approximately half the exposure time each) of $\sim$ 14 minutes each, and where NIRCam tiles overlap give a total exposure time of $\sim$ 28 minutes.  Approximately 75\% of the mosaic has 28 minute or greater coverage.

The MINERVA observations are based on a copy of the PRIMER APT footprint, but using different filters and a different dither/read strategy. With MIRI, MINERVA will obtain imaging in the F1280W and F1500W filters.  In order to improve data reduction in MIRI, specifically background subtraction and masking of bad pixels, MINERVA uses a 3 point dither strategy for the MIRI observations compared to the 2 point dither strategy used for MIRI in PRIMER.  This requires slightly longer observing time and therefore total integrations in these filters are slightly longer, $\sim$ 38 minutes per filter.  By using the same PA as PRIMER in parallel, MINERVA obtains medium band imaging in the F140M, F162M, F182M, F210M, F250M, F300M, F360M, and F460M filters with the same footprint as the PRIMER broadbands.

Table \ref{tab:UDS_info} also shows some of the major spectroscopic campaigns within the UDS field.  These include CAPERS (PID: 6368, PI: Dickinson), EXCELS \citep{carnall24}, RUBIES, \citep{degraaff25}, as well as the NIRSpec GTO Wide Survey \citep{maseda24}, PID: 2565 (PI: Glazebrook) and Mirage or Miracle (PID:5224, PI: Oesch \& Naidu).  Listed in the table are the total number of MSA slits placed, and in brackets are the number of class 3 (high confidence) spectroscopic redshifts currently on the DAWN JWST archive\footnote{{https://s3.amazonaws.com/msaexp-nirspec/extractions/public\_prelim\_v4.2.html}}.  Overall, these total an impressive $\sim$ 3100 NIRSpec spectroscopic redshifts in the field.

\begin{deluxetable*}{lcccccll}
    \tablecaption{Summary of Existing/Planned UDS Field Observations
  	     \label{tab:UDS_info}
    }     
    \tablewidth{0pt}
    \tablehead{
    \colhead{Instrument} & \colhead{Filters} & \colhead{Exptime} & \colhead{N} & \colhead{Coverage} & \colhead{Proposal ID} & \colhead{Survey}  & \colhead{PI} \\ 
    \colhead{} & \colhead{} & \colhead{(min)} & \colhead{Pointings} & \colhead{} & \colhead{} & \colhead{} & \colhead{} 
    }
    \startdata
ACS & F435W & 32 & 40 & High & 16872 & -- & Grogin \\
ACS & F606W, F814W & 21, 42 & 44 & High & 12064 & CANDELS & Faber \\
\hline
WFC3IR & F125W, F160W & 21, 42 & 44 & High & 12064 & CANDELS & Faber \\
WFC3IR & F140W & 15 & 28 & High & 12328 & 3D-HST & van Dokkum \\
\hline
NIRCam & F090W, F115W, F150W, F200W, & 14 - 28 & 54 & High & 1837 & PRIMER & Dunlop \\
 &  F277W, F356W, F410M, F444W &  &  &  &  &  &  \\
 NIRCam & F140M, F162M, F182M, F210M, & 18 - 36 & 54 & High & 7814 & MINERVA & Muzzin, Suess \\
 &  F250M, F300M, F360M, F460M &  &  &  &  &  &  \& Marchesini \\
\hline
MIRI & F770W, F1800W & 30 & 54 & Medium &  1837 & PRIMER & Dunlop \\
MIRI & F1280W, F1500W & 38 & 54 & Medium & 7814 & MINERVA & Muzzin, Suess \\ 
 &  &  &  &  &  &  &  \& Marchesini  \\
 \hline \hline
 Instrument & Grating & Exptime & N$_{\rm{approx}}$ & & Proposal ID & Survey & PI \\ 
 & & (min) & Spectra & &&& \\ \hline
NIRSpec & PRISM, G235H, G395H & 26-40 & 710 (390) & &1215 & GTO-WIDE & Luetzgendorf \\
NIRSpec & PRISM & 22-66 & 240 (150) & & 2565 & & Glazebrook \\
NIRSpec & G140M, G235M, G395M & 240-655 & 350 (300) & & 3543& EXCELS & Carnall,Cullen \\
NIRSpec & PRISM, G395M & 48-95 & 3200 (1560) &  &4233 & RUBIES & de Graaff, Brammer\\
NIRSpec & PRISM & 264 & 230 & &5224 & MoM & Oesch, Naidu\\
NIRSpec & PRISM & 95-285 & 1100 (700) &  &  6368 & CAPERS & Dickinson \\ 
\hline
\enddata
\tablecomments{The number of spectra for each program corresponds to the total number of spectra in the public v4.2 DJA table at \url{https://s3.amazonaws.com/msaexp-nirspec/extractions/public_prelim_v4.2.html} as retrieved July 1 2025, rounded to the nearest ten. Numbers in parenthesis show the number with high-quality (grade$=$3) redshifts. Objects with spectra in multiple gratings may be double-counted; programs not yet available in DJA may not be shown. Exposure times show the minimum and maximum exposure times in the DJA across all gratings; some objects may be observed on multiple masks, leading to longer total times than typical objects in the survey.}
 \end{deluxetable*}

\subsubsection{The CANDELS/PRIMER COSMOS Field}

The CANDELS/PRIMER COSMOS field is the second largest field in MINERVA (see Table \ref{tab:field_info}). Like the UDS field, the MINERVA observations of COSMOS are a duplication of the PRIMER APT footprint and PA, to obtain similar coverage. Also similar to the UDS observations, MINERVA obtains MIRI imaging in the F1280W and F1500W filters, and with NIRCam in the F140M, F162M, F182M, F210M, F250M, F300M, F360M, and F460M filters.  Again a 3-point dither pattern with MIRI is chosen which results in slightly longer exposure times in both MIRI and NIRCam.  The layout of the MINERVA observations of the COSMOS field is shown in Figure \ref{fig:all_coverage} and all ancillary HST and JWST data is reported in Table \ref{tab:COSMOS_info}.

The $\sim$ 90~deg difference in major axis of the survey area between UDS and COSMOS means that the NIRCam mosaic of the two fields has quite different characteristics even though the total number of MIRI pointings and overall MIRI area are similar.  The COSMOS field has significant overlaps in NIRCam, and so while it has 48 unique non-overlapping MIRI pointings, compared to the 54 MIRI pointings in UDS, the NIRCam area is only 144 arcmin$^2$ compared to 234 arcmin$^2$ in UDS.  Naturally, with more overlaps the COSMOS NIRCam data is somewhat deeper.  

There is excellent supporting data in the COSMOS field.  Similar to UDS, it has NIRCam broadband coverage from PRIMER in F090W, F115W, F150W, F200W, F277W, F356W, F410M, and F444W.  The COSMOS-Web survey \citep{casey23} covers both the MINERVA region (144 arcmin$^2$) and a much larger surrounding region in F090W, F115W, F277W, and F444W at a shallower depth ($\sim$ 9 - 18 min/pointing, depth $\sim$ 27 - 28 mag).  Likewise, even more depth is obtained in F115W, F200W, and F356W as part of the COSMOS-3D survey (PID: 5893, PI: Kakiichi).  In particular, the F200W imaging from COSMOS-3D more than doubles the PRIMER exposure time in that filter as it is taken with the NIRCam Short Wavelength (SW) channel while the NIRCam Long Wavelength (LW) channel obtains WFSS spectroscopy.

The COSMOS field also has optical supporting data from HST in the F606W and F814W bands from the CANDELS survey, and the original COSMOS survey adds additional depth in the F814W filter.  Like the UDS field, there is also WFC3/IR data in F125W and F160W from CANDELS, as well as F140W imaging from 3D-HST.  Notably the COSMOS field is part of the UVCANDELS project \citep{wang24b,mehta24}.  This adds quite deep ($\sim$ 2 hour integration) observations in F275W with WFC3/UVIS with parallel F435W observations with ACS. Lastly, the MIRI coverage in the COSMOS footprint is well-augmented by coverage in F1000W and F2100W obtained as part of the COSMOS-3D survey (see Figure \ref{fig:all_coverage}).  

The COSMOS field also has multiple NIRSpec surveys covering the field, the largest of which is the NIRSpec GTO-WIDE \citep{maseda24}, but there is also the Blue Jay \citep{davies23} and AURORA \citep{shapley25} surveys.  There are also several other smaller MSA spectroscopic programs, PID: 1879 (PI: Curti), PID: 6585 (PI: Coulter) and PID: 2565 (PI: Glazebrook).  In total there are $\sim$ 1110 grade 3 NIRSpec spectroscopic redshifts from the DJA within the field.


\begin{deluxetable*}{lcccccll}
    \tablecaption{Summary of Existing/Planned COSMOS Field Observations
    \label{tab:COSMOS_info}
  	}
    \tablewidth{0pt}
    \tablehead{
    \colhead{Instrument} & \colhead{Filters} & \colhead{Exptime} & \colhead{N} & \colhead{Coverage} & \colhead{Proposal ID} & \colhead{Survey}  & \colhead{PI} \\ 
    \colhead{} & \colhead{} & \colhead{(min)} & \colhead{Pointings} & \colhead{} & \colhead{} & \colhead{} & \colhead{} 
    }
    \startdata
UVIS/ACS & F275W, F435W & 125, 120 & 18 & High & 15647 & UVCANDELS & Teplitz \\
ACS & F606W, F814W & 21, 42 & 44 & High & 12440 & CANDELS & Faber \\
ACS & F814W & 34 & 30 & High & 9822 & COSMOS & Scoville \\
\hline
WFC3IR & F125W, F160W & 21, 42 & 44 & High & 12440 & CANDELS & Faber \\
WFC3IR & F140W & 15 & 28 & High & 12328 & 3D-HST & van Dokkum \\
\hline
NIRCam & F090W, F115W, F150W, F200W, & 14 - 42 & 48 & High & 1837 & PRIMER & Dunlop \\
 &  F277W, F356W, F410M, F444W &  &  &  &  &  &  \\
 NIRCam & F090W, F115W, F277W, F444W & 9 - 17 & 50 & High &  5893 & COSMOS-Web & Kartaltepe \\
NIRCam & F115W, F200W, F356W & 15, 30, 15 & 50 & Medium &  5893 & COSMOS-3D & Kakiichi \\
 NIRCam & F140M, F162M, F182M, F210M, & 18 - 54 & 54 & High & 7814 & MINERVA & Muzzin, Suess \\
 &  F250M, F300M, F360M, F460M &  &  &  &  &  &  \& Marchesini \\
\hline
MIRI & F770W, F1800W & 30 & 48 & Medium &  1837 & PRIMER & Dunlop \\
MIRI & F1000W, F2100W & 15, 30 & 50 & Medium &  5893 & COSMOS-3D & Kakiichi \\
MIRI & F1280W, F1500W & 38 & 54 & Medium & 7814 & MINERVA & Muzzin, Suess \\ 
 &  &  &  &  &  &  &  \& Marchesini \\
\hline  \hline
 Instrument & Grating & Exptime & N$_{\rm{approx}}$ & & Proposal ID & Survey & PI \\ 
 & & (min) & Spectra & &&& \\ \hline
NIRSpec & PRISM, G235H, G395H & 27-40 & 760 (420) & & 1214 & GTO-WIDE & Luetzgendorf \\
NIRSpec & G140M, G235M, G395M & 97-778 & 150 (140) & & 1810 & Blue Jay & Belli \\
NIRSpec & G140M, G235M, G235H & 175-1896 & 120 (110) & & 1879 &  & Curti \\
NIRSpec &  G140M, G235M, G395M & 248-730 & 50 (40) & & 1914 & AURORA & Shapley, Sanders \\
NIRSpec & PRISM & 33 & 310 (220) & &2565 & & Glazebrook \\
NIRSpec & PRISM & 264 &280 & &5224 & MoM & Oesch, Naidu\\
NIRSpec & G235H & 382 & & & 5427 && Davies \\
NIRSpec & PRISM & 47 & & & 5545 && Barrufet \\
NIRSpec & PRISM & 48-95 & & & 6368 & CAPERS & Dickinson \\
NIRSpec & PRISM & 98-295 & 310 (180) & &6585 & & Coulter \\
\hline 
\enddata
\tablecomments{The number of spectra for each program corresponds to the total number of spectra in the public v4.2 DJA table at \url{https://s3.amazonaws.com/msaexp-nirspec/extractions/public_prelim_v4.2.html} as retrieved July 1 2025, rounded to the nearest ten. Numbers in parenthesis show the number with high-quality (grade$=$3) redshifts. Objects with spectra in multiple gratings may be double-counted; programs not yet available in DJA may not be shown. Exposure times show the minimum and maximum exposure times in the DJA across all gratings; some objects may be observed on multiple masks, leading to longer total times than typical objects in the survey.}
 \end{deluxetable*}
 \begin{deluxetable*}{lcccccll}
    \tablecaption{Summary of Existing/Planned AEGIS Field Observations
    \label{tab:AEGIS_info}
  	}
    \tablewidth{0pt}
    \tablehead{
    \colhead{Instrument} & \colhead{Filters} & \colhead{Exptime} & \colhead{N} & \colhead{Coverage} & \colhead{Proposal ID} & \colhead{Survey}  & \colhead{PI} \\ 
    \colhead{} & \colhead{} & \colhead{(min)} & \colhead{Pointings} & \colhead{} & \colhead{} & \colhead{} & \colhead{} 
    }
    \startdata
UVIS/ACS & F275W, F435W & 125, 120 & 20 & High & 15647 & UVCANDELS & Teplitz \\
ACS & F606W, F814W & 9, 9 & 21 & High & 10134 & EGS & M. Davis \\
ACS & F606W, F814W & 21 - 84, 42 - 168 & 45 & High & 12063 & CANDELS & Faber \\
\hline
WFC3IR & F125W, F160W & 21, 42 & 45 & High & 12063 & CANDELS & Faber \\
WFC3IR & F140W & 15 & 30 & High & 12177 & 3D-HST & van Dokkum \\
\hline
NIRCam & F090W, F470N & 70 & 10 & High &  2234 & -- & Banados \\
NIRCam & F115W, F150W, F200W,  & 50 - 103 & 10 & High & 1345 & CEERS & Finkelstien \\
 &  F277W, F356W, F410M, F444W &  &  &  &  &  &  \\
 NIRCam & F115W, F150W, F200W & 60 & 2 & High &  2275 & -- & Arrabal Haro \\
  &  F277W, F356W, F444W &  &  &  &  &  &  \\
   NIRCam & F115W, F150W, F200W & 28 - 70 & 3 & High &  4287 & -- & Mason, Stark \\
  &  F277W, F356W, F444W &  &  &  &  &  &  \\
   NIRCam & F070W, F140M, F162M, F182M, & 42 - 56 & 10 & High & 8559 & SPAM &  K. Davis, \\
 &  F210M, F300M, F335M, F360M, &  &  &  &  &  &  Larson \\
  &  F430M, F480M &  &  &  &  &  &   \\
 NIRCam & F140M, F162M, F182M, F210M, & 31 - 62 & 10 & High & 7814 & MINERVA & Muzzin, Suess \\
 &  F250M, F460M &  &  &  &  &  &  \& Marchesini \\
\hline
MIRI & F560W, F770W, F1000W,   & 15 - 131 & 8 & Low &  1345 & CEERS & Finkelstein \\
 &  F1280W, F1500W &  &  &  &  &  &   \\
  &  F1800W, F2100W &  &  &  &  &  &   \\
 MIRI & F770W, F1000W & 19 - 34 & 27 & High &  3794 & MEGA & Kirkpatrick \\
   &  F1500W, F2100W &  &  &  &  &  &   \\
MIRI & F560W, F770W, & 63 - 126 & 10 & Medium & 7814 & MINERVA & Muzzin, Suess \\ 
 & F1280W, F1500W  &  &  &  &  &  &  \& Marchesini \\ \hline \hline
 Instrument & Grating & Exptime & N$_{\rm{approx}}$ & & Proposal ID & Survey & PI \\ 
 & & (min) & Spectra & &&& \\ \hline
NIRSpec & PRISM, G235H, G395H & 27-80 & 910 (440) & & 1213 & GTO-Wide & Luetzgendorf \\
NIRSpec & PRISM, G140M, F235M, G395M & 51-153 & 1080 (800) && 1345 & CEERS & Finkelstein \\
NIRSpec & PRISM & 33 & 120 (70) & & 2565 & & Glazebrook \\
NIRSpec & PRISM & 306 & 150 (120) & & 2750 &  & Arrabal Haro \\
NIRSpec & PRISM, G395M & 66-219 & 140 (100) & &4106&& Nelson, Labb\'e\\
NIRSpec & PRISM, G395M & 47 & 1500 (860) & & 4233& RUBIES & de Graaff, \\
 &  &  &  & & &  & Brammer \\
NIRSpec & G395M, F140H & 59-233 & 100 (60) && 4287 & & Mason, Stark \\
NIRSpec & PRISM & 47-285 & 2300 (1410) & &6368 & CAPERS & Dickinson
\enddata
\tablecomments{The number of spectra for each program corresponds to the total number of spectra in the public v4.2 DJA table at \url{https://s3.amazonaws.com/msaexp-nirspec/extractions/public_prelim_v4.2.html} as retrieved July 1 2025, rounded to the nearest ten. Numbers in parenthesis show the number with high-quality (grade$=$3) redshifts. Objects with spectra in multiple gratings may be double-counted; programs not yet available in DJA may not be shown. Exposure times show the minimum and maximum exposure times in the DJA across all gratings; some objects may be observed on multiple masks, leading to longer total times than typical objects in the survey.}
 \end{deluxetable*}
 \subsubsection{The CANDELS/CEERS AEGIS Field}
 The CANDELS/CEERS field is unique within the MINERVA survey in that it overlaps with many previous HST and JWST programs and therefore the supporting data is arguably the most extensive, and the NIRCam data is the deepest by combination with other programs.  However, this superiority in supporting data comes with the challenge that the footprint of each instrument in each filter is in some cases quite complex.  An overview of the field layout of the AEGIS field is shown in Figure \ref{fig:all_coverage}, and a compilation of these observations is shown in Table \ref{tab:AEGIS_info}.

 In terms of JWST imaging, the field is anchored by the CEERS ERS program \citep{finkelstein25}.  CEERS contains imaging in 7 NIRCam filters spanning 10 different pointings along the Extended Groth Strip (EGS or AEGIS) original project.  Additional F090W and F470N imaging in the same footprint was obtained in PID 2234 (PI: Banados) in Cycle-1 to bring the NIRCam broadband coverage to 8 filters, similar to PRIMER.  Subsequently, several other programs have obtained multi-filter NIRCam observations in parallel to spectroscopic observations (e.g., PID 2275, PI: Arrabal Haro, and PID 4287, PI: Mason \& Stark). The field has also been part of extensive pure parallel observations taken as part of the PANORAMIC \citep{williams25}, SAPPHIRES \citep{sun25} and BEACON \citep{morishita25} programs.  This leads to good coverage over much of the field with deep broadband imaging.  Many of the pure parallel imaging campaigns occupy a region not covered by the MINERVA MIRI or NIRCam medium band observations (see Figure \ref{fig:all_coverage}).

 In addition to the MINERVA medium band observations, the SPAM program (PID 8559, PI: K. Davis, Co-PI: R. Larsen) was approved in Cycle-4.  SPAM will obtain 9 filter medium band imaging (F140M, F162M, F182M, F210M, F300M, F335M, F360M, F430M, F480M), as well as one broadband filter (F070W).  These are observed to integration times of 42 - 56 minutes.  Due to the overlap in filter coverage between MINERVA and SPAM, we chose to modify the MINERVA observing strategy in this field.  Given the lack of additional filters on the NIRCam SW, MINERVA will observe the same 4 medium bands as the other fields (F140M, F162M, F182M, F210M).  When combined with the SPAM medium-band imaging, this will lead to imaging with $\sim$ 70 minutes integration time.  For the NIRCam LW observations, the MINERVA F300M and F360M filters have been removed, as they are observed in SPAM, and instead this time is used to double the integration time on the F250M and F460M filters to 62 minutes each.  These two filters are not observed as part of SPAM.  Therefore the combined MINERVA+SPAM observations of the AEGIS field will have integration times of 56 - 72 minutes, notably deeper than any other MINERVA field. AEGIS will also have observations in all 12 NIRCam medium bands, the only field in the survey to have full medium band coverage.

The HST UV, optical and NIR coverage of the AEGIS field is excellent.  The field was part of the original EGS project which obtained very wide field ACS observations in F606W and F814W \citep{davis07}.  It was subsequently observed with much deeper F606W and F814W ACS observations as part of the CANDELS survey.  CANDELS also obtained WFC3 F125W and F160W observations of most of the field \citep{grogin11,koekemoer11}.  F140W observations were also obtained as part of the 3D-HST survey \citep{brammer12,momcheva16}.  AEGIS was also part of the UVCANDELS survey which obtained very deep WFC3/UVIS imaging in F275W while obtaining ACS F435W imaging in parallel \citep{wang24b, mehta24}.

The MIRI data in the AEGIS field is also comprehensive. The original CEERS observations contain a complex structure of MIRI observations in nearly every MIRI filter, however, no pointings contain all filters \citep[see][]{finkelstein25}.  To complement this MINERVA will obtain MIRI imaging in the same 10 footprints as CEERS, however, it will use filters that have not yet been observed in those footprints as part of CEERS.  This results in very deep F560W (126 minutes integration time) in some regions, where in others it is split into two or more other filters: F770W, F1280W or F1500W, depending on where those filters have already been observed.  Figure \ref{fig:all_coverage} shows the MINERVA AEGIS field layout including the NIRCam and MIRI coverage.  Only approximately 50\% of the CEERS/MINERVA MIRI data have good overlap with the NIRCam observations (i.e., only $\sim$ 5 MIRI pointings).  

Although the CEERS/MINERVA MIRI coverage is somewhat disjoint with the NIRCam coverage, the MEGA survey \citep{backhaus25} contains 25 pointings with F770W, F1000W, F1500W, and F2100W imaging with excellent overlap of the NIRCam imaging.  The MEGA MIRI footprints are also shown in Figure \ref{fig:all_coverage}, and these combined with the CEERS/MINERVA MIRI data give the AEGIS field excellent coverage in MIRI.

There have also been several large NIRSpec campaigns in the AEGIS field and these are listed in Table \ref{tab:AEGIS_info}.   These include the original CEERS spectroscopy \citep{finkelstein25}, CAPERS (PID: 6368, PI: Dickinson), RUBIES, \citep{degraaff25}, as well as the NIRSpec GTO Wide Survey \citep{maseda24} and PID: 2565 (PI: Glazebrook).  Also in AEGIS are several smaller programs such as PID: 2750 (PI: Arrabal Haro), PID: 4287 (PI: Mason \& Stark), and PID: 4106 (PI:Nelson \& Labb\'e).  Overall, these total an impressive $\sim$ 3860 NIRSpec spectroscopic redshifts in the field, the largest total in any MINERVA field.

Overall, with many existing and planned programs in the AEGIS field it promises to be one of the most data-rich fields within the MINERVA survey: Remarkably, some regions of the field will have deep imaging in up to 34 space-based filters from 0.28 $\micron$ -- 21 $\micron$.


\subsubsection{The CANDELS/JADES GOODS-N Field}
The JADES GOODS-N field is the smallest field by number of pointings in MINERVA.  However, as it contains the HDF-N, it has been covered by many previous programs and has by a large margin the strongest supporting HST data.   The GOODS-N field has significant existing JWST data taken as part of both the JADES \citep{eisenstein23a,eisenstein23b} and FRESCO surveys \citep{oesch23}.  Information on coverage from these surveys is listed in Table \ref{tab:GOODSN_info}.

The main MINERVA footprint is defined based on the original JADES imaging and is shown in Figure \ref{fig:all_coverage}. In particular MINERVA covers the 7 pointings of broadband (F090W, F115W, F150W, F200W, F277W, F356W, and F444W) and 2 medium bands (F335M, F410M) imaging taken as JADES Medium NIRCam Prime.  Those pointings are in a footprint running mostly north-south in the middle of the field.  Like the other MINERVA survey fields, the APT layout and PA of these pointings was duplicated  to maximize the overlap between medium and broadband data.  Four additional NIRCam imaging fields were obtained in GOODS-N as parallel observations by JADES when NIRSpec spectroscopy was performed on the seven initial fields, however, these are not included as part of MINERVA.  

Also covering part of the MINERVA footprint are medium band observations in the F182M and F210M filters from the FRESCO survey.  These observations are  deep (up to 75 min per pixel) as they are taken in parallel with long exposure in the NIRCam LW grism.  Their overlap with the MINERVA footprint is good, $\sim$ 50\%, but not complete, so in some areas the medium band data in those filters have augmented depth from FRESCO.

The supporting HST data in the GOODS-N field is the best within the MINERVA survey (see Table \ref{tab:GOODSN_info}).  It benefits from deep UVIS observations in F275W, F336W, and F350LP taken as part of CANDELS but also as part of an HST UV initiative program (PID 13872, PI: Oesch).  Moreover, there is additional very deep ACS coverage in F435W taken in parallel with that program.  There is also deep coverage in F606W and F814W taken as part of CANDELS, as well as F775W and F850LP data taken in parallel as part of a WFC3 G141 grism proposal (PID 11600, PI: Weiner), and the PANS supernova program (PIDs 10189, 10339, PI: Riess).   

The MIRI coverage of the GOODS-N field from JADES and MINERVA is notably very modest, by far the least of the four MINERVA fields.  Prior to the MINERVA observations the only imaging data comes from JADES in parallel in the F770W and F1500W bands.  But this is taken for only 3 pointings of the 7, as NIRSpec spectroscopy was performed in parallel with the four other pointings.  Moreover, the footprint of the field is such that only one of the fields has overlap with the NIRCam medium band imaging.  With MINERVA we perform MIRI imaging in the F1280W and F1800W bands (note the PRIMER fields use F1280W and F1500W) so as to not duplicate the F1500W and add an additional filter. The layout of the MINERVA observations of GOODS-N is shown in Figure \ref{fig:all_coverage}.

Although the JADES/MINERVA MIRI coverage of GOODS-N is modest, the Cycle-3 program MEOW (PID: 5407, PI: Leung) obtained 30 MIRI pointings of F1000W and F2100W observations in the GOODS-N field that have good overlap with the MINERVA medium band observations.  The layout of the MEOW observations are shown in Figure \ref{fig:all_coverage}.

Like the other MINERVA fields, there are also several NIRSpec campaigns conducted in the GOODS-N field.  The largest of these are the JADES survey \citep{eisenstein23a}, the NIRSpec GTO-WIDE \citep{maseda24}, and the AURORA survey \citep{shapley25}.  In total there are $\sim$ 2710 class 3 spectroscopic redshifts in the field in the DJA.

\subsection{NIRCam Filter Choice}

\begin{deluxetable*}{lcccccll}
    \tablecaption{Summary of Existing/Planned GOODS-N Field Observations
  	     \label{tab:GOODSN_info}
    }
    \tablewidth{0pt}
    \tablehead{
    \colhead{Instrument} & \colhead{Filters} & \colhead{Exptime} & \colhead{N} & \colhead{Coverage} & \colhead{Proposal ID} & \colhead{Survey}  & \colhead{PI} \\ 
    \colhead{} & \colhead{} & \colhead{(min)} & \colhead{Pointings} & \colhead{} & \colhead{} & \colhead{} & \colhead{} 
    }
    \startdata
UVIS & F200LP & 4 & 24 & High & 12479 & -- & Hu \\
UVIS & F275W, F336W, F350LP & 24 - 96, 24 - 96, 43 & 130 & High & 12444, 12445 & CANDELS & Faber \\
UVIS/ACS & F275W, F336W, F435W & 159, 318, 396 & 5 & High & 13872 & -- & Oesch \\
ACS & F606W, F814W & 42 - 189 & 130 & High & 12444, 12445 & CANDELS & Faber \\
ACS & F775W & 85 & 28 & High & 11600 & -- & Weiner \\
ACS & F775W, F850LP & 12 - 46 & 31 & High & 10189, 10339 & PANS & Riess \\
\hline
WFC3IR & F125W, F160W & 21, 42 & 44 & High & 12440 & CANDELS & Faber \\
WFC3IR & F140W & 15 & 28 & High & 12328 & 3D-HST & van Dokkum \\
\hline
NIRCam & F090W, F115W, F150W, & 52 - 189 & 7 & High & 1181 & JADES & Eisenstein \\
 &  F200W, F277W, F335M, &  &  &  &  &  &  \\
  &   F356W, F410M, F444W &  &  &  &  &  &  \\
 NIRCam & F182M, F210M, F444W & 15 - 75 & 8 & High &  1895 & FRESCO & Oesch \\
 NIRCam & F140M, F162M, F182M, & 32 & 7 & High & 7814 & MINERVA & Muzzin, Suess \\
 &  F210M, F250M, F300M &  &  &  &  &  &  \& Marchesini \\
  &  F360M, F460M &  &  &  &  &  &  \\
\hline
MIRI & F770W, F1280W & 100 - 150 & 3 & Low &  1181 & JADES & Eisenstein \\
MIRI & F1000W, F2100W & 12, 51 & 3 & High &  1181 & MEOW & Leung \\
MIRI & F1280W, F1800W & 60 & 7 & Low & 7814 & MINERVA & Muzzin, Suess \\ 
 &  &  &  &  &  &  &  \& Marchesini \\ \hline  \hline
 Instrument & Grating & Exptime & N$_{\rm{approx}}$ & & Proposal ID & Survey & PI \\ 
 & & (min) & Spectra & &&& \\ \hline
NIRSpec & PRISM, G140M, G235M, & 51-306 & 2130 (1190) & & 1181& JADES & Eisenstein \\
 &  G395M, G395H \\
NIRSpec & PRISM, G235H, G395H & 27-40 & 1230 (740) & &1211& GTO-WIDE & Isaak \\
NIRSpec & G235H, G395H & 161-875 & 20 (10) & &1871& & Chisholm\\
NIRSpec & G140M, G235M, G395M & 248-729 & 50 (50) & &1914& AURORA & Shapley, Sanders \\
NIRSpec & G395M & 116-233 & 240 (120) & & 2674 & & Arrabal Haro \\
NIRSpec & G395M & 125 & & & 7935 & & Sun
\enddata
\tablecomments{The number of spectra for each program corresponds to the total number of spectra in the public v4.2 DJA table at \url{https://s3.amazonaws.com/msaexp-nirspec/extractions/public_prelim_v4.2.html} as retrieved July 1 2025, rounded to the nearest ten. Numbers in parenthesis show the number with high-quality (grade$=$3) redshifts. Objects with spectra in multiple gratings may be double-counted; programs not yet available in DJA may not be shown. Exposure times show the minimum and maximum exposure times in the DJA across all gratings; some objects may be observed on multiple masks, leading to longer total times than typical objects in the survey.}

 \end{deluxetable*}
 
After fields for MINERVA were chosen, designing the survey required choosing an optimum set of medium bands to observe.
NIRCam has 4 medium bands available on the SW, but 8 medium bands available on the LW.  The unequal distribution of filters between SW and LW means that not all filters in the LW can be used, assuming roughly equal exposure time in all filters.  In order to determine which filters would provide the best improvement in photometric redshifts and stellar masses when added to the existing broadband data, we performed several simulations of potential MINERVA datasets combined with the PRIMER broadband data, as the majority of the area of MINERVA is in the PRIMER fields.  Of the 8 medium bands on the LW only 7 were considered as F410M has already been observed by previous programs in all MINERVA fields.

To perform these simulations we used the extensive medium band catalogs of the UNCOVER/MegaScience \citep{suess24} and CANUCS/Technicolor \citep{sarrouh25} surveys.  These catalogs have deep imaging available in all 8 NIRCam broadband filters (F070W, F090W, F115W, F150W, F200W, F277W, F356W, F444W), all 12 NIRCam medium band filters (F140M, F162M, F182M, F210M, F250M, F300M, F335M, F360M, F410M, F430M, F460M, F480M) as well as 3 deep HST ACS filters (F435W, F606W, F814W), totaling 23 filters and covering a total area of $\sim$ 60 arcmin$^2$.  At $\sim$ 1.5 - 3.0 hours integration time per filter, they are also deeper than the PRIMER and MINERVA exposure times, which are $\sim$ 0.5 hours.  

The simulations make the assumption that the UNCOVER/MegaScience and CANUCS/Technicolor photometry is the ``ground truth," as are the photometric redshifts and stellar masses calculated from it.  This assumption is well justified given that the $z_{phot}$ outlier fraction in both surveys has been shown to be $<$ 5\% over a wide range of stellar mass and redshift with a $\sim$ 1\% scatter \citep[see, e.g.,][]{suess24,sarrouh25}.  This impressive performance is sufficient for a fiducial study of $z_{phot}$ and stellar mass improvements with the addition of medium bands in a shallower survey like MINERVA.

\subsubsection{Filter Choice Effects on Photometric Redshifts}
\label{sec:filter_choice_z}
\begin{figure*}
    \centering
\plottwo{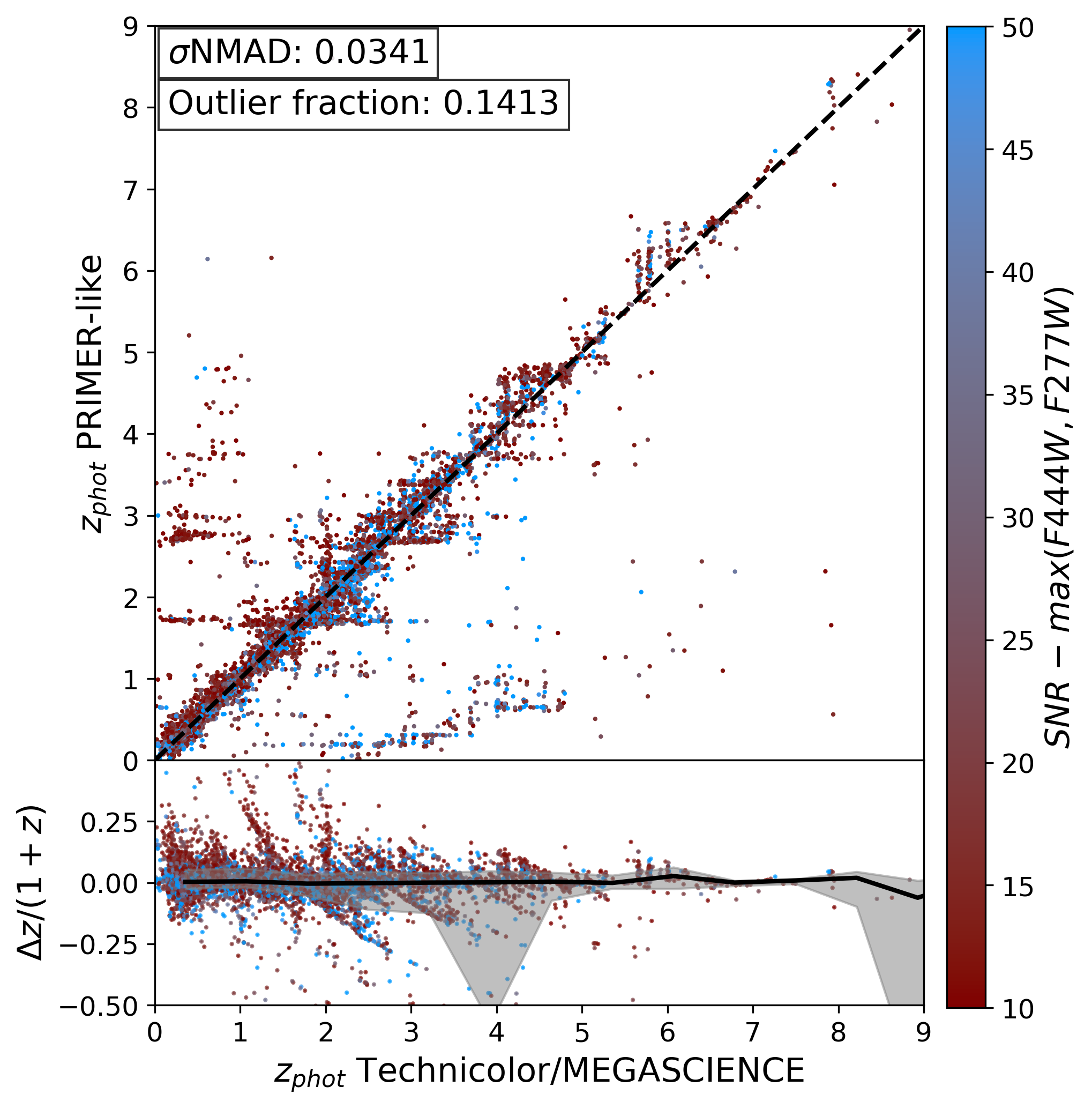}{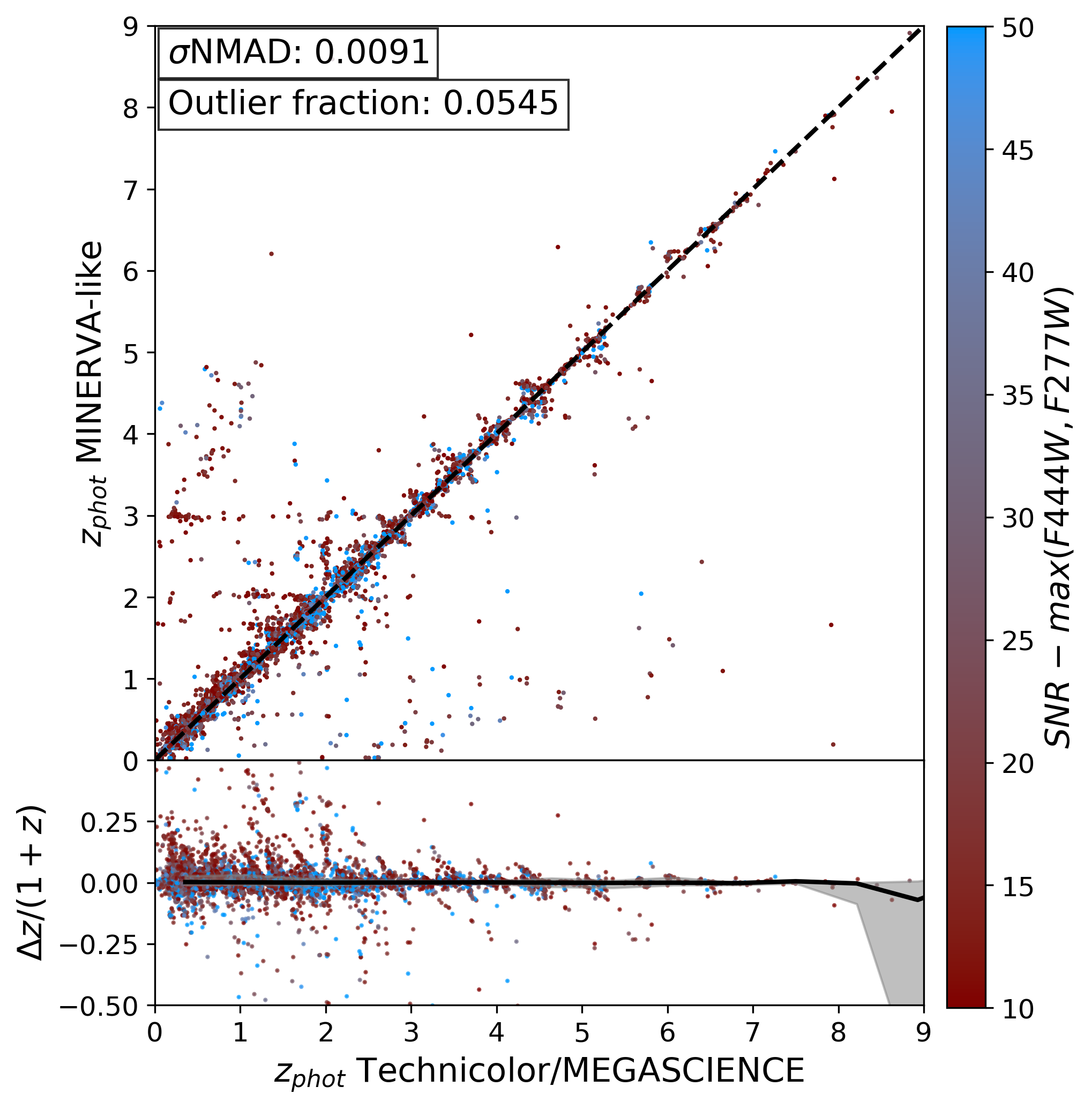}\epsscale{2.5}
\caption{Left Panels: Photometric redshift results from simulating a PRIMER-like survey by degrading existing deep photometry from UNCOVER/MegaScience and CANUCS/Technicolor to PRIMER depth and filter coverage (8 JWST broadband filters + 3 HST ACS filters).  PRIMER-like is plotted on the Y-axis and the "ground truth" from UNCOVER/MegaScience and CANUCS/Technicolor is plotted on the X-axis (23 filters).  Right Panels: The same simulation but adding the 8 MINERVA medium bands at simulated MINERVA depth to the 8 PRIMER broadbands and 3 ACS bands (16 JWST filters + 3 HST filters).  With the additional medium bands MINERVA's photometric redshifts are estimated to have a 3.7x reduction in $\sigma$NMAD scatter, and 2.6x reduction in outlier fraction.}
    \label{fig:PvM_zphot}
\end{figure*}

The first simulation performed was to ascertain the effect on $z_{phot}$ when adding additional medium bands to a PRIMER-like survey.  To simulate PRIMER-like photometry, noise was added to the photometry of both the UNCOVER/MegaScience and CANUCS/Technicolor catalogs such that the photometry has similar depth in each filter to PRIMER, and therefore galaxies of a given magnitude would have equivalent S/N if observed in the PRIMER survey.  In order to benchmark the performance of a PRIMER-like survey {\it without} medium bands the same 8 JWST filters used in PRIMER were selected (F090W, F115W, F150W, F200W, F277W, F356W, F410M, F444W) as well as the 3 HST filters in PRIMER (F435W, F606W, F814W) from the catalogs simulated to PRIMER depth.  From those data $z_{phot}$ were computed using the EAZY code \citep{brammer09} using the modified template set used in the CANUCS survey \citep[see][]{sarrouh25} and then compared to the original ``ground truth" $z_{phot}$ from the full depth 23-band UNCOVER/MegaScience and CANUCS/Technicolor catalogs.  This simulation was labeled ``PRIMER-like", and the $z_{phot}$ compared to those from UNCOVER/MegaScience and CANUCS/Technicolor are shown in the left panels of Figure \ref{fig:PvM_zphot}.  In order to have a fair comparison of the effects of filter differences, and not be dominated by low S/N uncertainties, only galaxies with a S/N $>$ 10 in F277W or F444W (when scaled to PRIMER depth) are plotted.  As Figure \ref{fig:PvM_zphot} shows, the performance of just the broadband photometry is good, with a $\sigma$ Normalized Median Absolute Deviation ($\sigma$NMAD) scatter in $\delta z/(1+z)$ = 3.41\%.  We define the catastrophic outlier fraction as the fraction of galaxies with $(\delta z/(1+z) >$ 3*$\sigma$NMAD.  By this definition, the broadband photometry has an outlier fraction of 14.1\%, driven primarily by the common Lyman/Balmer break degeneracy in fainter galaxies.

To simulate a MINERVA-like survey, the identical 8 JWST broadbands and 3 HST ACS filters were selected from the UNCOVER/MegaScience and CANUCS/Technicolor catalogs, as well as 8 medium bands, also at the PRIMER depth (integration time $\sim$ 0.5 hr per filter).  All four available medium bands were selected in the SW: F140M, F162M, F182M, and F210M, and on the LW:  F250M, F300M, F360M and F460M were selected. These were chosen because when combined with the existing F410M  in PRIMER, they give an even wavelength sampling of the spectral energy distribution (SED) of galaxies.   We combined the 8 broadbands, 8 medium bands and 3 HST bands and determined $z_{phot}$ for each galaxy with EAZY.  The results of the simulated $z_{phot}$ are shown in the right panels of Figure \ref{fig:PvM_zphot}.  We note that in some simulations we explored exchanging F460M for F480M, and F335M for F300M; however, differences from those combinations compared to the current filter choice were insignificant and therefore they are not discussed here.

The simulations in Figure \ref{fig:PvM_zphot} show that adding our chosen 8 medium bands to the PRIMER filters reduces the $\sigma$NMAD and outlier fraction of the $z_{phot}$ to 0.91\% and 5.5\%, respectively\footnote{Note in this and future comparisons the outlier fraction is calculated with respect to the PRIMER-like 3*$\sigma$NMAD, not the much improved MINERVA-like 3*$\sigma$NMAD to give a quantitative assessment of the improved outlier fraction.}. When compared to the broadband PRIMER-like survey alone, this is an impressive improvement in $\sigma$NMAD and outlier fraction by factors of 3.7, and 2.6, respectively.  This shows that the additional spectral resolution of medium bands is extremely valuable for computing $z_{phot}$, particularly for a medium-depth survey like PRIMER.  Notably, the {\it total} integration time of MINERVA is 2x longer than PRIMER alone.  This is because MINERVA is 8 broad plus 8 medium bands, whereas PRIMER is just 8 broadbands.  Therefore it is notable that  the improvement in $\sigma$NMAD and outlier fraction are factors of 3.7 and 2.6, even though the total integration time is only 2x longer.  This shows the power of the medium bands to markedly improve $z_{phot}$ quality without excessive integration time increases, primarily due to the higher spectral resolution.  This typoe of improvement has also been shown in previous medium band surveys \citep{whitaker11,straatman16}.

\begin{figure*}
    \centering
\plottwo{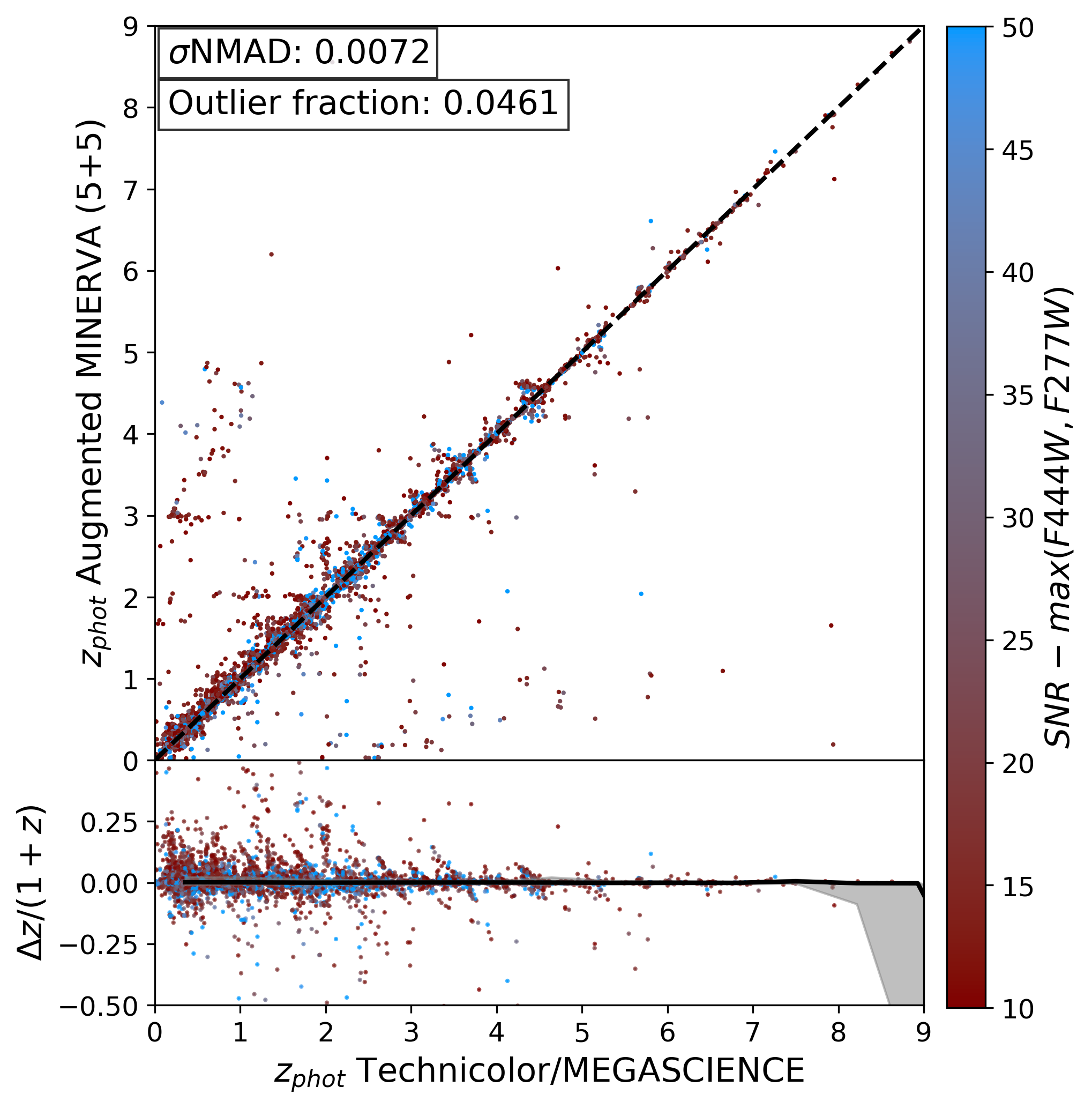}{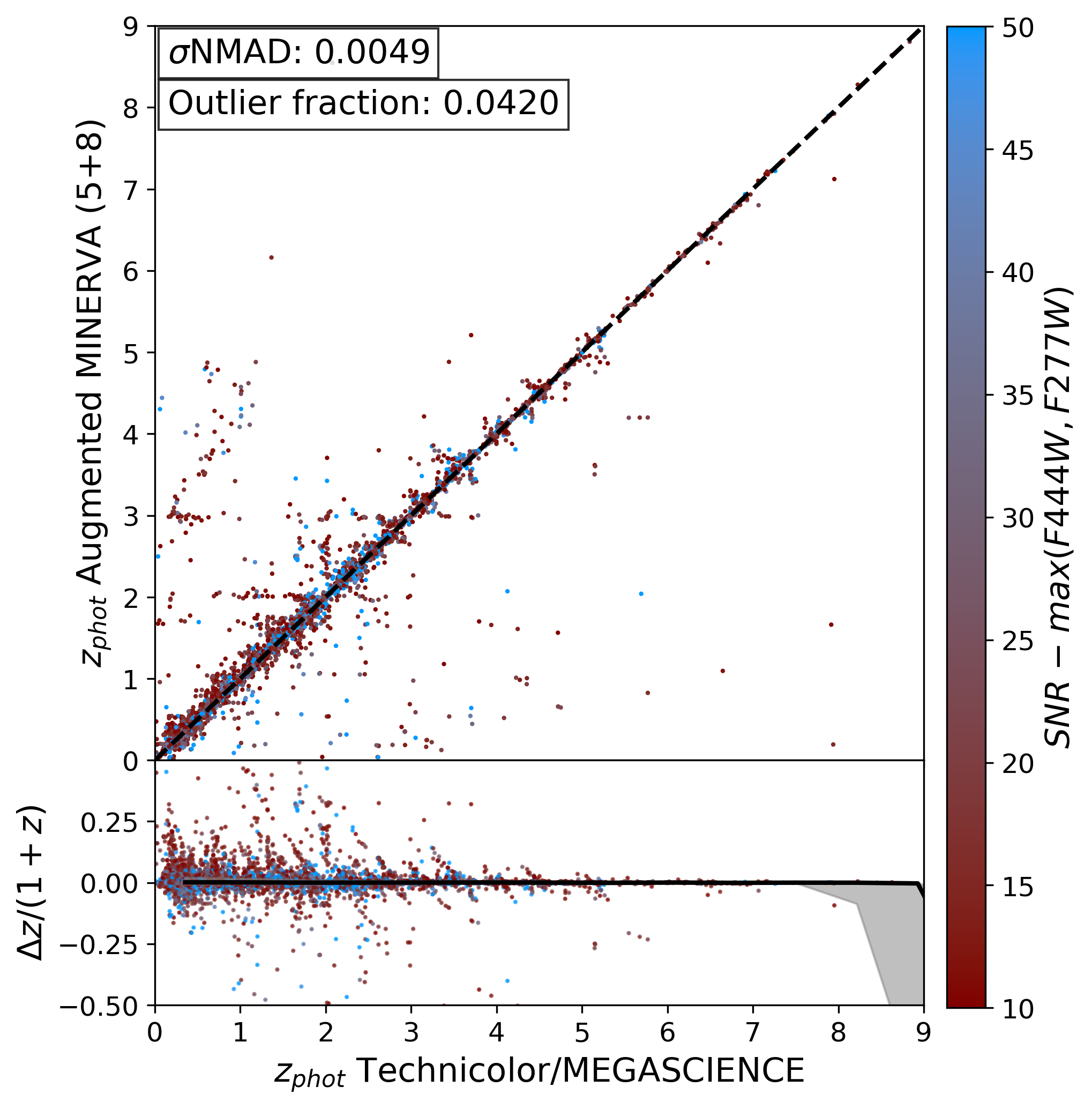}
\caption{Left Panels: Photometric redshift results from simulating a 10 filter MINERVA survey from UNCOVER/MegaScience and CANUCS/Technicolor to PRIMER/MINERVA depth and including the PRIMER filters (18 JWST filters + 3 HST filters).  This is plotted on the Y-axis, and the "ground truth" from UNCOVER/MegaScience and CANUCS/Technicolor is plotted on the X-axis. In this case, MINERVA's $z_{phot}$ are estimated to have had a $\sim$ 1.26x improvement in $\sigma$NMAD and 1.18x reduction in outlier fraction.  Right Panels: The same simulation but for a MINERVA survey that would include 5 filters in the SW (4 medium bands + F070W) and all 8 LW medium bands at simulated MINERVA depth, and including PRIMER broadbands (21 JWST filters + 3 HST filters).  With all medium bands included, MINERVA's $z_{phot}$ would have had a 1.86x improvement in $\sigma$NMAD and 1.30x reduction in outlier fraction.  Given that the required observing time increase for each dataset would be a factor of 1.25 and 1.75, respectively, this shows that the improvement in $z_{phot}$ scales approximately linearly with exposure time. }
    \label{fig:5v4_zphot}
\end{figure*}
When designing the MINERVA observations,  the effect of adding additional filters (at the cost of additional observing time) to the improvement of the $z_{phot}$ was also tested.  Two additional simulated surveys were created, the first was effectively an ``Augmented MINERVA", with 5 filters in the SW and 5 filters in the LW, instead of the previous 4+4.  Given there is not an additional medium band on the SW,  F070W was added as it has not yet been observed in the PRIMER fields.  On the LW side F480M, the reddest medium band, was added.  This filter combination was labeled ``Augemented MINERVA 5+5" and is plotted in the left panels of Figure \ref{fig:5v4_zphot}.  This shows that the addition of two filters to MINERVA would result in an $\sigma$NMAD and outlier fraction that is 0.72\% and 4.6\%, respectively.  If compared to the nominal MINERVA-like survey these are improvements of a factor of 1.26 and 1.18 in $\sigma$NMAD and outlier fraction, respectively.  Given that obtaining the additional filters would require a factor of 1.25 in integration time, it shows that $z_{phot}$ after 8 medium bands are added, the performance improvement scales linearly with additional exposure time on additional medium bands.

The last simulation performed was an ``Idealized MINERVA" that employed all 7 remaining medium bands on the LW, and all 4 medium bands on the SW as well as F070W, plus the additional 8 PRIMER broadbands and existing 3 HST filters.  This is effectively all available NIRCam broad and medium band filters, so is the equivalent of UNCOVER/MegaScience and CANUCS/Technicolor at the reduced MINERVA/PRIMER depth. The results of that simulation are shown in the right panels of Figure \ref{fig:5v4_zphot}.  The $\sigma$NMAD and outlier fraction are 0.49\% and 4.2\%, respectively.  When compared to the MINERVA-like survey these are improvements in NMAD and outlier fraction by factors of 1.86 and 1.30, respectively. This is quite impressive and desirable, however, obtaining such data would require a 1.75x increase in the total integration time of MINERVA.  Once again this shows that the improvement in $z_{phot}$ scales approximately linearly with increased exposure time in additional medium bands.

Overall, the simulations show that the most efficient improvement in $z_{phot}$ compared to a PRIMER-like survey alone is the current MINERVA setup with 4 medium bands in both the SW and LW of NIRCam.  This is a factor of 2 increase in total integration time compared to PRIMER, however it is a factor of 3.7 improvement in $\sigma$NMAD scatter and factor of 2.6 improvement in outlier fraction.  The simulations show that adding additional medium bands to MINERVA could further reduce the scatter and outlier fraction at a rate that scales with integration time.  While this is valuable, it is expensive, and therefore for MINERVA it was determined that the 4+4 SW+LW filter combination was an efficient use of telescope time in Cycle 4 that allowed us to reach our science goals and therefore was chosen as an optimum setup for the survey.  However, the simulations do indicate that obtaining additional medium bands in future cycles could still be a worthwhile use of telescope time as they do provide notable gains in $z_{phot}$ accuracy and precision. 

\begin{figure*}
    \centering
\plottwo{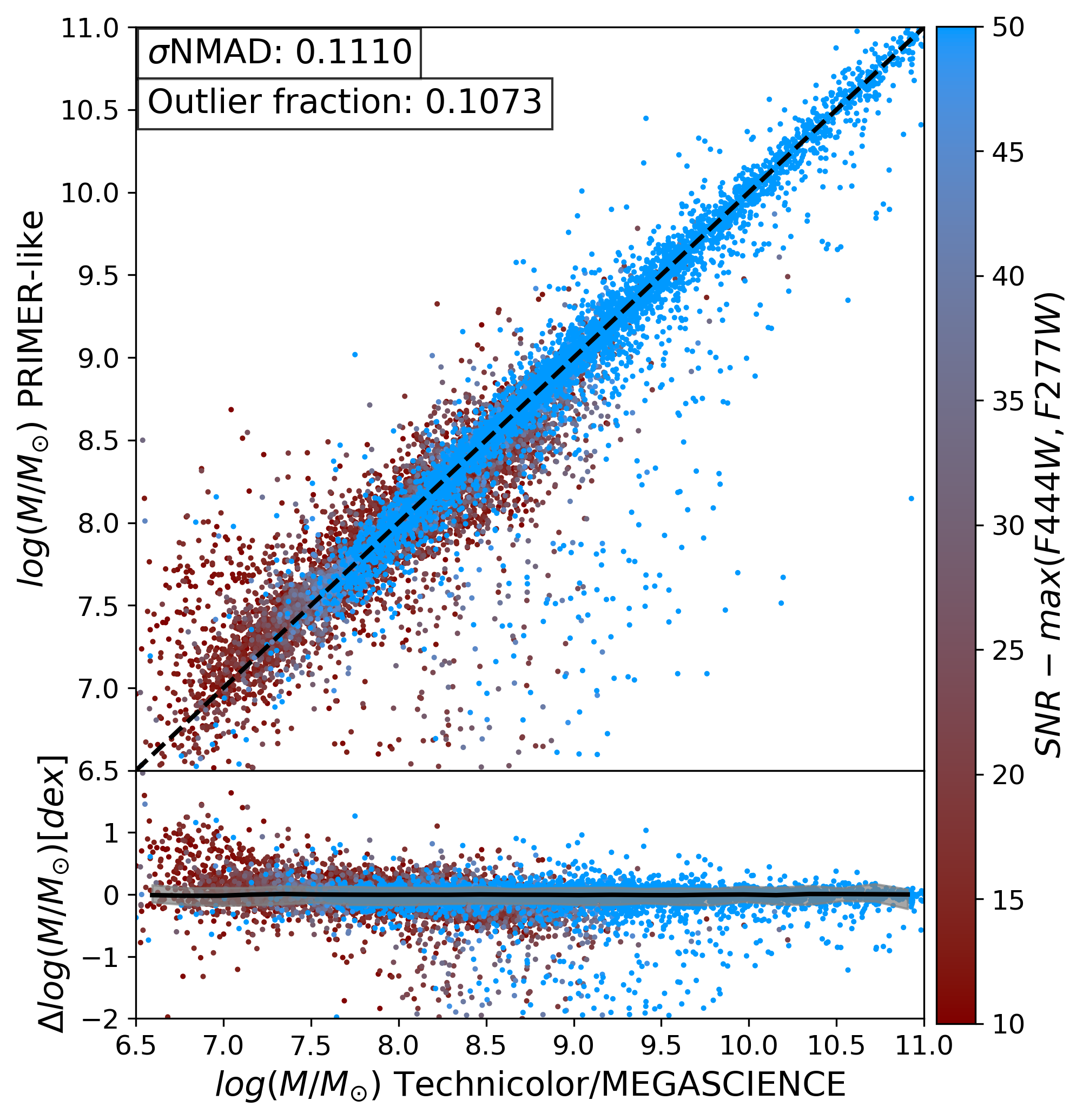}{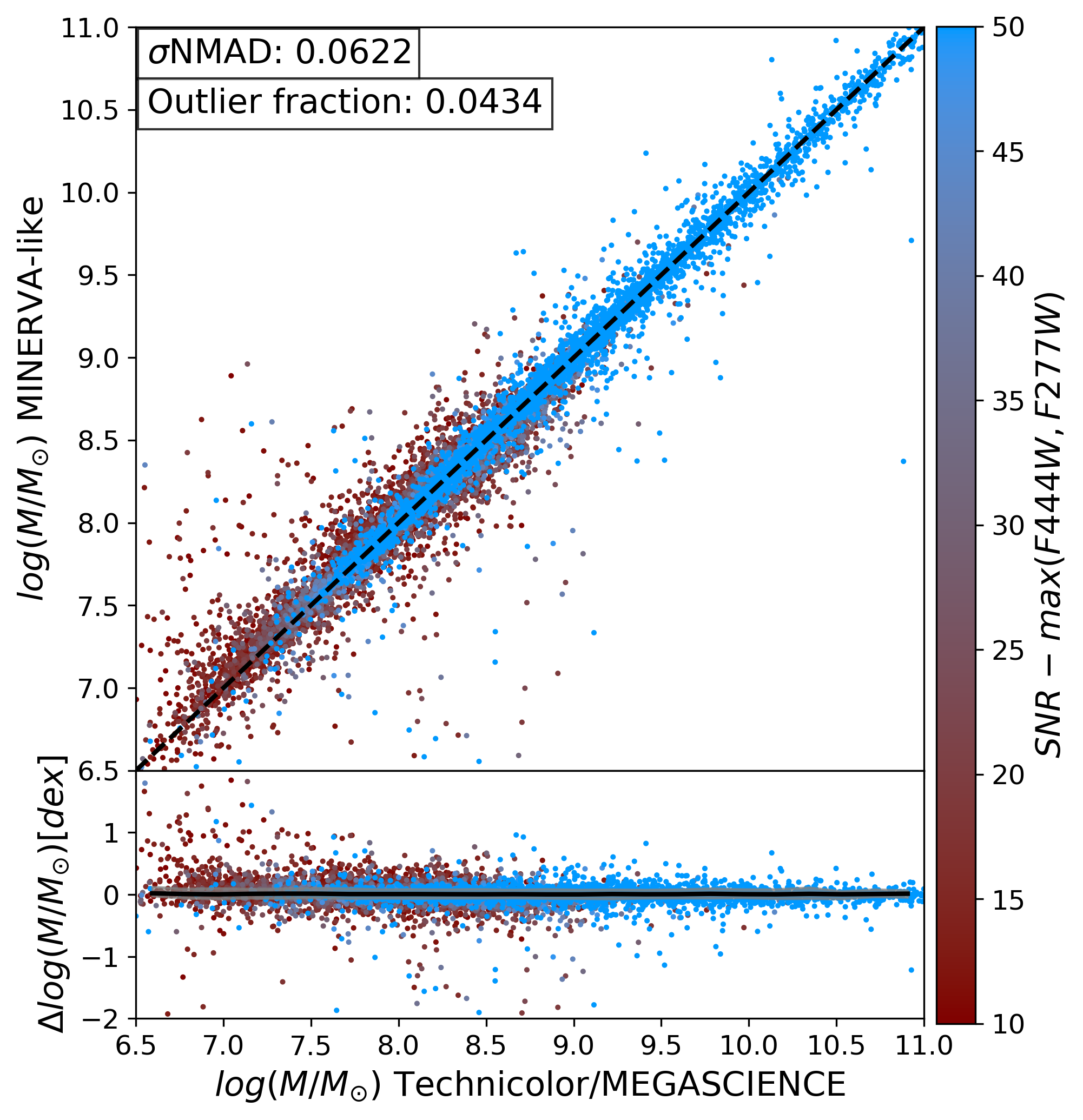}
\caption{Left Panels: Stellar mass results from simulating a PRIMER-like survey by degrading existing deep photometry from UNCOVER/MegaScience and CANUCS/Technicolor to PRIMER depth and filter coverage (8 JWST broadband filters + 3 HST ACS filters).  Right Panels: The same simulation this time adding the 8 MINERVA medium bands at MINERVA depth to the 8 PRIMER broadbands and 3 ACS bands (16 JWST filters + 3 HST filters).  With the additional medium bands MINERVA's stellar masses are estimated to have a 1.78x reduction in $\sigma$NMAD and  2.47x reduction in outlier fraction compared to PRIMER alone.}
    \label{fig:PvM_mstar}
\end{figure*}
\begin{figure*}
    \centering
\plottwo{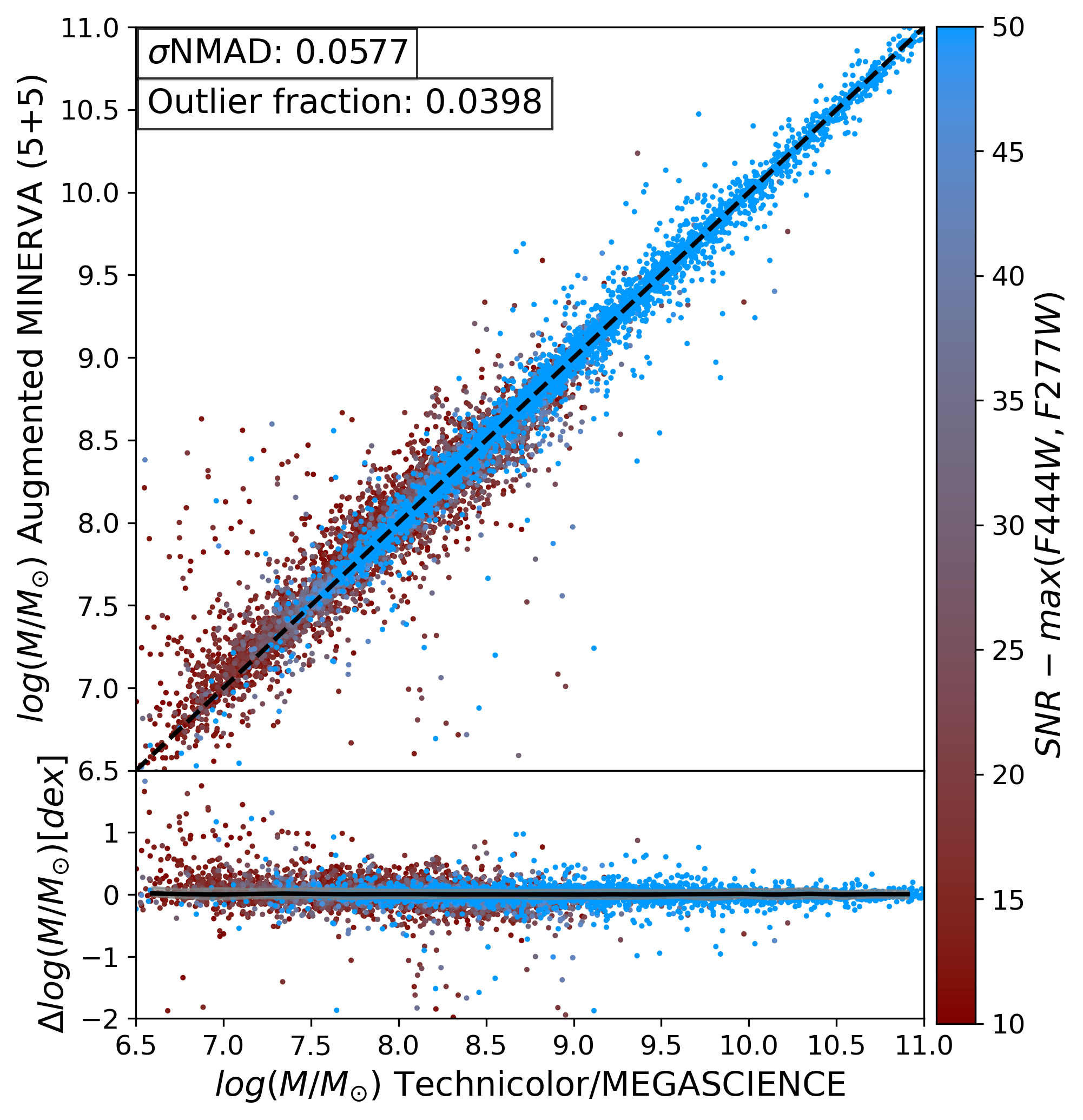}{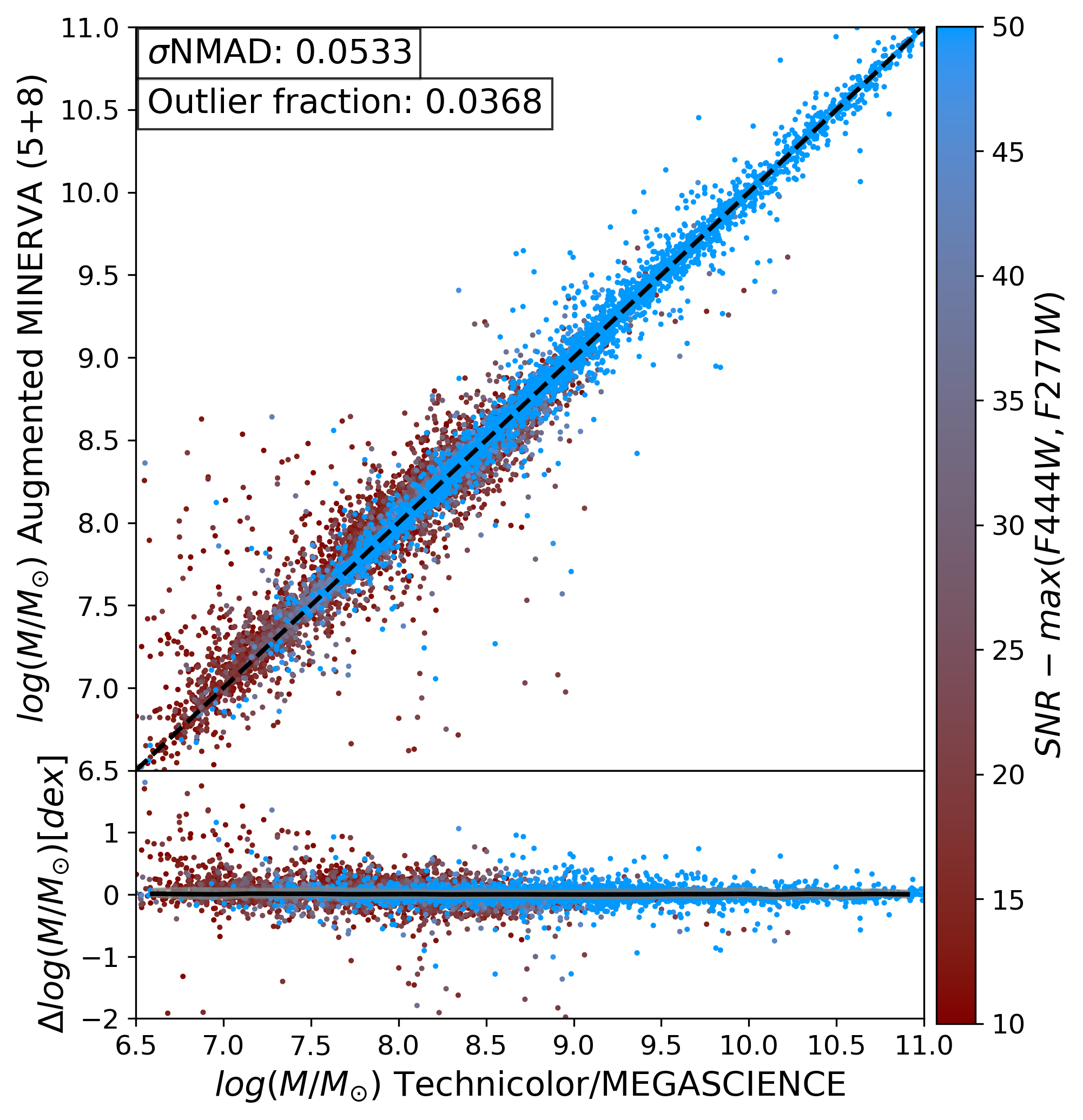}
\caption{Left Panels: Stellar mass results from simulating a 10 filter MINERVA survey from UNCOVER/MegaScience and CANUCS/Technicolor to PRIMER/MINERVA depth and including the PRIMER filters (18 JWST filters + 3 HST filters).  In this case, MINERVA's $M/M_{\odot}$ are estimated to have had a $\sim$ 1.19x improvement in $\sigma$NMAD and 1.23x reduction in outlier fraction (note the plot and inset statistics are $\log(M/M_{\odot)}$ but factors quoted are converted to linear).  Right Panels: The same simulation but for a MINERVA survey that would include 5 filters in the SW (4 medium bands + F070W) and all 8 LW medium bands at simulated MINERVA depth, and including PRIMER broadbands (21 JWST filters + 3 HST filters).  If the maximum number of medium bands were employed MINERVA's stellar masses are estimated to further improve by a factor of 1.47 in $\sigma$NMAD and 1.51 in outlier fraction. Given that the required observing time increase for each dataset would be a factor of 1.25 and 1.75, respectively, this shows that, like $z_{phot}$, the improvement in $M/M_{\odot}$ scales approximately linearly with exposure time.}
    \label{fig:5v4_mstar}
\end{figure*}

\subsubsection{Filter Choice Effects on Stellar Masses}
In addition to considering the effects of medium bands on $z_{phot}$, their effect on the determination of stellar masses was also considered.  This was done by performing SED fitting of the previously simulated PRIMER and MINERVA photometry using the DENSE BASIS code \citep{iyer17, iyer19}.  As with the $z_{phot}$ simulations, the first consideration was a benchmark of the stellar masses (hereafter $\log(M/M_{\odot})$) in a  PRIMER-like survey with only 8 JWST broadbands and 3 HST bands.  Figure \ref{fig:PvM_mstar} shows the results of that fitting, with PRIMER-like stellar masses plotted on the Y-axis of the left panel vs. the UNCOVER/MegaScience and CANUCS/Technicolor ``ground truth" stellar masses on the X-axis.  

Figure \ref{fig:PvM_mstar} shows that the $\log(M/M_{\odot})$ from the PRIMER-like survey have a $\sigma$NMAD in $\delta z / (1 + z)$ that is 0.1110 compared to UNCOVER/MegaScience and CANUCS/Technicolor and an outlier fraction of 10.7\%.  This is good performance, and the outlier fraction is primarily driven by galaxies that have catastrophic failures in the determination of $z_{phot}$, as was also found in the study of \cite{sarrouh24}.

The right panel of Figure \ref{fig:PvM_mstar} shows the performance of the simulated MINERVA-like survey with 8 broadbands, 3 HST ACS bands and 8 medium bands.  The $\sigma$NMAD $\log(M/M_{\odot})$ is a factor of 1.78 lower than the PRIMER-like survey alone, and has an outlier fraction that is an impressive factor of 2.47 times lower.  The performance is notably better at both the high and low mass end, both in the $\sigma$NMAD but also in the reduction of the outlier fraction.  Notably, even at $\log(M/M_{\odot})$ $>$ 10.5 where galaxies typically have S/N $>$ 50 in the broadbands, the number of catastrophic outliers is reduced by adding medium bands, showing that the finer spectral resolution of medium bands cannot be compensated for by higher S/N broadbands.  

The left and right panels of Figure \ref{fig:5v4_mstar} show the performance of $\log(M/M_{\odot})$ for the ``Augmented MINERVA" with 5+5 filters on the SW and LW, and the ``Idealized MINERVA" with all medium and broadbands.  Similar to the comparison of the $z_{phot}$, when comparing to the MINERVA-like survey (not the PRIMER-like) the improvement in the $\sigma$NMAD scatter is a factor of 1.08 and 1.17, respectively.  The corresponding reductions in outlier fraction are 1.09 and 1.18, respectively.  Given that the quantities computed are logarithmic, the improved performance of $\sigma$NMAD in M/$M_{\odot}$ are factors of 1.19 and 1.47, for the two surveys, respectively.  For the outlier fraction it is 1.23 and 1.51, respectively. As with the $z_{phot}$ comparison this is desirable, however, it requires 1.25x and 1.75x additional integration time, and so similar to $z_{phot}$, the improvements in $\log(M/M_{\odot})$ precision scale approximately linearly with more integration on additional medium bands. 

Overall, these simulations show that the additional medium bands in a MINERVA survey with 4+4 medium bands in the SW and LW channels provide a factor of 1.78 and 2.47 improvement in the $\sigma$NMAD and outlier fraction of $\log(M/M_{\odot})$ compared to a PRIMER-like survey.  The improvement occurs both at high and low stellar masses, and is therefore essential for key measurements such as the stellar mass function and stellar mass density of galaxies as these are particularly sensitive to uncertainties near $\log(M^*/M_{\odot})$ $\sim$ 10.8.


\subsection{MINERVA Compared to Existing Medium Band Surveys}\label{subsec-compare}
The analysis in the previous section demonstrates the value in $z_{phot}$ and $\log(M/M_{\odot})$ precision when adding medium bands to a broadband-only PRIMER-like survey.  However, several deep medium band surveys have already been, or are currently being conducted, so in this section we discuss the parameter space of MINERVA relative to those existing or upcoming surveys.

\begin{figure}
    \centering
     \includegraphics[width=1.0\linewidth]{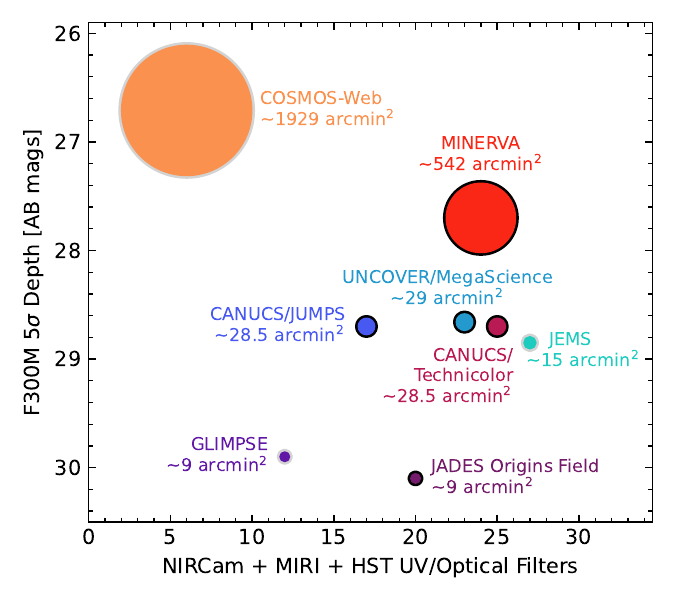}
    \caption{Comparison of various deep/wide JWST imaging surveys.  The Y-axis represents the 5$\sigma$ depth of the surveys in the F300M filter where available, or if unavailable an extrapolation to the equivalent depth using the identical exposure time on the nearest broadband filter.  Surveys without F300M are plotted with a grey outline.  The X-axis shows the typical number of filters with good coverage of the field. The size of the circles represents the total area of the surveys. In terms of surveys with good medium band coverage ($>$ 14 filters) MINERVA occupies the niche as the widest but shallowest medium band survey.}
    \label{fig:mb_surveys}
\end{figure}

Table \ref{tab:survey_info} shows some basic parameters of existing or planned medium band surveys.  It includes the upcoming Cycle-4 SPAM survey (PID 8559, PI: K. Davis, R. Larson), the UNCOVER/MegaScience Survey \citep{suess24}, the CANUCS/Technicolor Survey \citep{sarrouh25}, the CANUCS/JUMPS Survey (PID 5890, PI: Withers), the JEMS Survey \citep{williams23} as well as the JADES Origins Field \citep{eisenstein23b}.  Also included are the two CANUCS flanking fields (CANUCS-NCF) that are not part of the Technicolor Survey \citep[see][]{sarrouh25}, as well as the GLIMPSE survey \citep{kokorev25}.  The depth, filter coverage, and area of these surveys are also shown in graphical form in Figure \ref{fig:mb_surveys}.  On the Y-axis of Figure \ref{fig:mb_surveys} the estimated 5$\sigma$ depth of these surveys is plotted in the F300M filter.  F300M is chosen as a reference filter as it lies in the middle of the wavelength coverage of the medium bands (1.4 $\micron - 4.8\micron$).  If a given survey does not contain F300M, the depth in the nearest filter in wavelength space is used.  For reference, in Figure \ref{fig:mb_surveys} we also show the COSMOS-Web survey, which does not contain medium bands but is the widest-field JWST imaging survey to date, $\sim$ 4x more area than MINERVA.  

Table \ref{tab:survey_info} and Figure \ref{fig:mb_surveys} show that JWST is accumulating an impressive amount of medium band imaging.  It also shows the emerging ``wedding cake" nature of those data in the combination of depth/area.  In particular, Table \ref{tab:survey_info} and Figure \ref{fig:mb_surveys} illustrate that prior to MINERVA, most of the existing medium band data has been accumulated in deep fields from just a handful of NIRCam pointings.  This shows the key role of MINERVA as the wide-field component of the overall JWST extragalactic medium-band imaging wedding cake.  Although it is a factor of a few shallower in integration time compared to previous medium band surveys, MINERVA covers a total area $\sim$ 7x larger than the existing UNCOVER/MegaScience, CANUCS/Technicolor, JEMS and JOF combined.

Perhaps one of the most notable aspects of MINERVA compared to the other medium band fields is the extensive amount of MIRI data available.  The MINERVA MIRI imaging in F1280W and F1500W combined with existing MIRI imaging from the previous PRIMER, CEERS and JADES surveys totals 4 - 8 filters in each field and covers an impressive area of 275 arcmin$^2$.  This total does not include the additional MIRI imaging available in the COSMOS-3D (PID: 5893, PI: Kakiichi), MEGA \citep{backhaus25} and MEOW (PID: 5407, PI: Leung) surveys.  This adds another $\sim$ 150 arcmin$^2$ of MIRI photometry in various bands and this large combined area containing both NIRCam broad and medium bands and MIRI imaging ($\sim$ 425 arcmin$^2$) is extremely promising for opening up new observational parameter spaces.

\begin{deluxetable*}{lcccccc}
    \tablecaption{Existing and Proposed NIRCam Medium Band Surveys
  	\label{tab:survey_info}}
    \tablewidth{0pt}
    \tablehead{
    \colhead{Field} & \colhead{NIRCam} & \colhead{NIRCam Area} & \colhead{NIRCam Exptime} & \colhead{MIRI Area}  & \colhead{MIRI Exptime} & \colhead{Number of} \\
    \colhead{} & \colhead{Pointings} & \colhead{(arcmin$^2$)} & \colhead{(min)} & \colhead{(arcmin$^2$)} & \colhead{(min)} & \colhead{Filters\tablenotemark{a}}
    }
    \startdata
    MINERVA  & 119 & 542 & $\sim 30 - 60$ & 275\tablenotemark{b} & $\sim 35$ & 26 - 35\\
    SPAM\tablenotemark{c} & 10 & 96 & $\sim 45 - 65$ & 23\tablenotemark{d} & $\sim 35$ & 34  \\ 
    UNCOVER/MegaScience & 3 & 29 & $\sim 75 - 120$ & -- & -- & 27 \\
    CANUCS/Technicolor & 3 & 28.5 & $\sim 75 - 120$ & -- & -- & 29  \\ 
    CANUCS/JUMPS & 3 & 28.5 & $\sim 75 - 120$ & -- & -- & 15  \\ 
    CANUCS-NCF & 2 & 19 & $\sim 75 - 120$ & -- & -- & 21  \\ 
    JEMS\tablenotemark{e} & 2 & 15 & $\sim 200 - 400 $ & 15 & $\sim$ 10 - 35 & 27 \\
    JADES Origins Field& 1 & 9 & $\sim 600 - 2800 $ & 5 & $\sim$ 2100 & 20
\enddata
    \tablenotetext{a}{Total number of  filters in the survey footprint from HST (UV+optical, NIR excluded), NIRCam and MIRI.}
     \tablenotetext{b}{MINERVA shares the footprint of COSMOS-3D, MEGA and MEOW MIRI surveys.  With those included the total MIRI area would be $\sim$ 425 arcmin$^2$.}
    \tablenotetext{c}{SPAM covers the same footprint as CEERS/AEGIS and MINERVA/AEGIS.  Its area is included in MINERVA and the MINERVA filters are included in SPAM.}
    \tablenotetext{d}{SPAM covers the same footprint as the MEGA survey.  With that included the total MIRI area would be $\sim$ 85 arcmin$^2$.}
    \tablenotetext{e}{The JEMS footprint is covered by the SMILES survey.  The quoted 8 MIRI filter coverage for JEMS comes from SMILES.}

\end{deluxetable*}

\subsection{Scheduling}
The long range plan scheduling windows for MINERVA observations as of July 2025 are shown in Table \ref{tab:schedule_info}.  The UDS and COSMOS fields are divided into two windows as they are taken with PA separated by $\sim$ 180 degrees to facilitate maximum overlap between MIRI and NIRCam observations.  Overall, Table \ref{tab:schedule_info} shows that the MINERVA observations should be obtained at relatively regular intervals over the course of JWST Cycle-4.

\begin{deluxetable}{lr}
    \tablecaption{MINERVA Scheduling Windows
  	\label{tab:schedule_info}}
    \tablewidth{0pt}
    \tablehead{
    \colhead{Field} & \colhead{Observability Window} 
    }
    \startdata
    UDS-2         & July 18 - August 10, 2025 \\
    COSMOS-2 & November 18 - January 7, 2026  \\
    UDS-1         & January 4 - January 25, 2026 \\
    GOODS-N & January 31 - February 10, 2027  \\
    COSMOS-1 & April 16 - April 27, 2026  \\
    AEGIS & June 16 - June 27, 2026
    \enddata
\end{deluxetable}


\section{Science Objectives}
\label{sec:science}

Here we describe the main science goals of the MINERVA survey: identifying rare populations (Section~\ref{sec:rare}), statistical constraints on the stellar mass function (Section~\ref{sec:smf}), and resolved studies of galaxies (Section~\ref{sec:resolved}).

\subsection{Identifying and characterizing rare populations}
\label{sec:rare}

Our key goal is to accurately measure the number density of unique populations and understand their role in galaxy evolution. Identifying and characterizing these populations requires large area to sample rare objects, high spectral resolution to accurately characterize redshifts and physical properties, high spatial resolution to deblend sources, and a well-understood completeness function in order to characterize survey volume. As described in Section~\ref{sec:observations}, these are exactly the main design goals for MINERVA: It covers a wide area of $\sim$ 542 arcmin$^2$ at an effective spectral resolution of $R\sim15-30$, and as an imaging survey its magnitude completeness limits can be well-characterized. Not only are these rare sources interesting in their own right: we must understand them, otherwise key observables such as the SMF and SFRD are biased by factors of 2-3 \citep[e.g.,][see also Section~\ref{sec:smf}]{robertsborsani21,sarrouh24}. Below, we describe several known classes of objects where MINERVA medium bands are expected to significantly improve our constraints on physical properties. However, we note that MINERVA also has the potential to discover ``unknown unknowns" currently hiding in our broadband catalogs.


\subsubsection{Robust $z > 13$ Candidates}
Despite three years of JWST observations, the first $\sim$300Myr of our cosmos remains a  mystery. Numerous galaxies have been confirmed at $z{ \sim} 12$ \citep{finkelstein22,curtis-lake23,castellano24,wang23}; however, just three galaxies have been confirmed at $z {>} 14$ \citep{carniani24,naidu25}, and none at $z {>} 15$. Finding the first galaxies in existing data is difficult in part due to contamination from low-redshift Balmer breaks or strong line emission, both of which can mimic a high-redshift Lyman break in broadband photometry \citep[e.g.,][]{donnan23,naidu22,zavala23}. The large wavelength gap between NIRCam's F200W and F277W filters also introduces selection biases such that, if following traditional Lyman-break selection techniques, only bright high-redshift candidates with blue UV slopes are identified at $z\gtrsim13$. 

Deep medium band observations -- both at 2$\mu$m to pin down the Lyman break and 4$\mu$m to filter out low-redshift contaminants -- have been identified as the most promising way to robustly ID high-redshift galaxies rather than discarding them as contaminants \citep{eisenstein23b,asada25}. While MINERVA is shallower than existing small-area medium band surveys, the extraordinarily bright rest-frame UV luminosities of $z\gtrsim14$ galaxies confirmed to date means that 2/3 would be confidently detected at MINERVA depths (both MoM-z14, \citealt{naidu25} and JADES-GS-z14-0, \citealt{carniani24}). MINERVA will increase the area with this critical medium-band data by a factor of $\sim$ 7, potentially allowing us to identify rare bright high-redshift candidates that were previously deemed contaminants. Our large on-sky coverage, precise photometric redshifts, and multiple sightlines are also expected to reduce uncertainties on the bright-end UV luminosity density at $z>10$ by a factor of $\sim$4 \citep{Bouwens2023, willott24,weibel25_pano}, enabling confident discrimination of models of early galaxy formation \citep[e.g.][]{kannan23}. Finally, MINERVA's medium bands will sample multiple points in the rest-frame UV, allowing us to accurately measure rest-frame UV $\beta$ slopes of galaxies at $z\gtrsim10$, allowing us to explore the intrinsic diversity of stellar populations in high-redshift galaxies \citep[e.g.][]{morales24,cullen24,narayanan25}. 


\subsubsection{Balmer Breaks and the Quenched Galaxy Population}

One surprising discovery from the first several years of JWST observations is the ubiquity of high-redshift quiescent galaxies. Massive (${>}10^{10}M_\odot$) quiescent galaxies have now been observed out to $z\sim7$, so early that their rapid assembly may be in tension with $\Lambda$CDM \citep[e.g.,][]{carnall24,degraaff24,glazebrook24,setton24,weibel25}. At the other end of the mass function, low-mass ``mini-quenched" or ``napping" galaxies have been seen out to $z{\sim}8$ \citep[e.g.,][]{looser23,baker25}. The relatively strong Balmer breaks and weak emission lines in these galaxies point towards extremely bursty star formation or even temporary quiescence in the first billion years of the Universe \citep[e.g.,][]{trussler24,endsley24}.

\begin{figure}[ht]
    \centering
    \includegraphics[width=\linewidth]{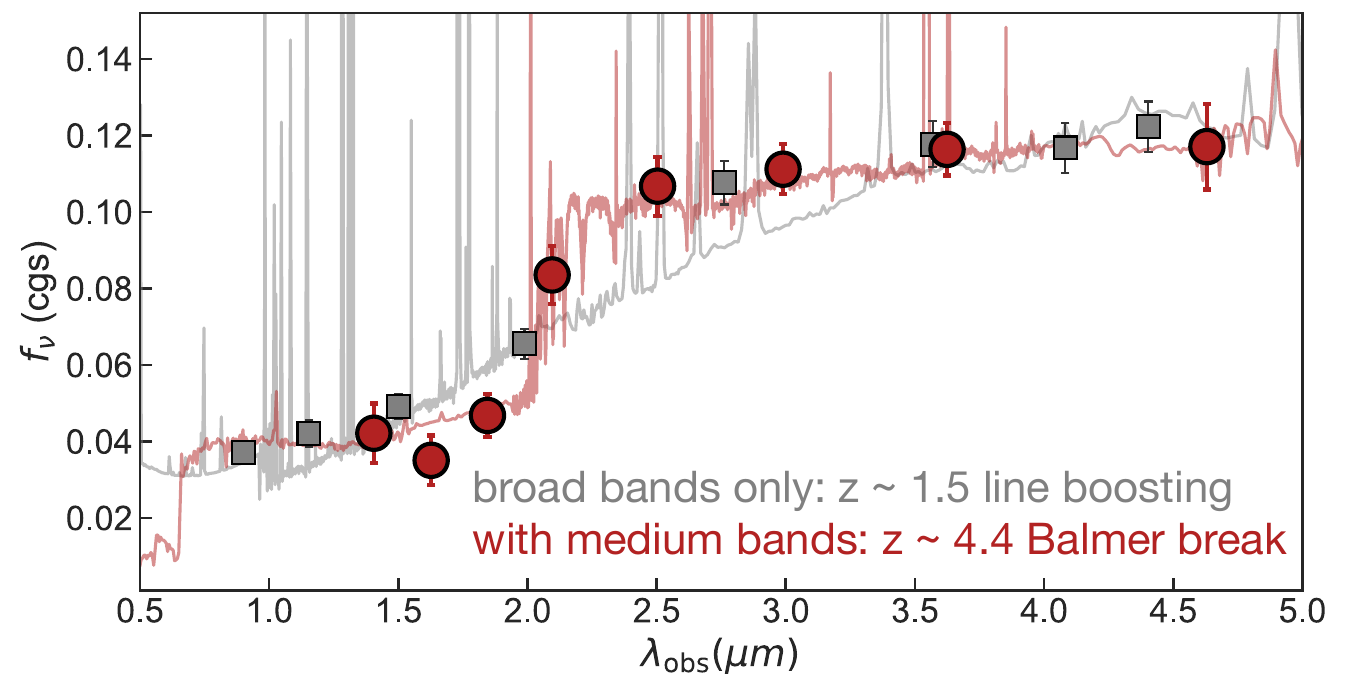}
    \caption{Example of constraints on Balmer break galaxies with (red) and without (grey) medium-band data. Galaxy and SED fits taken from the UNCOVER/MegaScience public catalog \citep{bezanson24,suess24}, scaled to the depth and filter coverage of PRIMER/MINERVA. Without medium bands, the photometric redshift (as well as Balmer break and emission line strengths) are mischaracterized in existing data.}
    \label{fig:quiescent}
\end{figure}

Accurately selecting and characterizing these (perhaps temporarily) quiescent systems at all masses requires accurate photometric redshifts, precise measurements of Balmer break strengths, and the ability to rule out contamination from strong emission lines. Medium band observations are an efficient way to achieve this goal \citep[e.g.,][]{trussler24}. Figure~\ref{fig:quiescent} shows an example ``napping" galaxy at $z\sim4.4$ from the UNCOVER/MegaScience survey, fit with broadbands only (grey) and including medium bands (red). Photometric error and included filters are matched to MINERVA-UDS depth as described in Section~\ref{sec:filter_choice_z}. With broadbands alone, the fit returns a $z\sim1.5$ line-boosting solution; the strong Balmer break is clear only with medium bands. Scaling from public UNCOVER/MegaScience catalogs \citep{suess24}, we expect to find $\sim$150 low-mass ($10^{8-9.5}M_\odot$) quiescent galaxies at $z>6$, and $\sim1500$ at ($4<z<6$), decreasing the error bars on current number density estimates of these galaxies by a factor of $\sim10$. Our medium band measurements will also allow us to calculate the UV-to-H$\alpha$ luminosity ratio \citep[e.g.,][Mitsuhashi in prep.]{asada24}, enabling us to measure robust star formation timescales at $3<z<8$ that can be compared with simulations to quantify the role of feedback \citep[e.g.,][]{faucher-giguere18} and assess the impact of burstiness on completeness in flux-limited surveys \citep[e.g.,][]{iyer20,wang25}. 

We also expect to observe tens of massive quiescent galaxies at $z>3$; while the majority of these galaxies are likely already known from previous HST/JWST imaging, our medium band measurements will enable us to estimate ages from accurate Balmer/4000$\AA$~break measurements \citep[e.g.,][]{trussler24,mintz25}. Furthermore, our multi-band MIRI measurements may significantly improve the selection of massive quiescent sources at high redshift \citep[e.g.,][]{alberts24_quiescent} as well as mass estimates for red quiescent and/or dusty sources \citep[which may be overestimated by a full dex even with deep medium-band observations, e.g.,][]{williams24,wang24}.

\subsubsection{Characterizing the Extreme Emission Line Population at $2 < z < 10$}

\begin{figure}
    \centering
    \includegraphics[width=\linewidth]{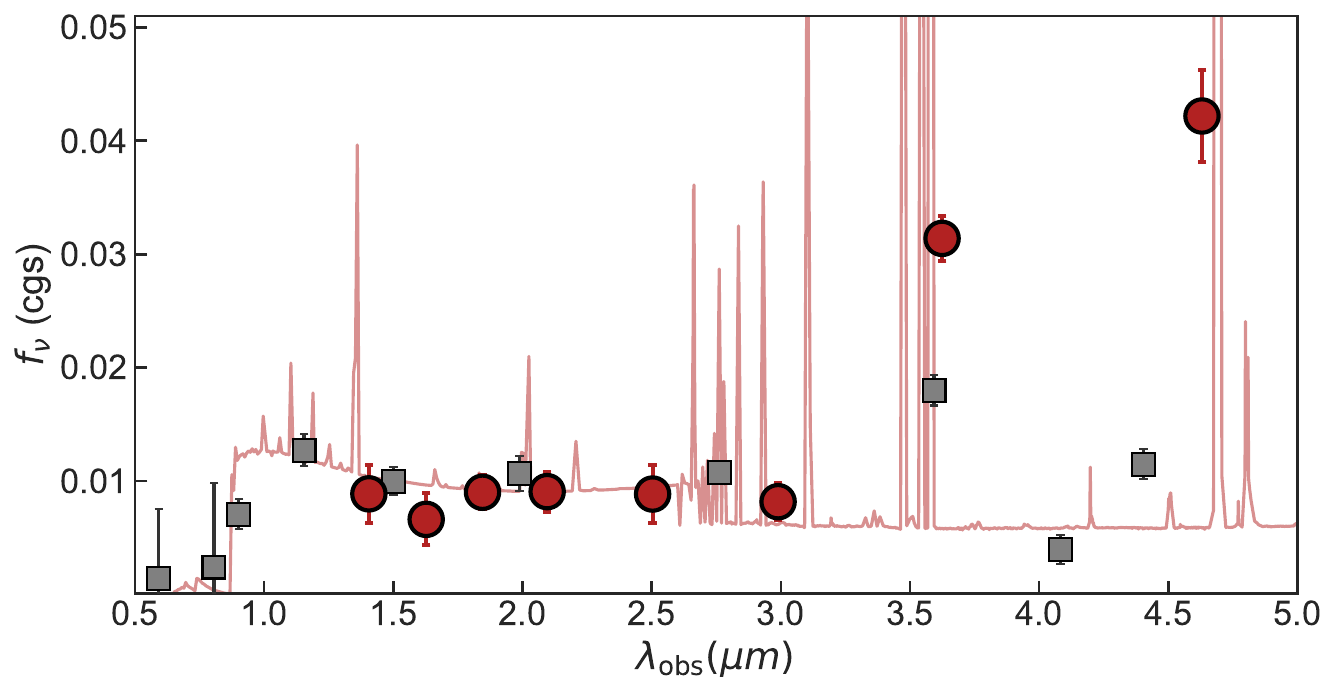}
    \caption{Example of constraints on extreme emission line galaxies with (red) and without (grey) medium-band data. Galaxy and SED fits taken from CANUCS/Technicolor (CITE), scaled to the depth and filter coverage of PRIMER/MINERVA. The medium bands allow for significantly more accurate estimates of the emission line strengths, help identify possible Balmer jumps, and improve measurement accuracy in the UV slope $\beta$.}
    \label{fig:eelg}
\end{figure}

Existing narrow-field medium-band surveys have shown that JWST can efficiently identify hundreds of extreme emission line galaxies \citep[EELGs, see e.g.,][]{vanderwel11} across a wide redshift range \citep[e.g.,][]{rinaldi2023,withers23,wold2024}. Furthermore, medium-band photometry can measure the equivalent widths of emission line complexes such as H$\alpha$+[NII], H$\beta$+[OIII], and Pa-$\beta$ with accuracy $\sim0.15$~dex \citep[]{withers23,lorenz25}. These medium band emission line strengths can be used to infer (dust-corrected) star formation rates and nebular dust attenuation values \citep{lorenz25}, search for extreme H$\alpha$/[OIII]+H$\beta$ ratios that imply very low metallicities or even Population III candidates \citep[e.g.,][]{fujimoto25}, and identify ``Balmer Jump" or nebular-dominated galaxies \cite[]{cameron23,katz24,tacchella25} that are extremely metal poor with potentially top-heavy IMFs. An example SED scaled to MINERVA depth is shown in Figure~\ref{fig:eelg}, demonstrating how adding medium band data helps accurately constrain emission line strengths. Extrapolating the detections in \citet{withers23} to the area of MINERVA, we expect to detect $\sim$ 8,000 EELGs, of which $\sim$400 will be Balmer Jump galaxies and $\sim$1200 will have EW $>$ 2000$\AA$. We will also search for ultra metal-poor candidates by identifying galaxies with anomalously low ([OIII]$+$H$\beta$)/(H$\alpha$+[NII]) ratios, implying low contributions of [OIII] due to extremely low metallicities (or large amounts of dust, which is unlikely for sources that have significant UV emission) \citep[e.g.,][]{fujimoto25,morishita25_Z}. This factor of $\sim60$ increase in sample size of EELGs will allow population demographics of these extreme galaxies for the first time. 

\subsubsection{The extremely red and dust-obscured Universe}

JWST has already revealed significant populations of sources previously invisible to HST and Spitzer \citep[e.g.,][]{williams23,nelson23,barrufet23,gibson24,williams24,sun24}. Some exotic objects may even be dark to NIRCam and only visible in MIRI \citep{ppg24}. 
These previously-invisible galaxies contribute significantly to the stellar mass density at $z\gtrsim4$ \citep[e.g.,][]{gottumukkala24}. 
Often dusty, these galaxies can also be morphologically complex, with JWST revealing features such as spiral arms and radial color gradients \citep[e.g.,][]{price25,kokorev23,setton24, rujopakarn23}. MINERVA's area and MIRI multi-band coverage will allow us additional discovery space. For example, multi-color MIRI observations from GTO-SMILES revealed a population of dusty AGN that is almost double previous estimates; $\sim80\%$ were undiscovered with pre-MIRI searches despite extremely deep legacy data \citep{lyu24}.

Due to the relatively small footprint of MIRI compared to NIRCam, existing multi-filter MIRI data sets are primarily in small fields (e.g., MIDIS, \cite{ostlin25}, 2 arcmin$^2$; SMILES, \cite{alberts24,rieke24}, 34 arcmin$^2$; MEGA, \cite{backhaus25}, 65 arcmin$^2$). MINERVA will cover an area of $\sim277$ arcmin$^2$ with at least four MIRI filters, and is thus expected to reduce the uncertainties on the bright end of the AGN fraction and the contribution of ultra-dusty galaxies to the SFRD by a factor of $\sim 4$.

\subsection{A Robust Measurement of the Stellar Mass Function and Star-formation Rate Function at $6<z<10$}
\label{sec:smf}

Another key goal of the MINERVA survey is to produce robust measurements of the cosmic evolution of the integrated growth of the stellar content of the Universe (i.e., the stellar mass density, SMD) and of the star-formation rate density (SFRD) during the first Gyr of cosmic history. The SMD is obtained from measurements of the stellar mass function (SMF) of galaxies, i.e., the number density of galaxies per unit stellar mass interval as a function of stellar mass. The SMF is a fundamental cosmological observable in the study of the statistical properties of galaxies, and its shape and evolution provide insights on the growth of the stellar content of the Universe, making the SMF one of the most commonly adopted observational benchmarks for theoretical simulations (e.g., \citealt{fontanot09}; \citealt{lovell21}). In particular, the SMF of massive galaxies puts the most stringent constraints on theoretical models. However, the relative rarity of massive galaxies at all redshifts and the exponential cutoff of the high-mass end of the SMF combine to make these measurements very challenging and the resulting comparisons uncertain.  
Similarly, the SFRD can be obtained from measurements of the SFR function (SFRF) of galaxies (e.g., \citealt{picouet23}). Measurements of the SFRF can be used to connect the growth of galaxies at high $z$ to galaxies at later epochs, and also provide insight into the way star formation turns on in dark matter halos (e.g., \citealt{smit12}). 

\begin{figure}
    \centering
    \includegraphics[width=\linewidth]{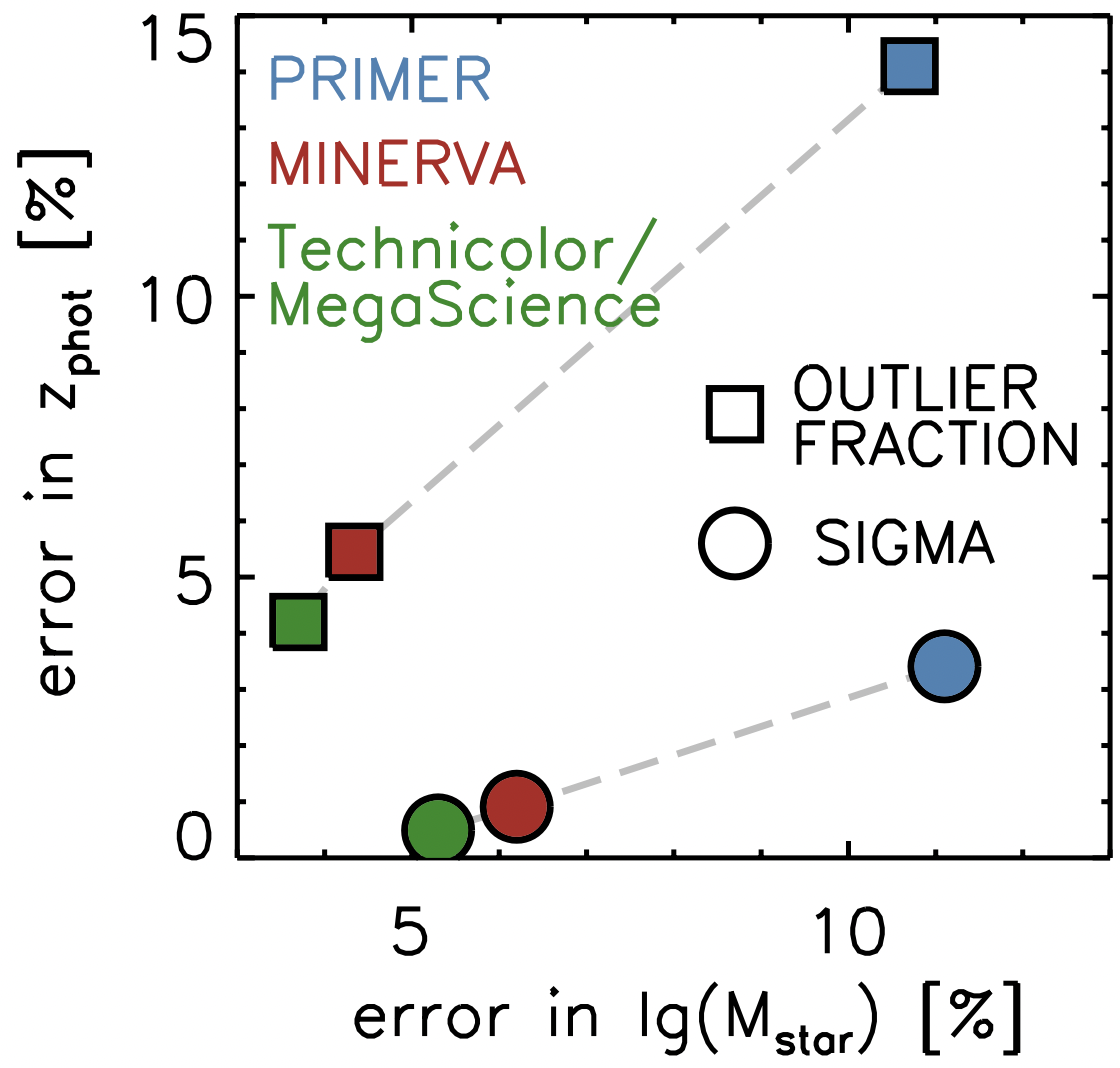}
    \caption{Random errors (circles) and catastrophic outlier fractions (squares) on photometric redshifts (y-axis) and stellar mass (x-axis) when using only broadband NIRCam filters (blue;``PRIMER-like'' simulations), as delivered by the MINERVA survey (red; ``MINERVA-like'' simulations), and for the case with all NIRCam broad- and medium-band filters (green; ``Idealized MINERVA'' simulations), similar to the CANUCS/Technicolor and UNCOVER/MegaScience surveys. MINERVA improves uncertainties on $z_{\rm phot}$ and M$_{\rm star}$ by an average factor of 2-4 over existing broadband data, with even more dramatic improvements for the rare galaxy populations presented in \S3.1.}
    \label{fig:DM_fig1}
\end{figure}

Furthermore, over past decades, the quenching of star formation has been identified as a key aspect of galaxy evolution. Several quenching mechanisms have been proposed (see \citealt{manbelli18} for a review), and understanding the dominant mechanism(s) turning off star formation as a function of redshift and stellar mass is still an unsolved challenge. Robustly measuring the evolution with cosmic time of the SMF of quiescent galaxies presents the strongest observable of the first appearance and the subsequent growth of the population of quiescent galaxies, providing a very powerful statistical tool to identify quenching mechanisms (e.g., \citealt{delucia25}).

Impressive observational progress has been made over the past quarter century in the measurements of the SMFs of galaxies (see, e.g., \citealt{weaver23} for a recent compilation of citations). However, despite many hundreds of hours of JWST broadband imaging, our ability to ``count'' galaxies and our measurements of the evolution of the SMD and SFRD in the first 1.5 Gyr (i.e., $z$$>$4) have not yet made transformative improvements in accuracy \citep{harvey24,weibel24}. This problem persists even at $z\sim1$ for low-mass ($M_{\star}$$<$10$^{9}$~M$_{\odot}$) galaxies (e.g., \citealt{santini22,hamadouche25}). This is in part due to the inability to remove systematics (such as catastrophic $z_{phot}$ outliers, emission line contamination, AGN, and degeneracies with dust obscuration) when only broadband photometry is available. These biases are dramatically amplified for quiescent galaxies, since dusty star formation and/or emission line contamination produce apparent breaks in broad-band photometry that are misidentified as the Balmer/4000A breaks. 

\begin{figure*}
    \centering
    \includegraphics[width=\linewidth]{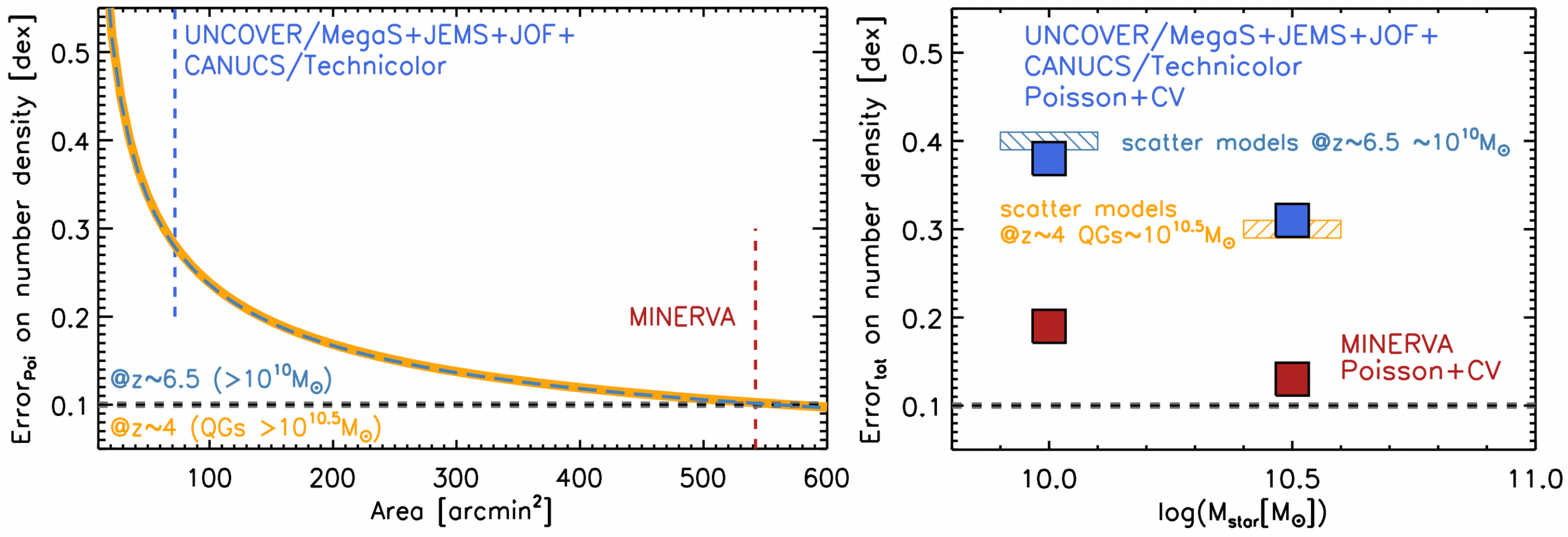}
    \caption{{\it Left:} Poisson error vs area for z$\sim$6 galaxies with $M_{\rm star}$$>$10$^{10}$~M$_{\odot}$ (blue dashed curve) and z$\sim$4 massive ($M_{\rm star}$$>$10$^{10.5}$~M$_{\odot}$) quiescent galaxies (orange curve). The vertical dashed lines show the combined area in current JWST imaging surveys with MBs (blue) and MINERVA (red). {\it Right:} Total (Poisson and cosmic variance) error on the number density of z$\sim$6 galaxies with $M_{\rm star}$$>$10$^{10}$~M$_{\odot}$ and massive quiescent galaxies at z$\sim$4 with $M_{\rm star}$$>$10$^{10.5}$~M$_{\odot}$. Hatched light blue and orange regions show the typical scatter among theoretical model predictions for these populations, respectively. Filled blue squares show the estimated total errors from current JWST imaging surveys with MBs, while filled red squares show the estimated total errors from the MINERVA survey. MINERVA will allow us to distinguish between different theoretical models of early galaxy evolution, with errors a factor of $\sim$2-3 smaller than the current scatter between different models.}
    \label{fig:DM_fig2}
\end{figure*}

\begin{figure*}
    \centering
    \includegraphics[width=.8\linewidth]{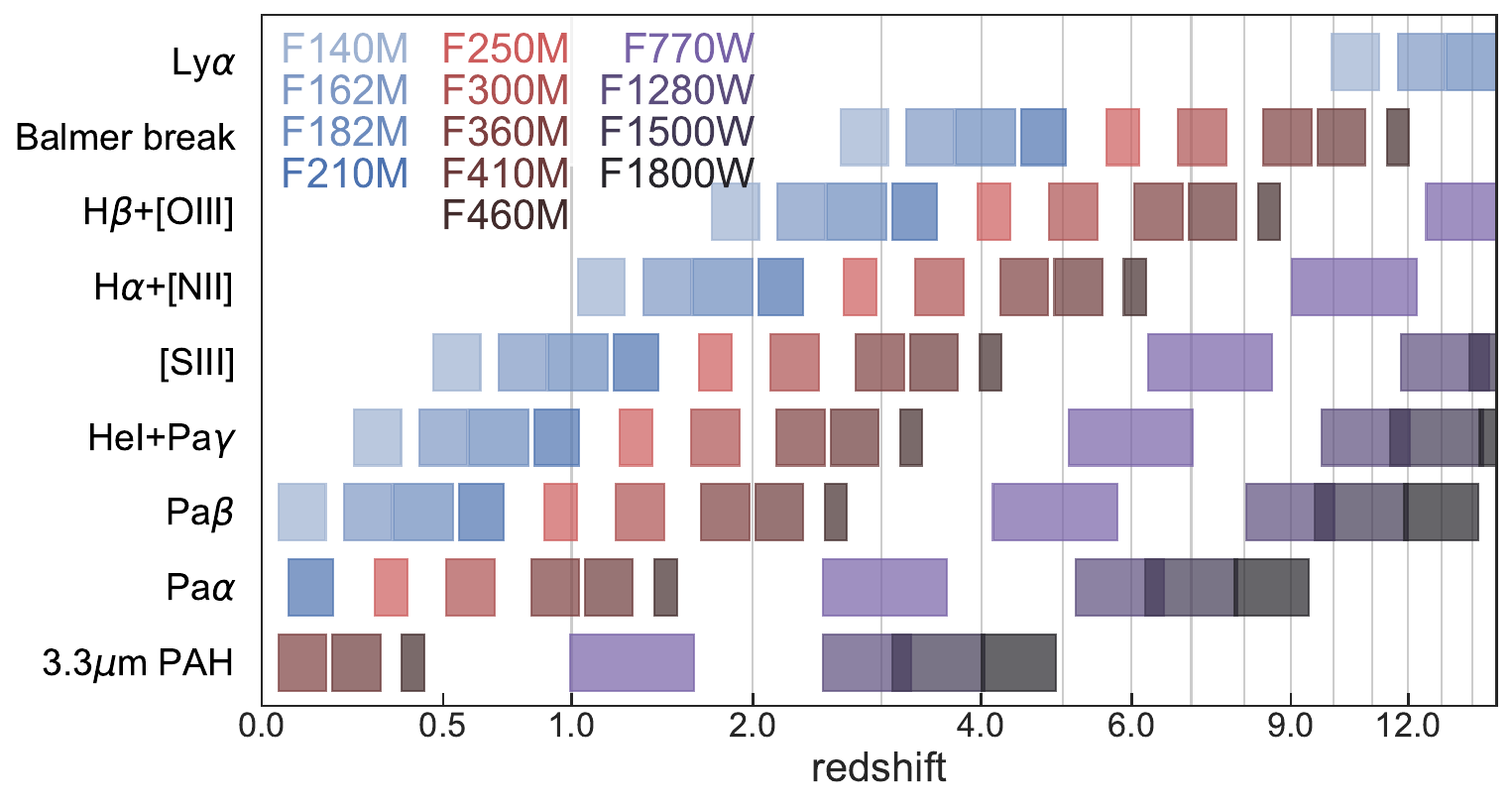}
    \caption{MINERVA filter coverage as a function of redshift for several strong spectral features. We show only the four MIRI filters that are available in all of the MINERVA fields; however, we note that additional MIRI imaging is available in COSMOS, GOODS-N, and AEGIS (see Tables~\ref{tab:UDS_info}-\ref{tab:GOODSN_info}).}
    \label{fig:lines}
\end{figure*}

\begin{figure}
    \centering
    \includegraphics[width=.8\linewidth]{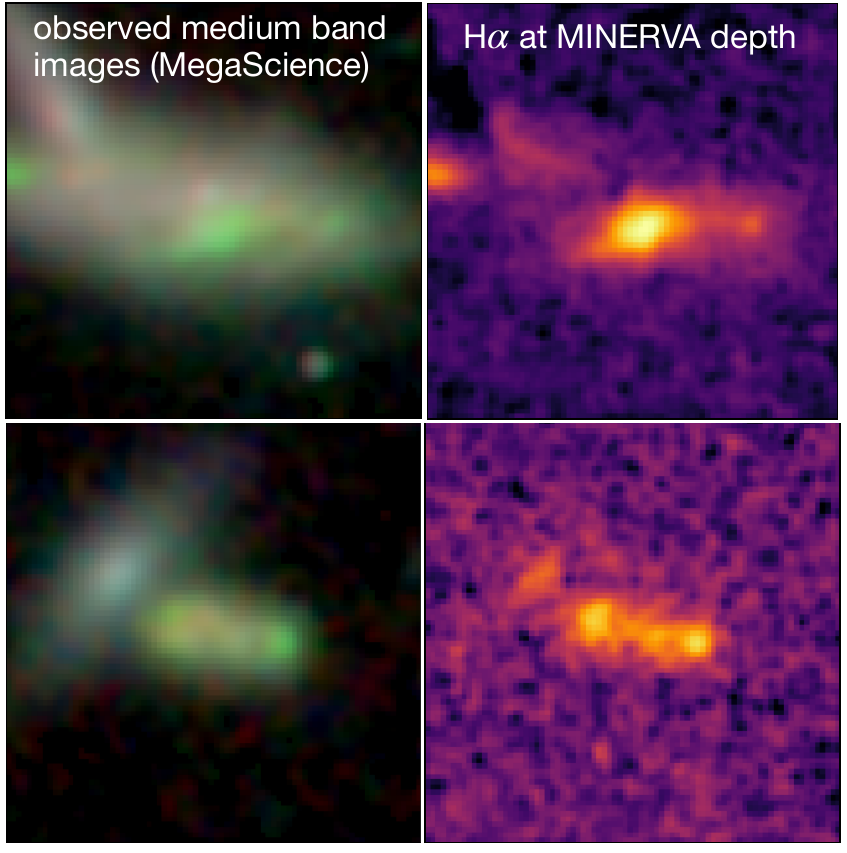}
    \caption{Three-color medium-band images of two $z\sim2.3$ galaxies (H$\alpha$ in green; continuum in blue and red) as well as continuum-subtracted H$\alpha$ maps following the methods in \citet{lorenz25}. Data is from UNCOVER/MegaScience, with additional noise added to simulate the shallower depth of MINERVA.}
    \label{fig:halpha}
\end{figure}

The augmentation of broadband imaging surveys with medium-band photometry enables complete and unbiased studies of the high-$z$ Universe, as demonstrated at $z<3$ by ground-based MB surveys (i.e., NMBS, \citealt{whitaker11,whitaker12}; zFOURGE, \citealt{tomczack14,tomczak16,straatman16}) and at $z>4$ by space-based JWST MB surveys (i.e., UNCOVER/Technicolor, \citealt{sarrouh24}). As shown in Figure~\ref{fig:DM_fig1}, MINERVA will dramatically improve $z_{phot}$ and stellar properties, reducing by a factor of $\sim$3-4 random uncertainties and fractions of catastrophic outliers for $\sim$13,300(5,200)\footnote{These numbers were obtained using the CANUCS/Technicolor and UNCOVER/MegaScience catalogs and scaling by a factor of $\sim$9.4, accounting for the area difference with respect to MINERVA.} galaxies with $M_{\star}$$>$10$^{9(10)}$M$_{\odot}$ at $z$$>$4, enabling a complete and unbiased picture of the $z>3$ Universe, targeting the high-mass regime and robustly measuring the growth of quiescent galaxies in the first 2 Gyr, as well as returning a complete census of unobscured and dust-obscured star formation. 

In addition to systematic-free photometric redshifts and stellar masses, a large surveyed volume is the other critical aspect of a survey aiming to probe relatively rare objects, like massive galaxies and distant quiescent galaxies. Current JWST MB imaging programs have only surveyed $\sim$76~arcmin$^{2}$ to-date (JEMS, JOF, CANUCS/Technicolor, and UNCOVER/MegaScience; see \S\ref{subsec-compare}). This area is too small to 1) measure the SMF/SMD of massive (i.e., $M_{\star}$$>$10$^{10-10.5}$M$_{\odot}$) galaxies with small enough errors to discriminate among theoretical model predictions \citep{weaver23,weibel24}, and 2) measure the buildup of the quiescent galaxies' SMF at $z$=3-8. As shown in Figure~\ref{fig:DM_fig2}, the combined Poisson and cosmic variance error on the number density of galaxies with $M_{\star}$$>$10$^{10}$M$_{\odot}$ at $z\sim6.5$ from JEMS+JOF+Technicolor+MegaScience is $\sim$0.38~dex, similar to the typical scatter ($\sim0.4$~dex) in their predicted densities from theoretical models. Similarly, the total error on the number density of massive ($M_{\star}$$>$10$^{10.5}$M$_{\odot}$) quiescent galaxies at $z\sim4$ is $\sim0.31$~dex, similar to the scatter in their theoretically predicted densities. 

With a surveyed area of $\sim$542~arcmin$^{2}$ over 4 independent fields, MINERVA will dramatically reduce both Poisson errors and cosmic variance (see Fig.~\ref{fig:DM_fig2}, filled red squares). Therefore, the MINERVA survey will measure 1) the evolution of the high-mass end of the SMFs of star-forming and quiescent galaxies at $z>3$ with unprecedented precision and accuracy, enabling discrimination among theoretical model predictions, and 2) the evolution of the SMD and SFRD at $3<z<10$, free of systematics.

\subsection{Galaxy Formation in 2D: Resolved Continuum \& Line Mapping of Galaxy Growth}
\label{sec:resolved}

In addition to the spatially-integrated science described above, MINERVA's medium bands will spatially resolve the properties of distant galaxies, allowing us to map out stellar populations at ${\sim 500}$pc physical scales and measure multi-band galaxy sizes and morphologies across a wide redshift range. Figure~\ref{fig:lines} shows how MINERVA's eight medium bands and four MIRI bands\footnote{we show the four MIRI filters that will be observed in all four fields; some fields have additional MIRI coverage, often with complex footprints} trace different spectral features from $z\sim0$ through $z\sim10$. 

Medium bands can be used to make direct line maps of relatively strong isolated emission features (e.g., H$\alpha$, H$\beta$+[OIII]), with accurate continuum subtraction from adjacent medium bands \citep[e.g.,][]{withers23,lorenz25}. 
These direct line maps can be used, for example, to study star formation and dust in galaxies at cosmic noon \citep{lorenz25}, to trace the buildup of stellar mass in galaxies from $z\sim1$ through $z\sim6$ with H$\alpha$ \citep[e.g.,][]{tacchella15,nelson16}, to map ionization cones from active galactic nuclei \citep{lebowitz25}, or to study ionizing radiation escape near the epoch of reionization \citep[e.g.,][]{simmonds23}. Figure~\ref{fig:halpha} shows example H$\alpha$ maps of two galaxies at cosmic noon, created using medium-band data from UNCOVER/MegaScience and techniques from \citet{lorenz25}, but scaled to MINERVA/PRIMER depth. Given the depth and area of MINERVA, we can create H$\alpha$ maps of this quality for $>10,000$ galaxies at cosmic noon and above. Maps of [OIII]+H$\beta$ can similarly be used to search for outflows \citep[e.g.,][]{zhu24} or very high equivalent-width regions that may leak Lyman continuum emission \citep[e.g.,][]{chen25}.

Even if no strong isolated emission features are present in a galaxy, medium bands are sensitive to continuum emission and can be used in spatially-resolved stellar population modeling to make maps of properties such as stellar mass, stellar age, and dust attenuation. These maps can be used to chart the evolution of color gradients and growth of galaxy half-mass radii over cosmic time \citep[e.g.,][]{suess19,suess19b,mosleh20,miller22,miller23,jin24,tan24}, to understand bursty star formation histories and the effects of outshining \citep[e.g.,][]{gimenezarteaga23,lines25,harvey25_outshining}, and the properties of star-forming clumps \citep[e.g.,][]{tanaka24,ji24b,claeyssens25}.


\section{Discussion and Conclusions}
\label{sec:discussion}
In this paper we have presented an overview of the fields, observational layout, and primary science goals of the MINERVA Survey, a large treasury cycle 4 NIRCam medium band imaging survey in 8 bands (F140M, F162M, F182M, F210M, F250M, F300M, F360M, F460M) with MIRI parallels in F1280W and F1500W covering four of the five CANDELS fields and totaling 542 arcmin$^2$ of new NIRCam imaging and 275 arcmin$^2$ of MIRI imaging.  

This paper also showed simulations of the relative improvement of adding medium bands to existing broadband datasets such as PRIMER.  It showed that the MINERVA observations, which require 2x the integration time on the PRIMER field decrease the $\sigma$NMAD scatter in photometric redshifts and log(M/M$_{\odot}$) by factors 3.7 and 1.78, respectively, and are therefore highly efficient at improving our characterization of distant galaxies.  The simulations show that increasing the number of medium band filters in potential future surveys does continue to increase the quality of photometric redshifts and stellar masses at a rate proportional to the increase in integration time.

This paper also discussed MINERVA in the context of other medium band surveys, showing clearly that there is an emerging ``wedding cake" structure to these projects in terms of area/depth.  At 542 arcmin$^2$ MINERVA is $\sim$ 7x larger than all existing medium band coverage and clearly occupies the wide-shallow parameter space of extragalactic medium band imaging.

Lastly, we discussed the primary science goals of MINERVA.  These include 1) the discovery of a robust sample of $z > 13$ galaxy candidates, 2) the discovery of evolved galaxies with Balmer breaks at $z > 6$, 3) the discovery of extreme emission line galaxies at $z >$ 6 as well as the identification of candidate ultra-metal-poor galaxies, 4) improved constraints on the stellar mass function and stellar mass density measurements by factors of $\sim$ 3, and 5) resolved mapping of the stellar mass and star formation rates of galaxies at $1 < z < 6$ using the extensive medium band data.

Overall, with its location in the prime extragalactic deep fields and extensive NIRCam medium band imaging and supporting MIRI images, the MINERVA survey is poised to become the final component in the legacy extragalactic imaging from the HST and JWST telescopes, enabling unique science with high-precision photometric redshifts and stellar masses for decades to come.

\acknowledgements 
AM acknowledges support via grants 23JWGO2A13 and 23JWGO2B15 from the Canadian Space Agency (CSA).
DM acknowledges support from the Leonard and Jane Holmes Bernstein Professorship in Evolutionary Science. MB, NM, RT, and VM acknowledge support from the ERC Grant FIRSTLIGHT and Slovenian national research agency ARIS through grants N1-0238 and P1-0188. MST and MSH acknowledge support from the European Research Commission Consolidator Grant 101088789 (SFEER). MST also acknowledges support from the CIDEGENT/2021/059 grant by Generalitat Valenciana, and from project PID2023-149420NB-I00 funded by MICIU/AEI/10.13039/501100011033 and by ERDF/EU, and from the MCIN with funding from the European Union NextGenerationEU and Generalitat Valenciana in the call Programa de Planes Complementarios de I+D+i (PRTR 2022) Project (VAL-JPAS), reference ASFAE/2022/025. DJS and JRW acknowledge support for this work was provided by The Brinson Foundation through a Brinson Prize Fellowship grant. HA acknowledges support from CNES, focused on the JWST mission, the Programme National Cosmology and Galaxies (PNCG) of CNRS/INSU with INP and IN2P3, co-funded by CEA and CNES, and the French National Research Agency (ANR) under grant ANR-21-CE31-0838. The work of CCW is supported by NOIRLab, which is managed by the Association of Universities for Research in Astronomy (AURA) under a cooperative agreement with the National Science Foundation. TBM was supported by a CIERA Postdoctoral Fellowship.

\software{Astropy \citep{astropy22}, Grizli \citep{brammer23}, Seaborn \citep{waskom23}
}
\bibliographystyle{aasjournal}
\bibliography{minervabib}

\begin{thebibliography}{}
\expandafter\ifx\csname natexlab\endcsname\relax\def\natexlab#1{#1}\fi
\providecommand{\url}[1]{\href{#1}{#1}}
\providecommand{\dodoi}[1]{doi:~\href{http://doi.org/#1}{\nolinkurl{#1}}}
\providecommand{\doeprint}[1]{\href{http://ascl.net/#1}{\nolinkurl{http://ascl.net/#1}}}
\providecommand{\doarXiv}[1]{\href{https://arxiv.org/abs/#1}{\nolinkurl{https://arxiv.org/abs/#1}}}

\bibitem[{{Alberts} {et~al.}(2024{\natexlab{a}}){Alberts}, {Williams}, {Helton}, {Suess}, {Ji}, {Shivaei}, {Lyu}, {Rieke}, {Baker}, {Bonaventura}, {Bunker}, {Carniani}, {Charlot}, {Curtis-Lake}, {D'Eugenio}, {Eisenstein}, {de Graaff}, {Hainline}, {Hausen}, {Johnson}, {Maiolino}, {Parlanti}, {Rieke}, {Robertson}, {Sun}, {Tacchella}, {Willmer}, \& {Willott}}]{alberts24_quiescent}
{Alberts}, S., {Williams}, C.~C., {Helton}, J.~M., {et~al.} 2024{\natexlab{a}}, \apj, 975, 85, \dodoi{10.3847/1538-4357/ad66cc}

\bibitem[{{Alberts} {et~al.}(2024{\natexlab{b}}){Alberts}, {Lyu}, {Shivaei}, {Rieke}, {P{\'e}rez-Gonz{\'a}lez}, {Bonaventura}, {Zhu}, {Helton}, {Ji}, {Morrison}, {Robertson}, {Stone}, {Sun}, {Williams}, \& {Willmer}}]{alberts24}
{Alberts}, S., {Lyu}, J., {Shivaei}, I., {et~al.} 2024{\natexlab{b}}, \apj, 976, 224, \dodoi{10.3847/1538-4357/ad7396}

\bibitem[{{Argyriou} {et~al.}(2023){Argyriou}, {Glasse}, {Law}, {Labiano}, {{\'A}lvarez-M{\'a}rquez}, {Patapis}, {Kavanagh}, {Gasman}, {Mueller}, {Larson}, {Vandenbussche}, {Glauser}, {Royer}, {Dicken}, {Harkett}, {Sargent}, {Engesser}, {Jones}, {Kendrew}, {Noriega-Crespo}, {Brandl}, {Rieke}, {Wright}, {Lee}, \& {Wells}}]{argyriou23}
{Argyriou}, I., {Glasse}, A., {Law}, D.~R., {et~al.} 2023, \aap, 675, A111, \dodoi{10.1051/0004-6361/202346489}

\bibitem[{{Arnouts} {et~al.}(2007){Arnouts}, {Walcher}, {Le F{\`e}vre}, {Zamorani}, {Ilbert}, {Le Brun}, {Pozzetti}, {Bardelli}, {Tresse}, {Zucca}, {Charlot}, {Lamareille}, {McCracken}, {Bolzonella}, {Iovino}, {Lonsdale}, {Polletta}, {Surace}, {Bottini}, {Garilli}, {Maccagni}, {Picat}, {Scaramella}, {Scodeggio}, {Vettolani}, {Zanichelli}, {Adami}, {Cappi}, {Ciliegi}, {Contini}, {de la Torre}, {Foucaud}, {Franzetti}, {Gavignaud}, {Guzzo}, {Marano}, {Marinoni}, {Mazure}, {Meneux}, {Merighi}, {Paltani}, {Pell{\`o}}, {Pollo}, {Radovich}, {Temporin}, \& {Vergani}}]{arnouts07}
{Arnouts}, S., {Walcher}, C.~J., {Le F{\`e}vre}, O., {et~al.} 2007, \aap, 476, 137, \dodoi{10.1051/0004-6361:20077632}

\bibitem[{{Arrabal Haro} {et~al.}(2023){Arrabal Haro}, {Dickinson}, {Finkelstein}, {Kartaltepe}, {Donnan}, {Burgarella}, {Carnall}, {Cullen}, {Dunlop}, {Fern{\'a}ndez}, {Fujimoto}, {Jung}, {Krips}, {Larson}, {Papovich}, {P{\'e}rez-Gonz{\'a}lez}, {Amor{\'\i}n}, {Bagley}, {Buat}, {Casey}, {Chworowsky}, {Cohen}, {Ferguson}, {Giavalisco}, {Huertas-Company}, {Hutchison}, {Kocevski}, {Koekemoer}, {Lucas}, {McLeod}, {McLure}, {Pirzkal}, {Seill{\'e}}, {Trump}, {Weiner}, {Wilkins}, \& {Zavala}}]{arrabal-haro23}
{Arrabal Haro}, P., {Dickinson}, M., {Finkelstein}, S.~L., {et~al.} 2023, \nat, 622, 707, \dodoi{10.1038/s41586-023-06521-7}

\bibitem[{{Asada} {et~al.}(2024){Asada}, {Sawicki}, {Abraham}, {Brada{\v{c}}}, {Brammer}, {Desprez}, {Estrada-Carpenter}, {Iyer}, {Martis}, {Matharu}, {Mowla}, {Muzzin}, {Noirot}, {Sarrouh}, {Strait}, {Willott}, \& {Harshan}}]{asada24}
{Asada}, Y., {Sawicki}, M., {Abraham}, R., {et~al.} 2024, \mnras, 527, 11372, \dodoi{10.1093/mnras/stad3902}

\bibitem[{{Asada} {et~al.}(2025){Asada}, {Willott}, {Muzzin}, {Brada{\v{c}}}, {Brammer}, {Desprez}, {Iyer}, {Marchesini}, {Martis}, {Noirot}, {Sarrouh}, {Sawicki}, {Withers}, {Fujimoto}, {Felicioni}, {Goovaerts}, {Jude{\v{z}}}, {Jagga}, {Merchant}, {M{\'e}rida}, \& {Robbins}}]{asada25}
{Asada}, Y., {Willott}, C., {Muzzin}, A., {et~al.} 2025, arXiv e-prints, arXiv:2507.03124, \dodoi{10.48550/arXiv.2507.03124}

\bibitem[{{Astropy Collaboration} {et~al.}(2022){Astropy Collaboration}, {Price-Whelan}, {Lim}, {Earl}, {Starkman}, {Bradley}, {Shupe}, {Patil}, {Corrales}, {Brasseur}, {N{\"o}the}, {Donath}, {Tollerud}, {Morris}, {Ginsburg}, {Vaher}, {Weaver}, {Tocknell}, {Jamieson}, {van Kerkwijk}, {Robitaille}, {Merry}, {Bachetti}, {G{\"u}nther}, {Aldcroft}, {Alvarado-Montes}, {Archibald}, {B{\'o}di}, {Bapat}, {Barentsen}, {Baz{\'a}n}, {Biswas}, {Boquien}, {Burke}, {Cara}, {Cara}, {Conroy}, {Conseil}, {Craig}, {Cross}, {Cruz}, {D'Eugenio}, {Dencheva}, {Devillepoix}, {Dietrich}, {Eigenbrot}, {Erben}, {Ferreira}, {Foreman-Mackey}, {Fox}, {Freij}, {Garg}, {Geda}, {Glattly}, {Gondhalekar}, {Gordon}, {Grant}, {Greenfield}, {Groener}, {Guest}, {Gurovich}, {Handberg}, {Hart}, {Hatfield-Dodds}, {Homeier}, {Hosseinzadeh}, {Jenness}, {Jones}, {Joseph}, {Kalmbach}, {Karamehmetoglu}, {Ka{\l}uszy{\'n}ski}, {Kelley}, {Kern}, {Kerzendorf}, {Koch}, {Kulumani}, {Lee}, {Ly}, {Ma}, {MacBride}, {Maljaars}, {Muna}, {Murphy}, {Norman},
  {O'Steen}, {Oman}, {Pacifici}, {Pascual}, {Pascual-Granado}, {Patil}, {Perren}, {Pickering}, {Rastogi}, {Roulston}, {Ryan}, {Rykoff}, {Sabater}, {Sakurikar}, {Salgado}, {Sanghi}, {Saunders}, {Savchenko}, {Schwardt}, {Seifert-Eckert}, {Shih}, {Jain}, {Shukla}, {Sick}, {Simpson}, {Singanamalla}, {Singer}, {Singhal}, {Sinha}, {Sip{\H{o}}cz}, {Spitler}, {Stansby}, {Streicher}, {{\v{S}}umak}, {Swinbank}, {Taranu}, {Tewary}, {Tremblay}, {de Val-Borro}, {Van Kooten}, {Vasovi{\'c}}, {Verma}, {de Miranda Cardoso}, {Williams}, {Wilson}, {Winkel}, {Wood-Vasey}, {Xue}, {Yoachim}, {Zhang}, {Zonca}, \& {Astropy Project Contributors}}]{astropy22}
{Astropy Collaboration}, {Price-Whelan}, A.~M., {Lim}, P.~L., {et~al.} 2022, \apj, 935, 167, \dodoi{10.3847/1538-4357/ac7c74}

\bibitem[{{Backhaus} {et~al.}(2025){Backhaus}, {Kirkpatrick}, {Yang}, {Troiani}, {Hamblin}, {Kartaltepe}, {Kocevski}, {Koekemoer}, {Lambrides}, {Papovich}, \& {Ronayne}}]{backhaus25}
{Backhaus}, B.~E., {Kirkpatrick}, A., {Yang}, G., {et~al.} 2025, arXiv e-prints, arXiv:2503.19078, \dodoi{10.48550/arXiv.2503.19078}

\bibitem[{Baker {et~al.}(2025)Baker, D’Eugenio, Maiolino, Bunker, Simmonds, Tacchella, Witstok, Arribas, Carniani, Charlot, Chevallard, Curti, Curtis-Lake, Jones, Kumari, Rinaldi, Robertson, Williams, Willott, \& Zhu}]{baker25}
Baker, W.~M., D’Eugenio, F., Maiolino, R., {et~al.} 2025, Astronomy \&; Astrophysics, 697, A90, \dodoi{10.1051/0004-6361/202553766}

\bibitem[{{Barrufet} {et~al.}(2023){Barrufet}, {Oesch}, {Weibel}, {Brammer}, {Bezanson}, {Bouwens}, {Fudamoto}, {Gonzalez}, {Gottumukkala}, {Illingworth}, {Heintz}, {Holden}, {Labbe}, {Magee}, {Naidu}, {Nelson}, {Stefanon}, {Smit}, {van Dokkum}, {Weaver}, \& {Williams}}]{barrufet23}
{Barrufet}, L., {Oesch}, P.~A., {Weibel}, A., {et~al.} 2023, \mnras, 522, 449, \dodoi{10.1093/mnras/stad947}

\bibitem[{{Beckwith} {et~al.}(2006){Beckwith}, {Stiavelli}, {Koekemoer}, {Caldwell}, {Ferguson}, {Hook}, {Lucas}, {Bergeron}, {Corbin}, {Jogee}, {Panagia}, {Robberto}, {Royle}, {Somerville}, \& {Sosey}}]{beckwith96}
{Beckwith}, S. V.~W., {Stiavelli}, M., {Koekemoer}, A.~M., {et~al.} 2006, \aj, 132, 1729, \dodoi{10.1086/507302}

\bibitem[{{Bell} {et~al.}(2004){Bell}, {Wolf}, {Meisenheimer}, {Rix}, {Borch}, {Dye}, {Kleinheinrich}, {Wisotzki}, \& {McIntosh}}]{bell04}
{Bell}, E.~F., {Wolf}, C., {Meisenheimer}, K., {et~al.} 2004, \apj, 608, 752, \dodoi{10.1086/420778}

\bibitem[{{Bezanson} {et~al.}(2024){Bezanson}, {Labbe}, {Whitaker}, {Leja}, {Price}, {Franx}, {Brammer}, {Marchesini}, {Zitrin}, {Wang}, {Weaver}, {Furtak}, {Atek}, {Coe}, {Cutler}, {Dayal}, {van Dokkum}, {Feldmann}, {F{\"o}rster Schreiber}, {Fujimoto}, {Geha}, {Glazebrook}, {de Graaff}, {Greene}, {Juneau}, {Kassin}, {Kriek}, {Khullar}, {Maseda}, {Mowla}, {Muzzin}, {Nanayakkara}, {Nelson}, {Oesch}, {Pacifici}, {Pan}, {Papovich}, {Setton}, {Shapley}, {Smit}, {Stefanon}, {Taylor}, \& {Williams}}]{bezanson24}
{Bezanson}, R., {Labbe}, I., {Whitaker}, K.~E., {et~al.} 2024, \apj, 974, 92, \dodoi{10.3847/1538-4357/ad66cf}

\bibitem[{{Bouwens} {et~al.}(2023){Bouwens}, {Illingworth}, {Oesch}, {Stefanon}, {Naidu}, {van Leeuwen}, \& {Magee}}]{Bouwens2023}
{Bouwens}, R., {Illingworth}, G., {Oesch}, P., {et~al.} 2023, \mnras, 523, 1009, \dodoi{10.1093/mnras/stad1014}

\bibitem[{{Brammer}(2023)}]{brammer23}
{Brammer}, G. 2023, {grizli}, 1.9.11,  Zenodo, \dodoi{10.5281/zenodo.8370018}

\bibitem[{{Brammer} {et~al.}(2009){Brammer}, {Whitaker}, {van Dokkum}, {Marchesini}, {Labb{\'e}}, {Franx}, {Kriek}, {Quadri}, {Illingworth}, {Lee}, {Muzzin}, \& {Rudnick}}]{brammer09}
{Brammer}, G.~B., {Whitaker}, K.~E., {van Dokkum}, P.~G., {et~al.} 2009, \apjl, 706, L173, \dodoi{10.1088/0004-637X/706/1/L173}

\bibitem[{{Brammer} {et~al.}(2012){Brammer}, {van Dokkum}, {Franx}, {Fumagalli}, {Patel}, {Rix}, {Skelton}, {Kriek}, {Nelson}, {Schmidt}, {Bezanson}, {da Cunha}, {Erb}, {Fan}, {F{\"o}rster Schreiber}, {Illingworth}, {Labb{\'e}}, {Leja}, {Lundgren}, {Magee}, {Marchesini}, {McCarthy}, {Momcheva}, {Muzzin}, {Quadri}, {Steidel}, {Tal}, {Wake}, {Whitaker}, \& {Williams}}]{brammer12}
{Brammer}, G.~B., {van Dokkum}, P.~G., {Franx}, M., {et~al.} 2012, \apjs, 200, 13, \dodoi{10.1088/0067-0049/200/2/13}

\bibitem[{{Cameron} {et~al.}(2023){Cameron}, {Katz}, {Rey}, \& {Saxena}}]{cameron23}
{Cameron}, A.~J., {Katz}, H., {Rey}, M.~P., \& {Saxena}, A. 2023, \mnras, 523, 3516, \dodoi{10.1093/mnras/stad1579}

\bibitem[{{Cardamone} {et~al.}(2010){Cardamone}, {van Dokkum}, {Urry}, {Taniguchi}, {Gawiser}, {Brammer}, {Taylor}, {Damen}, {Treister}, {Cobb}, {Bond}, {Schawinski}, {Lira}, {Murayama}, {Saito}, \& {Sumikawa}}]{cardamone10}
{Cardamone}, C.~N., {van Dokkum}, P.~G., {Urry}, C.~M., {et~al.} 2010, \apjs, 189, 270, \dodoi{10.1088/0067-0049/189/2/270}

\bibitem[{{Carnall} {et~al.}(2024){Carnall}, {Cullen}, {McLure}, {McLeod}, {Begley}, {Donnan}, {Dunlop}, {Shapley}, {Rowlands}, {Almaini}, {Arellano-C{\'o}rdova}, {Barrufet}, {Cimatti}, {Ellis}, {Grogin}, {Hamadouche}, {Illingworth}, {Koekemoer}, {Leung}, {Lovell}, {P{\'e}rez-Gonz{\'a}lez}, {Santini}, {Stanton}, \& {Wild}}]{carnall24}
{Carnall}, A.~C., {Cullen}, F., {McLure}, R.~J., {et~al.} 2024, \mnras, 534, 325, \dodoi{10.1093/mnras/stae2092}

\bibitem[{{Carniani} {et~al.}(2024){Carniani}, {Hainline}, {D'Eugenio}, {Eisenstein}, {Jakobsen}, {Witstok}, {Johnson}, {Chevallard}, {Maiolino}, {Helton}, {Willott}, {Robertson}, {Alberts}, {Arribas}, {Baker}, {Bhatawdekar}, {Boyett}, {Bunker}, {Cameron}, {Cargile}, {Charlot}, {Curti}, {Curtis-Lake}, {Egami}, {Giardino}, {Isaak}, {Ji}, {Jones}, {Kumari}, {Maseda}, {Parlanti}, {P{\'e}rez-Gonz{\'a}lez}, {Rawle}, {Rieke}, {Rieke}, {Del Pino}, {Saxena}, {Scholtz}, {Smit}, {Sun}, {Tacchella}, {{\"U}bler}, {Venturi}, {Williams}, \& {Willmer}}]{carniani24}
{Carniani}, S., {Hainline}, K., {D'Eugenio}, F., {et~al.} 2024, \nat, 633, 318, \dodoi{10.1038/s41586-024-07860-9}

\bibitem[{{Casey} {et~al.}(2023){Casey}, {Kartaltepe}, {Drakos}, {Franco}, {Harish}, {Paquereau}, {Ilbert}, {Rose}, {Cox}, {Nightingale}, {Robertson}, {Silverman}, {Koekemoer}, {Massey}, {McCracken}, {Rhodes}, {Akins}, {Allen}, {Amvrosiadis}, {Arango-Toro}, {Bagley}, {Bongiorno}, {Capak}, {Champagne}, {Chartab}, {Ch{\'a}vez Ortiz}, {Chworowsky}, {Cooke}, {Cooper}, {Darvish}, {Ding}, {Faisst}, {Finkelstein}, {Fujimoto}, {Gentile}, {Gillman}, {Gould}, {Gozaliasl}, {Hayward}, {He}, {Hemmati}, {Hirschmann}, {Jahnke}, {Jin}, {Khostovan}, {Kokorev}, {Lambrides}, {Laigle}, {Larson}, {Leung}, {Liu}, {Liaudat}, {Long}, {Magdis}, {Mahler}, {Mainieri}, {Manning}, {Maraston}, {Martin}, {McCleary}, {McKinney}, {McPartland}, {Mobasher}, {Pattnaik}, {Renzini}, {Rich}, {Sanders}, {Sattari}, {Scognamiglio}, {Scoville}, {Sheth}, {Shuntov}, {Sparre}, {Suzuki}, {Talia}, {Toft}, {Trakhtenbrot}, {Urry}, {Valentino}, {Vanderhoof}, {Vardoulaki}, {Weaver}, {Whitaker}, {Wilkins}, {Yang}, \& {Zavala}}]{casey23}
{Casey}, C.~M., {Kartaltepe}, J.~S., {Drakos}, N.~E., {et~al.} 2023, \apj, 954, 31, \dodoi{10.3847/1538-4357/acc2bc}

\bibitem[{{Castellano} {et~al.}(2024){Castellano}, {Napolitano}, {Fontana}, {Roberts-Borsani}, {Treu}, {Vanzella}, {Zavala}, {Arrabal Haro}, {Calabr{\`o}}, {Llerena}, {Mascia}, {Merlin}, {Paris}, {Pentericci}, {Santini}, {Bakx}, {Bergamini}, {Cupani}, {Dickinson}, {Filippenko}, {Glazebrook}, {Grillo}, {Kelly}, {Malkan}, {Mason}, {Morishita}, {Nanayakkara}, {Rosati}, {Sani}, {Wang}, \& {Yoon}}]{castellano24}
{Castellano}, M., {Napolitano}, L., {Fontana}, A., {et~al.} 2024, \apj, 972, 143, \dodoi{10.3847/1538-4357/ad5f88}

\bibitem[{{Chen} {et~al.}(2025){Chen}, {Motohara}, {Spitler}, \& {Malkan}}]{chen25}
{Chen}, N., {Motohara}, K., {Spitler}, L., \& {Malkan}, M.~A. 2025, \apj, 981, 96, \dodoi{10.3847/1538-4357/adad69}

\bibitem[{{Claeyssens} {et~al.}(2025){Claeyssens}, {Adamo}, {Messa}, {Dessauges-Zavadsky}, {Richard}, {Kramarenko}, {Matthee}, \& {Naidu}}]{claeyssens25}
{Claeyssens}, A., {Adamo}, A., {Messa}, M., {et~al.} 2025, \mnras, 537, 2535, \dodoi{10.1093/mnras/staf058}

\bibitem[{{Cullen} {et~al.}(2024){Cullen}, {McLeod}, {McLure}, {Dunlop}, {Donnan}, {Carnall}, {Keating}, {Magee}, {Arellano-Cordova}, {Bowler}, {Begley}, {Flury}, {Hamadouche}, \& {Stanton}}]{cullen24}
{Cullen}, F., {McLeod}, D.~J., {McLure}, R.~J., {et~al.} 2024, \mnras, 531, 997, \dodoi{10.1093/mnras/stae1211}

\bibitem[{{Curtis-Lake} {et~al.}(2023){Curtis-Lake}, {Carniani}, {Cameron}, {Charlot}, {Jakobsen}, {Maiolino}, {Bunker}, {Witstok}, {Smit}, {Chevallard}, {Willott}, {Ferruit}, {Arribas}, {Bonaventura}, {Curti}, {D'Eugenio}, {Franx}, {Giardino}, {Looser}, {L{\"u}tzgendorf}, {Maseda}, {Rawle}, {Rix}, {Rodr{\'\i}guez del Pino}, {{\"U}bler}, {Sirianni}, {Dressler}, {Egami}, {Eisenstein}, {Endsley}, {Hainline}, {Hausen}, {Johnson}, {Rieke}, {Robertson}, {Shivaei}, {Stark}, {Tacchella}, {Williams}, {Willmer}, {Bhatawdekar}, {Bowler}, {Boyett}, {Chen}, {de Graaff}, {Helton}, {Hviding}, {Jones}, {Kumari}, {Lyu}, {Nelson}, {Perna}, {Sandles}, {Saxena}, {Suess}, {Sun}, {Topping}, {Wallace}, \& {Whitler}}]{curtis-lake23}
{Curtis-Lake}, E., {Carniani}, S., {Cameron}, A., {et~al.} 2023, Nature Astronomy, 7, 622, \dodoi{10.1038/s41550-023-01918-w}

\bibitem[{{Davies} {et~al.}(2024){Davies}, {Belli}, {Park}, {Mendel}, {Johnson}, {Conroy}, {Benton}, {Bugiani}, {Emami}, {Leja}, {Li}, {Maheson}, {Mathews}, {Naidu}, {Nelson}, {Tacchella}, {Terrazas}, \& {Weinberger}}]{davies23}
{Davies}, R.~L., {Belli}, S., {Park}, M., {et~al.} 2024, \mnras, 528, 4976, \dodoi{10.1093/mnras/stae327}

\bibitem[{{Davis} {et~al.}(2007){Davis}, {Guhathakurta}, {Konidaris}, {Newman}, {Ashby}, {Biggs}, {Barmby}, {Bundy}, {Chapman}, {Coil}, {Conselice}, {Cooper}, {Croton}, {Eisenhardt}, {Ellis}, {Faber}, {Fang}, {Fazio}, {Georgakakis}, {Gerke}, {Goss}, {Gwyn}, {Harker}, {Hopkins}, {Huang}, {Ivison}, {Kassin}, {Kirby}, {Koekemoer}, {Koo}, {Laird}, {Le Floc'h}, {Lin}, {Lotz}, {Marshall}, {Martin}, {Metevier}, {Moustakas}, {Nandra}, {Noeske}, {Papovich}, {Phillips}, {Rich}, {Rieke}, {Rigopoulou}, {Salim}, {Schiminovich}, {Simard}, {Smail}, {Small}, {Weiner}, {Willmer}, {Willner}, {Wilson}, {Wright}, \& {Yan}}]{davis07}
{Davis}, M., {Guhathakurta}, P., {Konidaris}, N.~P., {et~al.} 2007, \apjl, 660, L1, \dodoi{10.1086/517931}

\bibitem[{{de Graaff} {et~al.}(2024){de Graaff}, {Setton}, {Brammer}, {Cutler}, {Suess}, {Labbe}, {Leja}, {Weibel}, {Maseda}, {Whitaker}, {Bezanson}, {Boogaard}, {Cleri}, {De Lucia}, {Franx}, {Greene}, {Hirschmann}, {Matthee}, {McConachie}, {Naidu}, {Oesch}, {Price}, {Rix}, {Valentino}, {Wang}, \& {Williams}}]{degraaff24}
{de Graaff}, A., {Setton}, D.~J., {Brammer}, G., {et~al.} 2024, arXiv e-prints, arXiv:2404.05683, \dodoi{10.48550/arXiv.2404.05683}

\bibitem[{{de Graaff} {et~al.}(2025){de Graaff}, {Brammer}, {Weibel}, {Lewis}, {Maseda}, {Oesch}, {Bezanson}, {Boogaard}, {Cleri}, {Cooper}, {Gottumukkala}, {Greene}, {Hirschmann}, {Hviding}, {Katz}, {Labb{\'e}}, {Leja}, {Matthee}, {McConachie}, {Miller}, {Naidu}, {Price}, {Rix}, {Setton}, {Suess}, {Wang}, {Whitaker}, \& {Williams}}]{degraaff25}
{de Graaff}, A., {Brammer}, G., {Weibel}, A., {et~al.} 2025, \aap, 697, A189, \dodoi{10.1051/0004-6361/202452186}

\bibitem[{{De Lucia} {et~al.}(2025){De Lucia}, {Fontanot}, {Hirschmann}, \& {Xie}}]{delucia25}
{De Lucia}, G., {Fontanot}, F., {Hirschmann}, M., \& {Xie}, L. 2025, arXiv e-prints, arXiv:2502.01724, \dodoi{10.48550/arXiv.2502.01724}

\bibitem[{{Desprez} {et~al.}(2024){Desprez}, {Martis}, {Asada}, {Sawicki}, {Willott}, {Muzzin}, {Abraham}, {Brada{\v{c}}}, {Brammer}, {Estrada-Carpenter}, {Iyer}, {Matharu}, {Mowla}, {Noirot}, {Sarrouh}, {Strait}, {Gledhill}, \& {Rihtar{\v{s}}i{\v{c}}}}]{desprez24}
{Desprez}, G., {Martis}, N.~S., {Asada}, Y., {et~al.} 2024, \mnras, 530, 2935, \dodoi{10.1093/mnras/stae1084}

\bibitem[{{Donnan} {et~al.}(2023){Donnan}, {McLeod}, {Dunlop}, {McLure}, {Carnall}, {Begley}, {Cullen}, {Hamadouche}, {Bowler}, {Magee}, {McCracken}, {Milvang-Jensen}, {Moneti}, \& {Targett}}]{donnan23}
{Donnan}, C.~T., {McLeod}, D.~J., {Dunlop}, J.~S., {et~al.} 2023, \mnras, 518, 6011, \dodoi{10.1093/mnras/stac3472}

\bibitem[{{Donnan} {et~al.}(2024){Donnan}, {McLure}, {Dunlop}, {McLeod}, {Magee}, {Arellano-C{\'o}rdova}, {Barrufet}, {Begley}, {Bowler}, {Carnall}, {Cullen}, {Ellis}, {Fontana}, {Illingworth}, {Grogin}, {Hamadouche}, {Koekemoer}, {Liu}, {Mason}, {Santini}, \& {Stanton}}]{donnan24}
{Donnan}, C.~T., {McLure}, R.~J., {Dunlop}, J.~S., {et~al.} 2024, \mnras, 533, 3222, \dodoi{10.1093/mnras/stae2037}

\bibitem[{{Doyon} {et~al.}(2023){Doyon}, {Willott}, {Hutchings}, {Sivaramakrishnan}, {Albert}, {Lafreni{\`e}re}, {Rowlands}, {Bego{\~n}a Vila}, {Martel}, {LaMassa}, {Aldridge}, {Artigau}, {Cameron}, {Chayer}, {Cook}, {Cooper}, {Darveau-Bernier}, {Dupuis}, {Earnshaw}, {Espinoza}, {Filippazzo}, {Fullerton}, {Gaudreau}, {Gawlik}, {Goudfrooij}, {Haley}, {Kammerer}, {Kendall}, {Lambros}, {Ignat}, {Maszkiewicz}, {McColgan}, {Morishita}, {Ouellette}, {Pacifici}, {Philippi}, {Radica}, {Ravindranath}, {Rowe}, {Roy}, {Roy}, {Saad}, {Sohn}, {Talens}, {Touahri}, {Thatte}, {Taylor}, {Vandal}, {Volk}, {Wander}, {Warner}, {Zheng}, {Zhou}, {Abraham}, {Beaulieu}, {Benneke}, {Ferrarese}, {Jayawardhana}, {Johnstone}, {Kaltenegger}, {Meyer}, {Pipher}, {Rameau}, {Rieke}, {Salhi}, \& {Sawicki}}]{doyon23}
{Doyon}, R., {Willott}, C.~J., {Hutchings}, J.~B., {et~al.} 2023, \pasp, 135, 098001, \dodoi{10.1088/1538-3873/acd41b}

\bibitem[{{Eisenstein} {et~al.}(2023{\natexlab{a}}){Eisenstein}, {Willott}, {Alberts}, {Arribas}, {Bonaventura}, {Bunker}, {Cameron}, {Carniani}, {Charlot}, {Curtis-Lake}, {D'Eugenio}, {Endsley}, {Ferruit}, {Giardino}, {Hainline}, {Hausen}, {Jakobsen}, {Johnson}, {Maiolino}, {Rieke}, {Rieke}, {Rix}, {Robertson}, {Stark}, {Tacchella}, {Williams}, {Willmer}, {Baker}, {Baum}, {Bhatawdekar}, {Boyett}, {Chen}, {Chevallard}, {Circosta}, {Curti}, {Danhaive}, {DeCoursey}, {de Graaff}, {Dressler}, {Egami}, {Helton}, {Hviding}, {Ji}, {Jones}, {Kumari}, {L{\"u}tzgendorf}, {Laseter}, {Looser}, {Lyu}, {Maseda}, {Nelson}, {Parlanti}, {Perna}, {Pusk{\'a}s}, {Rawle}, {Rodr{\'\i}guez Del Pino}, {Sandles}, {Saxena}, {Scholtz}, {Sharpe}, {Shivaei}, {Silcock}, {Simmonds}, {Skarbinski}, {Smit}, {Stone}, {Suess}, {Sun}, {Tang}, {Topping}, {{\"U}bler}, {Villanueva}, {Wallace}, {Whitler}, {Witstok}, \& {Woodrum}}]{eisenstein23a}
{Eisenstein}, D.~J., {Willott}, C., {Alberts}, S., {et~al.} 2023{\natexlab{a}}, arXiv e-prints, arXiv:2306.02465, \dodoi{10.48550/arXiv.2306.02465}

\bibitem[{{Eisenstein} {et~al.}(2023{\natexlab{b}}){Eisenstein}, {Johnson}, {Robertson}, {Tacchella}, {Hainline}, {Jakobsen}, {Maiolino}, {Bonaventura}, {Bunker}, {Cameron}, {Cargile}, {Curtis-Lake}, {Hausen}, {Pusk{\'a}s}, {Rieke}, {Sun}, {Willmer}, {Willott}, {Alberts}, {Arribas}, {Baker}, {Baum}, {Bhatawdekar}, {Carniani}, {Charlot}, {Chen}, {Chevallard}, {Curti}, {DeCoursey}, {D'Eugenio}, {de Graaff}, {Egami}, {Helton}, {Ji}, {Jones}, {Kumari}, {L{\"u}tzgendorf}, {Laseter}, {Looser}, {Lyu}, {Maseda}, {Nelson}, {Parlanti}, {Rauscher}, {Rawle}, {Rieke}, {Rix}, {Rujopakarn}, {Sandles}, {Saxena}, {Scholtz}, {Sharpe}, {Shivaei}, {Simmonds}, {Smit}, {Topping}, {{\"U}bler}, {Venturi}, {Williams}, {Witstok}, \& {Woodrum}}]{eisenstein23b}
{Eisenstein}, D.~J., {Johnson}, B.~D., {Robertson}, B., {et~al.} 2023{\natexlab{b}}, arXiv e-prints, arXiv:2310.12340, \dodoi{10.48550/arXiv.2310.12340}

\bibitem[{{Endsley} {et~al.}(2024){Endsley}, {Chisholm}, {Stark}, {Topping}, \& {Whitler}}]{endsley24}
{Endsley}, R., {Chisholm}, J., {Stark}, D.~P., {Topping}, M.~W., \& {Whitler}, L. 2024, arXiv e-prints, arXiv:2410.01905, \dodoi{10.48550/arXiv.2410.01905}

\bibitem[{{Faucher-Gigu{\`e}re}(2018)}]{faucher-giguere18}
{Faucher-Gigu{\`e}re}, C.-A. 2018, \mnras, 473, 3717, \dodoi{10.1093/mnras/stx2595}

\bibitem[{{Finkelstein} {et~al.}(2022){Finkelstein}, {Bagley}, {Arrabal Haro}, {Dickinson}, {Ferguson}, {Kartaltepe}, {Papovich}, {Burgarella}, {Kocevski}, {Huertas-Company}, {Iyer}, {Koekemoer}, {Larson}, {P{\'e}rez-Gonz{\'a}lez}, {Rose}, {Tacchella}, {Wilkins}, {Chworowsky}, {Medrano}, {Morales}, {Somerville}, {Yung}, {Fontana}, {Giavalisco}, {Grazian}, {Grogin}, {Kewley}, {Kirkpatrick}, {Kurczynski}, {Lotz}, {Pentericci}, {Pirzkal}, {Ravindranath}, {Ryan}, {Trump}, {Yang}, {Almaini}, {Amor{\'\i}n}, {Annunziatella}, {Backhaus}, {Barro}, {Behroozi}, {Bell}, {Bhatawdekar}, {Bisigello}, {Bromm}, {Buat}, {Buitrago}, {Calabr{\`o}}, {Casey}, {Castellano}, {Ch{\'a}vez Ortiz}, {Ciesla}, {Cleri}, {Cohen}, {Cole}, {Cooke}, {Cooper}, {Cooray}, {Costantin}, {Cox}, {Croton}, {Daddi}, {Dav{\'e}}, {de La Vega}, {Dekel}, {Elbaz}, {Estrada-Carpenter}, {Faber}, {Fern{\'a}ndez}, {Finkelstein}, {Freundlich}, {Fujimoto}, {Garc{\'\i}a-Argum{\'a}nez}, {Gardner}, {Gawiser}, {G{\'o}mez-Guijarro}, {Guo}, {Hamblin}, {Hamilton},
  {Hathi}, {Holwerda}, {Hirschmann}, {Hutchison}, {Jaskot}, {Jha}, {Jogee}, {Juneau}, {Jung}, {Kassin}, {Le Bail}, {Leung}, {Lucas}, {Magnelli}, {Mantha}, {Matharu}, {McGrath}, {McIntosh}, {Merlin}, {Mobasher}, {Newman}, {Nicholls}, {Pandya}, {Rafelski}, {Ronayne}, {Santini}, {Seill{\'e}}, {Shah}, {Shen}, {Simons}, {Snyder}, {Stanway}, {Straughn}, {Teplitz}, {Vanderhoof}, {Vega-Ferrero}, {Wang}, {Weiner}, {Willmer}, {Wuyts}, {Zavala}, \& {Ceers Team}}]{finkelstein22}
{Finkelstein}, S.~L., {Bagley}, M.~B., {Arrabal Haro}, P., {et~al.} 2022, \apjl, 940, L55, \dodoi{10.3847/2041-8213/ac966e}

\bibitem[{{Finkelstein} {et~al.}(2025){Finkelstein}, {Bagley}, {Arrabal Haro}, {Dickinson}, {Ferguson}, {Kartaltepe}, {Kocevski}, {Koekemoer}, {Lotz}, {Papovich}, {P{\'e}rez-Gonz{\'a}lez}, {Pirzkal}, {Somerville}, {Trump}, {Yang}, {Yung}, {Fontana}, {Grazian}, {Grogin}, {Kewley}, {Kirkpatrick}, {Larson}, {Pentericci}, {Ravindranath}, {Wilkins}, {Almaini}, {Amor{\'\i}n}, {Barro}, {Bhatawdekar}, {Bisigello}, {Brooks}, {Buat}, {Buitrago}, {Burgarella}, {Calabr{\`o}}, {Castellano}, {Cheng}, {Cleri}, {Cole}, {Cooper}, {Cooper}, {Costantin}, {Cox}, {Croton}, {Daddi}, {Davis}, {Dekel}, {Elbaz}, {Fern{\'a}ndez}, {Fujimoto}, {Gandolfi}, {Gardner}, {Gawiser}, {Giavalisco}, {G{\'o}mez-Guijarro}, {Guo}, {Gupta}, {Hathi}, {Harish}, {Henry}, {Hirschmann}, {Hu}, {Hutchison}, {Iyer}, {Jaskot}, {Jha}, {Jung}, {Kassin}, {Kokorev}, {Kurczynski}, {Leung}, {Llerena}, {Long}, {Lucas}, {Lu}, {McGrath}, {McIntosh}, {Merlin}, {Mobasher}, {Morales}, {Napolitano}, {Pacucci}, {Pandya}, {Rafelski}, {Rodighiero}, {Rose}, {Santini},
  {Seill{\'e}}, {Simons}, {Shen}, {Straughn}, {Tacchella}, {Taylor}, {Vanderhoof}, {Vega-Ferrero}, {Weiner}, {Willmer}, {Zhu}, {Bell}, {Wuyts}, {Holwerda}, {Wang}, {Wang}, {Zavala}, \& {CEERS Collaboration}}]{finkelstein25}
---. 2025, \apjl, 983, L4, \dodoi{10.3847/2041-8213/adbbd3}

\bibitem[{{Fontanot} {et~al.}(2009){Fontanot}, {De Lucia}, {Monaco}, {Somerville}, \& {Santini}}]{fontanot09}
{Fontanot}, F., {De Lucia}, G., {Monaco}, P., {Somerville}, R.~S., \& {Santini}, P. 2009, \mnras, 397, 1776, \dodoi{10.1111/j.1365-2966.2009.15058.x}

\bibitem[{{F{\"o}rster Schreiber} {et~al.}(2004){F{\"o}rster Schreiber}, {van Dokkum}, {Franx}, {Labb{\'e}}, {Rudnick}, {Daddi}, {Illingworth}, {Kriek}, {Moorwood}, {Rix}, {R{\"o}ttgering}, {Trujillo}, {van der Werf}, {van Starkenburg}, \& {Wuyts}}]{forster04}
{F{\"o}rster Schreiber}, N.~M., {van Dokkum}, P.~G., {Franx}, M., {et~al.} 2004, \apj, 616, 40, \dodoi{10.1086/424838}

\bibitem[{{Franco} {et~al.}(2024){Franco}, {Akins}, {Casey}, {Finkelstein}, {Shuntov}, {Chworowsky}, {Faisst}, {Fujimoto}, {Ilbert}, {Koekemoer}, {Liu}, {Lovell}, {Maraston}, {McCracken}, {McKinney}, {Robertson}, {Bagley}, {Champagne}, {Cooper}, {Ding}, {Drakos}, {Enia}, {Gillman}, {Gozaliasl}, {Harish}, {Hayward}, {Hirschmann}, {Jin}, {Kartaltepe}, {Kokorev}, {Laigle}, {Long}, {Magdis}, {Mahler}, {Martin}, {Massey}, {Mobasher}, {Paquereau}, {Renzini}, {Rhodes}, {Rich}, {Sheth}, {Silverman}, {Sparre}, {Talia}, {Trakhtenbrot}, {Valentino}, {Vijayan}, {Wilkins}, {Yang}, \& {Zavala}}]{Franco2024}
{Franco}, M., {Akins}, H.~B., {Casey}, C.~M., {et~al.} 2024, \apj, 973, 23, \dodoi{10.3847/1538-4357/ad5e6a}

\bibitem[{{Fujimoto} {et~al.}(2023{\natexlab{a}}){Fujimoto}, {Arrabal Haro}, {Dickinson}, {Finkelstein}, {Kartaltepe}, {Larson}, {Burgarella}, {Bagley}, {Behroozi}, {Chworowsky}, {Hirschmann}, {Trump}, {Wilkins}, {Yung}, {Koekemoer}, {Papovich}, {Pirzkal}, {Ferguson}, {Fontana}, {Grogin}, {Grazian}, {Kewley}, {Kocevski}, {Lotz}, {Pentericci}, {Ravindranath}, {Somerville}, {Wilkins}, {Amor{\'\i}n}, {Backhaus}, {Calabr{\`o}}, {Casey}, {Cooper}, {Fern{\'a}ndez}, {Franco}, {Giavalisco}, {Hathi}, {Harish}, {Hutchison}, {Iyer}, {Jung}, {Lucas}, \& {Zavala}}]{fujimoto2023a}
{Fujimoto}, S., {Arrabal Haro}, P., {Dickinson}, M., {et~al.} 2023{\natexlab{a}}, \apjl, 949, L25, \dodoi{10.3847/2041-8213/acd2d9}

\bibitem[{{Fujimoto} {et~al.}(2023{\natexlab{b}}){Fujimoto}, {Finkelstein}, {Burgarella}, {Carilli}, {Buat}, {Casey}, {Ciesla}, {Tacchella}, {Zavala}, {Brammer}, {Fudamoto}, {Ouchi}, {Valentino}, {Cooper}, {Dickinson}, {Franco}, {Giavalisco}, {Hutchison}, {Kartaltepe}, {Koekemoer}, {Kojima}, {Larson}, {Murphy}, {Papovich}, {P{\'e}rez-Gonz{\'a}lez}, {Somerville}, {Yoon}, {Wilkins}, {Akins}, {Amor{\'\i}n}, {Arrabal Haro}, {Bagley}, {Chworowsky}, {Cleri}, {Cooper}, {Costantin}, {Daddi}, {Ferguson}, {Grogin}, {Jim{\'e}nez-Andrade}, {Juneau}, {Kirkpatrick}, {Kocevski}, {Le Bail}, {Long}, {Lucas}, {Magnelli}, {McKinney}, {Rose}, {Seill{\'e}}, {Simons}, {Weiner}, \& {Yung}}]{fujimoto2023b}
{Fujimoto}, S., {Finkelstein}, S.~L., {Burgarella}, D., {et~al.} 2023{\natexlab{b}}, \apj, 955, 130, \dodoi{10.3847/1538-4357/aceb67}

\bibitem[{{Fujimoto} {et~al.}(2025){Fujimoto}, {Naidu}, {Chisholm}, {Atek}, {Endsley}, {Kokorev}, {Furtak}, {Pan}, {Liu}, {Bromm}, {Venditti}, {Visbal}, {Sarmento}, {Weibel}, {Oesch}, {Brammer}, {Schaerer}, {Adamo}, {Berg}, {Bezanson}, {Chemerynska}, {Claeyssens}, {Dessauges-Zavadsky}, {Frebel}, {Korber}, {Labbe}, {Marques-Chaves}, {Matthee}, {McQuinn}, {Mu{\~n}oz}, {Natarajan}, {Saldana-Lopez}, {Suess}, {Volonteri}, \& {Zitrin}}]{fujimoto25}
{Fujimoto}, S., {Naidu}, R.~P., {Chisholm}, J., {et~al.} 2025, arXiv e-prints, arXiv:2501.11678, \dodoi{10.48550/arXiv.2501.11678}

\bibitem[{{Furlanetto} \& {Mirocha}(2023)}]{furlanetto2023}
{Furlanetto}, S.~R., \& {Mirocha}, J. 2023, \mnras, 523, 5274, \dodoi{10.1093/mnras/stad1799}

\bibitem[{{Galametz} {et~al.}(2013){Galametz}, {Grazian}, {Fontana}, {Ferguson}, {Ashby}, {Barro}, {Castellano}, {Dahlen}, {Donley}, {Faber}, {Grogin}, {Guo}, {Huang}, {Kocevski}, {Koekemoer}, {Lee}, {McGrath}, {Peth}, {Willner}, {Almaini}, {Cooper}, {Cooray}, {Conselice}, {Dickinson}, {Dunlop}, {Fazio}, {Foucaud}, {Gardner}, {Giavalisco}, {Hathi}, {Hartley}, {Koo}, {Lai}, {de Mello}, {McLure}, {Lucas}, {Paris}, {Pentericci}, {Santini}, {Simpson}, {Sommariva}, {Targett}, {Weiner}, {Wuyts}, \& {CANDELS Team}}]{galametz13}
{Galametz}, A., {Grazian}, A., {Fontana}, A., {et~al.} 2013, \apjs, 206, 10, \dodoi{10.1088/0067-0049/206/2/10}

\bibitem[{{Gawiser} {et~al.}(2006){Gawiser}, {van Dokkum}, {Herrera}, {Maza}, {Castander}, {Infante}, {Lira}, {Quadri}, {Toner}, {Treister}, {Urry}, {Altmann}, {Assef}, {Christlein}, {Coppi}, {Dur{\'a}n}, {Franx}, {Galaz}, {Huerta}, {Liu}, {L{\'o}pez}, {M{\'e}ndez}, {Moore}, {Rubio}, {Ruiz}, {Toft}, \& {Yi}}]{gawiser06}
{Gawiser}, E., {van Dokkum}, P.~G., {Herrera}, D., {et~al.} 2006, \apjs, 162, 1, \dodoi{10.1086/497644}

\bibitem[{{Giavalisco} {et~al.}(2004){Giavalisco}, {Ferguson}, {Koekemoer}, {Dickinson}, {Alexander}, {Bauer}, {Bergeron}, {Biagetti}, {Brandt}, {Casertano}, {Cesarsky}, {Chatzichristou}, {Conselice}, {Cristiani}, {Da Costa}, {Dahlen}, {de Mello}, {Eisenhardt}, {Erben}, {Fall}, {Fassnacht}, {Fosbury}, {Fruchter}, {Gardner}, {Grogin}, {Hook}, {Hornschemeier}, {Idzi}, {Jogee}, {Kretchmer}, {Laidler}, {Lee}, {Livio}, {Lucas}, {Madau}, {Mobasher}, {Moustakas}, {Nonino}, {Padovani}, {Papovich}, {Park}, {Ravindranath}, {Renzini}, {Richardson}, {Riess}, {Rosati}, {Schirmer}, {Schreier}, {Somerville}, {Spinrad}, {Stern}, {Stiavelli}, {Strolger}, {Urry}, {Vandame}, {Williams}, \& {Wolf}}]{giavalisco04}
{Giavalisco}, M., {Ferguson}, H.~C., {Koekemoer}, A.~M., {et~al.} 2004, \apjl, 600, L93, \dodoi{10.1086/379232}

\bibitem[{{Gibson} {et~al.}(2024){Gibson}, {Nelson}, {Williams}, {Price}, {Whitaker}, {Suess}, {de Graaff}, {Johnson}, {Bunker}, {Baker}, {Bhatawdekar}, {Boyett}, {Charlot}, {Curtis-Lake}, {Eisenstein}, {Hainline}, {Hausen}, {Maiolino}, {Rieke}, {Rieke}, {Robertson}, {Tacchella}, \& {Willott}}]{gibson24}
{Gibson}, J.~L., {Nelson}, E., {Williams}, C.~C., {et~al.} 2024, \apj, 974, 48, \dodoi{10.3847/1538-4357/ad64c2}

\bibitem[{{Gim{\'e}nez-Arteaga} {et~al.}(2023){Gim{\'e}nez-Arteaga}, {Oesch}, {Brammer}, {Valentino}, {Mason}, {Weibel}, {Barrufet}, {Fujimoto}, {Heintz}, {Nelson}, {Strait}, {Suess}, \& {Gibson}}]{gimenezarteaga23}
{Gim{\'e}nez-Arteaga}, C., {Oesch}, P.~A., {Brammer}, G.~B., {et~al.} 2023, \apj, 948, 126, \dodoi{10.3847/1538-4357/acc5ea}

\bibitem[{{Glazebrook} {et~al.}(2024){Glazebrook}, {Nanayakkara}, {Schreiber}, {Lagos}, {Kawinwanichakij}, {Jacobs}, {Chittenden}, {Brammer}, {Kacprzak}, {Labbe}, {Marchesini}, {Marsan}, {Oesch}, {Papovich}, {Remus}, {Tran}, {Esdaile}, \& {Chandro-Gomez}}]{glazebrook24}
{Glazebrook}, K., {Nanayakkara}, T., {Schreiber}, C., {et~al.} 2024, \nat, 628, 277, \dodoi{10.1038/s41586-024-07191-9}

\bibitem[{{Gottumukkala} {et~al.}(2024){Gottumukkala}, {Barrufet}, {Oesch}, {Weibel}, {Allen}, {Alcalde Pampliega}, {Nelson}, {Williams}, {Brammer}, {Fudamoto}, {Gonz{\'a}lez}, {Heintz}, {Illingworth}, {Magee}, {Naidu}, {Shuntov}, {Stefanon}, {Toft}, {Valentino}, \& {Xiao}}]{gottumukkala24}
{Gottumukkala}, R., {Barrufet}, L., {Oesch}, P.~A., {et~al.} 2024, \mnras, 530, 966, \dodoi{10.1093/mnras/stae754}

\bibitem[{{Grogin} {et~al.}(2011){Grogin}, {Kocevski}, {Faber}, {Ferguson}, {Koekemoer}, {Riess}, {Acquaviva}, {Alexander}, {Almaini}, {Ashby}, {Barden}, {Bell}, {Bournaud}, {Brown}, {Caputi}, {Casertano}, {Cassata}, {Castellano}, {Challis}, {Chary}, {Cheung}, {Cirasuolo}, {Conselice}, {Roshan Cooray}, {Croton}, {Daddi}, {Dahlen}, {Dav{\'e}}, {de Mello}, {Dekel}, {Dickinson}, {Dolch}, {Donley}, {Dunlop}, {Dutton}, {Elbaz}, {Fazio}, {Filippenko}, {Finkelstein}, {Fontana}, {Gardner}, {Garnavich}, {Gawiser}, {Giavalisco}, {Grazian}, {Guo}, {Hathi}, {H{\"a}ussler}, {Hopkins}, {Huang}, {Huang}, {Jha}, {Kartaltepe}, {Kirshner}, {Koo}, {Lai}, {Lee}, {Li}, {Lotz}, {Lucas}, {Madau}, {McCarthy}, {McGrath}, {McIntosh}, {McLure}, {Mobasher}, {Moustakas}, {Mozena}, {Nandra}, {Newman}, {Niemi}, {Noeske}, {Papovich}, {Pentericci}, {Pope}, {Primack}, {Rajan}, {Ravindranath}, {Reddy}, {Renzini}, {Rix}, {Robaina}, {Rodney}, {Rosario}, {Rosati}, {Salimbeni}, {Scarlata}, {Siana}, {Simard}, {Smidt}, {Somerville}, {Spinrad},
  {Straughn}, {Strolger}, {Telford}, {Teplitz}, {Trump}, {van der Wel}, {Villforth}, {Wechsler}, {Weiner}, {Wiklind}, {Wild}, {Wilson}, {Wuyts}, {Yan}, \& {Yun}}]{grogin11}
{Grogin}, N.~A., {Kocevski}, D.~D., {Faber}, S.~M., {et~al.} 2011, \apjs, 197, 35, \dodoi{10.1088/0067-0049/197/2/35}

\bibitem[{{Guo} {et~al.}(2013){Guo}, {Ferguson}, {Giavalisco}, {Barro}, {Willner}, {Ashby}, {Dahlen}, {Donley}, {Faber}, {Fontana}, {Galametz}, {Grazian}, {Huang}, {Kocevski}, {Koekemoer}, {Koo}, {McGrath}, {Peth}, {Salvato}, {Wuyts}, {Castellano}, {Cooray}, {Dickinson}, {Dunlop}, {Fazio}, {Gardner}, {Gawiser}, {Grogin}, {Hathi}, {Hsu}, {Lee}, {Lucas}, {Mobasher}, {Nandra}, {Newman}, \& {van der Wel}}]{guo13}
{Guo}, Y., {Ferguson}, H.~C., {Giavalisco}, M., {et~al.} 2013, \apjs, 207, 24, \dodoi{10.1088/0067-0049/207/2/24}

\bibitem[{{Hamadouche} {et~al.}(2025){Hamadouche}, {McLure}, {Carnall}, {McLeod}, {Dunlop}, {Whitaker}, {Donnan}, {Begley}, {Stanton}, {Almaini}, {Aird}, {Cullen}, {Cutler}, {Grogin}, \& {Koekemoer}}]{hamadouche25}
{Hamadouche}, M.~L., {McLure}, R.~J., {Carnall}, A.~C., {et~al.} 2025, \mnras, 541, 463, \dodoi{10.1093/mnras/staf971}

\bibitem[{{Harvey} {et~al.}(2024){Harvey}, {Conselice}, {Adams}, {Austin}, {Juodzbalis}, {Trussler}, {Li}, {Ormerod}, {Ferreira}, {Duan}, {Westcott}, {Harris}, {Bhatawdekar}, {Coe}, {Cohen}, {Caruana}, {Cheng}, {Driver}, {Frye}, {Furtak}, {Grogin}, {Hathi}, {Holwerda}, {Jansen}, {Koekemoer}, {Lovell}, {Marshall}, {Nonino}, {Smail}, {Vijayan}, {Wilkins}, {Windhorst}, {Willmer}, {Yan}, \& {Zitrin}}]{harvey24}
{Harvey}, T., {Conselice}, C., {Adams}, N.~J., {et~al.} 2024, arXiv e-prints, arXiv:2403.03908, \dodoi{10.48550/arXiv.2403.03908}

\bibitem[{{Harvey} {et~al.}(2025{\natexlab{a}}){Harvey}, {Conselice}, {Adams}, {Austin}, {Juod{\v{z}}balis}, {Trussler}, {Li}, {Ormerod}, {Ferreira}, {Lovell}, {Duan}, {Westcott}, {Harris}, {Bhatawdekar}, {Coe}, {Cohen}, {Caruana}, {Cheng}, {Driver}, {Frye}, {Furtak}, {Grogin}, {Hathi}, {Holwerda}, {Jansen}, {Koekemoer}, {Marshall}, {Nonino}, {Vijayan}, {Wilkins}, {Windhorst}, {Willmer}, {Yan}, \& {Zitrin}}]{harvey25}
{Harvey}, T., {Conselice}, C.~J., {Adams}, N.~J., {et~al.} 2025{\natexlab{a}}, \apj, 978, 89, \dodoi{10.3847/1538-4357/ad8c29}

\bibitem[{{Harvey} {et~al.}(2025{\natexlab{b}}){Harvey}, {Conselice}, {Adams}, {Austin}, {Li}, {Rusakov}, {Westcott}, {Goolsby}, {Lovell}, {Cochrane}, {Vijayan}, \& {Trussler}}]{harvey25_outshining}
---. 2025{\natexlab{b}}, arXiv e-prints, arXiv:2504.05244, \dodoi{10.48550/arXiv.2504.05244}

\bibitem[{{Ilbert} {et~al.}(2006){Ilbert}, {Arnouts}, {McCracken}, {Bolzonella}, {Bertin}, {Le F{\`e}vre}, {Mellier}, {Zamorani}, {Pell{\`o}}, {Iovino}, {Tresse}, {Le Brun}, {Bottini}, {Garilli}, {Maccagni}, {Picat}, {Scaramella}, {Scodeggio}, {Vettolani}, {Zanichelli}, {Adami}, {Bardelli}, {Cappi}, {Charlot}, {Ciliegi}, {Contini}, {Cucciati}, {Foucaud}, {Franzetti}, {Gavignaud}, {Guzzo}, {Marano}, {Marinoni}, {Mazure}, {Meneux}, {Merighi}, {Paltani}, {Pollo}, {Pozzetti}, {Radovich}, {Zucca}, {Bondi}, {Bongiorno}, {Busarello}, {de La Torre}, {Gregorini}, {Lamareille}, {Mathez}, {Merluzzi}, {Ripepi}, {Rizzo}, \& {Vergani}}]{ilbert06}
{Ilbert}, O., {Arnouts}, S., {McCracken}, H.~J., {et~al.} 2006, \aap, 457, 841, \dodoi{10.1051/0004-6361:20065138}

\bibitem[{{Iyer} \& {Gawiser}(2017)}]{iyer17}
{Iyer}, K., \& {Gawiser}, E. 2017, \apj, 838, 127, \dodoi{10.3847/1538-4357/aa63f0}

\bibitem[{{Iyer} {et~al.}(2019){Iyer}, {Gawiser}, {Faber}, {Ferguson}, {Kartaltepe}, {Koekemoer}, {Pacifici}, \& {Somerville}}]{iyer19}
{Iyer}, K.~G., {Gawiser}, E., {Faber}, S.~M., {et~al.} 2019, \apj, 879, 116, \dodoi{10.3847/1538-4357/ab2052}

\bibitem[{{Iyer} {et~al.}(2020){Iyer}, {Tacchella}, {Genel}, {Hayward}, {Hernquist}, {Brooks}, {Caplar}, {Dav{\'e}}, {Diemer}, {Forbes}, {Gawiser}, {Somerville}, \& {Starkenburg}}]{iyer20}
{Iyer}, K.~G., {Tacchella}, S., {Genel}, S., {et~al.} 2020, \mnras, 498, 430, \dodoi{10.1093/mnras/staa2150}

\bibitem[{{Ji} {et~al.}(2024){Ji}, {Williams}, {Tacchella}, {Suess}, {Baker}, {Alberts}, {Bunker}, {Johnson}, {Robertson}, {Sun}, {Eisenstein}, {Rieke}, {Maseda}, {Hainline}, {Hausen}, {Rieke}, {Willmer}, {Egami}, {Shivaei}, {Carniani}, {Charlot}, {Chevallard}, {Curtis-Lake}, {Looser}, {Maiolino}, {Willott}, {Chen}, {Helton}, {Lyu}, {Nelson}, {Bhatawdekar}, {Boyett}, \& {Sandles}}]{ji24b}
{Ji}, Z., {Williams}, C.~C., {Tacchella}, S., {et~al.} 2024, \apj, 974, 135, \dodoi{10.3847/1538-4357/ad6e7f}

\bibitem[{{Jin} {et~al.}(2024){Jin}, {Sillassen}, {Hodge}, {Magdis}, {Rizzo}, {Casey}, {Koekemoer}, {Valentino}, {Kokorev}, {Magnelli}, {Gobat}, {Gillman}, {Franco}, {Faisst}, {Kartaltepe}, {Schinnerer}, {Toft}, {Algera}, {Harish}, {Lee}, {Liu}, {Shuntov}, {Talia}, \& {Vijayan}}]{jin24}
{Jin}, S., {Sillassen}, N.~B., {Hodge}, J., {et~al.} 2024, \aap, 690, L16, \dodoi{10.1051/0004-6361/202451445}

\bibitem[{{Kannan} {et~al.}(2023){Kannan}, {Springel}, {Hernquist}, {Pakmor}, {Delgado}, {Hadzhiyska}, {Hern{\'a}ndez-Aguayo}, {Barrera}, {Ferlito}, {Bose}, {White}, {Frenk}, {Smith}, \& {Garaldi}}]{kannan23}
{Kannan}, R., {Springel}, V., {Hernquist}, L., {et~al.} 2023, \mnras, 524, 2594, \dodoi{10.1093/mnras/stac3743}

\bibitem[{{Katz} {et~al.}(2024){Katz}, {Cameron}, {Saxena}, {Barrufet}, {Choustikov}, {Cleri}, {de Graaff}, {Ellis}, {Fosbury}, {Heintz}, {Maseda}, {Matthee}, {McConchie}, \& {Oesch}}]{katz24}
{Katz}, H., {Cameron}, A.~J., {Saxena}, A., {et~al.} 2024, arXiv e-prints, arXiv:2408.03189, \dodoi{10.48550/arXiv.2408.03189}

\bibitem[{{Kocevski} {et~al.}(2023){Kocevski}, {Onoue}, {Inayoshi}, {Trump}, {Arrabal Haro}, {Grazian}, {Dickinson}, {Finkelstein}, {Kartaltepe}, {Hirschmann}, {Aird}, {Holwerda}, {Fujimoto}, {Juneau}, {Amor{\'\i}n}, {Backhaus}, {Bagley}, {Barro}, {Bell}, {Bisigello}, {Calabr{\`o}}, {Cleri}, {Cooper}, {Ding}, {Grogin}, {Ho}, {Hutchison}, {Inoue}, {Jiang}, {Jones}, {Koekemoer}, {Li}, {Li}, {McGrath}, {Molina}, {Papovich}, {P{\'e}rez-Gonz{\'a}lez}, {Pirzkal}, {Wilkins}, {Yang}, \& {Yung}}]{kocevski2023}
{Kocevski}, D.~D., {Onoue}, M., {Inayoshi}, K., {et~al.} 2023, \apjl, 954, L4, \dodoi{10.3847/2041-8213/ace5a0}

\bibitem[{{Koekemoer} {et~al.}(2011){Koekemoer}, {Faber}, {Ferguson}, {Grogin}, {Kocevski}, {Koo}, {Lai}, {Lotz}, {Lucas}, {McGrath}, {Ogaz}, {Rajan}, {Riess}, {Rodney}, {Strolger}, {Casertano}, {Castellano}, {Dahlen}, {Dickinson}, {Dolch}, {Fontana}, {Giavalisco}, {Grazian}, {Guo}, {Hathi}, {Huang}, {van der Wel}, {Yan}, {Acquaviva}, {Alexander}, {Almaini}, {Ashby}, {Barden}, {Bell}, {Bournaud}, {Brown}, {Caputi}, {Cassata}, {Challis}, {Chary}, {Cheung}, {Cirasuolo}, {Conselice}, {Roshan Cooray}, {Croton}, {Daddi}, {Dav{\'e}}, {de Mello}, {de Ravel}, {Dekel}, {Donley}, {Dunlop}, {Dutton}, {Elbaz}, {Fazio}, {Filippenko}, {Finkelstein}, {Frazer}, {Gardner}, {Garnavich}, {Gawiser}, {Gruetzbauch}, {Hartley}, {H{\"a}ussler}, {Herrington}, {Hopkins}, {Huang}, {Jha}, {Johnson}, {Kartaltepe}, {Khostovan}, {Kirshner}, {Lani}, {Lee}, {Li}, {Madau}, {McCarthy}, {McIntosh}, {McLure}, {McPartland}, {Mobasher}, {Moreira}, {Mortlock}, {Moustakas}, {Mozena}, {Nandra}, {Newman}, {Nielsen}, {Niemi}, {Noeske}, {Papovich},
  {Pentericci}, {Pope}, {Primack}, {Ravindranath}, {Reddy}, {Renzini}, {Rix}, {Robaina}, {Rosario}, {Rosati}, {Salimbeni}, {Scarlata}, {Siana}, {Simard}, {Smidt}, {Snyder}, {Somerville}, {Spinrad}, {Straughn}, {Telford}, {Teplitz}, {Trump}, {Vargas}, {Villforth}, {Wagner}, {Wandro}, {Wechsler}, {Weiner}, {Wiklind}, {Wild}, {Wilson}, {Wuyts}, \& {Yun}}]{koekemoer11}
{Koekemoer}, A.~M., {Faber}, S.~M., {Ferguson}, H.~C., {et~al.} 2011, \apjs, 197, 36, \dodoi{10.1088/0067-0049/197/2/36}

\bibitem[{{Kokorev} {et~al.}(2023){Kokorev}, {Jin}, {Magdis}, {Caputi}, {Valentino}, {Dayal}, {Trebitsch}, {Brammer}, {Fujimoto}, {Bauer}, {Iani}, {Kohno}, {Bl{\'a}nquez Ses{\'e}}, {G{\'o}mez-Guijarro}, {Rinaldi}, \& {Navarro-Carrera}}]{kokorev23}
{Kokorev}, V., {Jin}, S., {Magdis}, G.~E., {et~al.} 2023, \apjl, 945, L25, \dodoi{10.3847/2041-8213/acbd9d}

\bibitem[{{Kokorev} {et~al.}(2025){Kokorev}, {Atek}, {Chisholm}, {Endsley}, {Chemerynska}, {Mu{\~n}oz}, {Furtak}, {Pan}, {Berg}, {Fujimoto}, {Oesch}, {Weibel}, {Adamo}, {Blaizot}, {Bouwens}, {Dessauges-Zavadsky}, {Khullar}, {Korber}, {Goovaerts}, {Jecmen}, {Labb{\'e}}, {Leclercq}, {Marques-Chaves}, {Mason}, {McQuinn}, {Naidu}, {Natarajan}, {Nelson}, {Rosdahl}, {Saldana-Lopez}, {Schaerer}, {Trebitsch}, {Volonteri}, \& {Zitrin}}]{kokorev25}
{Kokorev}, V., {Atek}, H., {Chisholm}, J., {et~al.} 2025, \apjl, 983, L22, \dodoi{10.3847/2041-8213/adc458}

\bibitem[{{Kriek} {et~al.}(2015){Kriek}, {Shapley}, {Reddy}, {Siana}, {Coil}, {Mobasher}, {Freeman}, {de Groot}, {Price}, {Sanders}, {Shivaei}, {Brammer}, {Momcheva}, {Skelton}, {van Dokkum}, {Whitaker}, {Aird}, {Azadi}, {Kassis}, {Bullock}, {Conroy}, {Dav{\'e}}, {Kere{\v{s}}}, \& {Krumholz}}]{kriek15}
{Kriek}, M., {Shapley}, A.~E., {Reddy}, N.~A., {et~al.} 2015, \apjs, 218, 15, \dodoi{10.1088/0067-0049/218/2/15}

\bibitem[{{Labb{\'e}} {et~al.}(2013){Labb{\'e}}, {Oesch}, {Bouwens}, {Illingworth}, {Magee}, {Gonz{\'a}lez}, {Carollo}, {Franx}, {Trenti}, {van Dokkum}, \& {Stiavelli}}]{labbe13}
{Labb{\'e}}, I., {Oesch}, P.~A., {Bouwens}, R.~J., {et~al.} 2013, \apjl, 777, L19, \dodoi{10.1088/2041-8205/777/2/L19}

\bibitem[{{Lawrence} {et~al.}(2007){Lawrence}, {Warren}, {Almaini}, {Edge}, {Hambly}, {Jameson}, {Lucas}, {Casali}, {Adamson}, {Dye}, {Emerson}, {Foucaud}, {Hewett}, {Hirst}, {Hodgkin}, {Irwin}, {Lodieu}, {McMahon}, {Simpson}, {Smail}, {Mortlock}, \& {Folger}}]{lawrence07}
{Lawrence}, A., {Warren}, S.~J., {Almaini}, O., {et~al.} 2007, \mnras, 379, 1599, \dodoi{10.1111/j.1365-2966.2007.12040.x}

\bibitem[{{Lebowitz} {et~al.}(2025){Lebowitz}, {Hainline}, {Juneau}, {Lyu}, {Williams}, {Alberts}, {Fan}, \& {Rieke}}]{lebowitz25}
{Lebowitz}, S., {Hainline}, K., {Juneau}, S., {et~al.} 2025, \apj, 984, 13, \dodoi{10.3847/1538-4357/adc07c}

\bibitem[{{Lines} {et~al.}(2025){Lines}, {Bowler}, {Adams}, {Fisher}, {Varadaraj}, {Nakazato}, {Aravena}, {Assef}, {Birkin}, {Ceverino}, {da Cunha}, {Cullen}, {De Looze}, {Donnan}, {Dunlop}, {Ferrara}, {Grogin}, {Herrera-Camus}, {Ikeda}, {Koekemoer}, {Killi}, {Li}, {McLeod}, {McLure}, {Mitsuhashi}, {P{\'e}rez-Gonz{\'a}lez}, {Relano}, {Solimano}, {Spilker}, {Villanueva}, \& {Yoshida}}]{lines25}
{Lines}, N.~E.~P., {Bowler}, R.~A.~A., {Adams}, N.~J., {et~al.} 2025, \mnras, 539, 2685, \dodoi{10.1093/mnras/staf627}

\bibitem[{{Looser} {et~al.}(2023){Looser}, {D'Eugenio}, {Maiolino}, {Tacchella}, {Curti}, {Arribas}, {Baker}, {Baum}, {Bonaventura}, {Boyett}, {Bunker}, {Carniani}, {Charlot}, {Chevallard}, {Curtis-Lake}, {Danhaive}, {Eisenstein}, {de Graaff}, {Hainline}, {Ji}, {Johnson}, {Kumari}, {Nelson}, {Parlanti}, {Rix}, {Robertson}, {Rodr{\'\i}guez Del Pino}, {Sandles}, {Scholtz}, {Smit}, {Stark}, {{\"U}bler}, {Williams}, {Willott}, \& {Witstok}}]{looser23}
{Looser}, T.~J., {D'Eugenio}, F., {Maiolino}, R., {et~al.} 2023, arXiv e-prints, arXiv:2306.02470, \dodoi{10.48550/arXiv.2306.02470}

\bibitem[{{Lorenz} {et~al.}(2025){Lorenz}, {Suess}, {Kriek}, {Price}, {Leja}, {Nelson}, {Atek}, {Bezanson}, {Brammer}, {Cutler}, {Dayal}, {de Graaff}, {Greene}, {Furtak}, {Labb{\'e}}, {Marchesini}, {Maseda}, {Miller}, {Mintz}, {Mitsuhashi}, {Pan}, {Porraz Barrera}, {Wang}, {Weaver}, {Williams}, \& {Whitaker}}]{lorenz25}
{Lorenz}, B., {Suess}, K.~A., {Kriek}, M., {et~al.} 2025, arXiv e-prints, arXiv:2505.10632, \dodoi{10.48550/arXiv.2505.10632}

\bibitem[{{Lotz} {et~al.}(2017){Lotz}, {Koekemoer}, {Coe}, {Grogin}, {Capak}, {Mack}, {Anderson}, {Avila}, {Barker}, {Borncamp}, {Brammer}, {Durbin}, {Gunning}, {Hilbert}, {Jenkner}, {Khandrika}, {Levay}, {Lucas}, {MacKenty}, {Ogaz}, {Porterfield}, {Reid}, {Robberto}, {Royle}, {Smith}, {Storrie-Lombardi}, {Sunnquist}, {Surace}, {Taylor}, {Williams}, {Bullock}, {Dickinson}, {Finkelstein}, {Natarajan}, {Richard}, {Robertson}, {Tumlinson}, {Zitrin}, {Flanagan}, {Sembach}, {Soifer}, \& {Mountain}}]{lotz17}
{Lotz}, J.~M., {Koekemoer}, A., {Coe}, D., {et~al.} 2017, \apj, 837, 97, \dodoi{10.3847/1538-4357/837/1/97}

\bibitem[{{Lovell} {et~al.}(2021){Lovell}, {Vijayan}, {Thomas}, {Wilkins}, {Barnes}, {Irodotou}, \& {Roper}}]{lovell21}
{Lovell}, C.~C., {Vijayan}, A.~P., {Thomas}, P.~A., {et~al.} 2021, \mnras, 500, 2127, \dodoi{10.1093/mnras/staa3360}

\bibitem[{{Lyu} {et~al.}(2024){Lyu}, {Alberts}, {Rieke}, {Shivaei}, {P{\'e}rez-Gonz{\'a}lez}, {Sun}, {Hainline}, {Baum}, {Bonaventura}, {Bunker}, {Egami}, {Eisenstein}, {Florian}, {Ji}, {Johnson}, {Morrison}, {Rieke}, {Robertson}, {Rujopakarn}, {Tacchella}, {Scholtz}, \& {Willmer}}]{lyu24}
{Lyu}, J., {Alberts}, S., {Rieke}, G.~H., {et~al.} 2024, \apj, 966, 229, \dodoi{10.3847/1538-4357/ad3643}

\bibitem[{{Man} \& {Belli}(2018)}]{manbelli18}
{Man}, A., \& {Belli}, S. 2018, Nature Astronomy, 2, 695, \dodoi{10.1038/s41550-018-0558-1}

\bibitem[{{Martis} {et~al.}(2025){Martis}, {Withers}, {Felicioni}, {Muzzin}, {Brada{\v{c}}}, {Abraham}, {Asada}, {Desprez}, {Iyer}, {Noirot}, {Sarrouh}, {Sawicki}, {Strait}, {Willot}, {Jagga}, {Jude{\v{z}}}, {Harshan}, {Marchesini}, {Markov}, {M{\'e}rida}, {Rihtar{\v{s}}i{\v{c}}}, \& {Tripodi}}]{martis25}
{Martis}, N., {Withers}, S., {Felicioni}, G., {et~al.} 2025, arXiv e-prints, arXiv:2503.01579, \dodoi{10.48550/arXiv.2503.01579}

\bibitem[{{Maseda} {et~al.}(2024){Maseda}, {de Graaff}, {Franx}, {Rix}, {Carniani}, {Laseter}, {Dudzevi{\v{c}}i{\={u}}t{\.{e}}}, {Rawle}, {Parlanti}, {Arribas}, {Bunker}, {Cameron}, {Charlot}, {Curti}, {D'Eugenio}, {Jones}, {Kumari}, {Maiolino}, {{\"U}bler}, {Saxena}, {Smit}, {Willott}, \& {Witstok}}]{maseda24}
{Maseda}, M.~V., {de Graaff}, A., {Franx}, M., {et~al.} 2024, \aap, 689, A73, \dodoi{10.1051/0004-6361/202449914}

\bibitem[{{McCracken} {et~al.}(2012){McCracken}, {Milvang-Jensen}, {Dunlop}, {Franx}, {Fynbo}, {Le F{\`e}vre}, {Holt}, {Caputi}, {Goranova}, {Buitrago}, {Emerson}, {Freudling}, {Hudelot}, {L{\'o}pez-Sanjuan}, {Magnard}, {Mellier}, {M{\o}ller}, {Nilsson}, {Sutherland}, {Tasca}, \& {Zabl}}]{mccracken12}
{McCracken}, H.~J., {Milvang-Jensen}, B., {Dunlop}, J., {et~al.} 2012, \aap, 544, A156, \dodoi{10.1051/0004-6361/201219507}

\bibitem[{{Mehta} {et~al.}(2024){Mehta}, {Rafelski}, {Sunnquist}, {Teplitz}, {Scarlata}, {Wang}, {Fontana}, {Hathi}, {Iyer}, {Alavi}, {Colbert}, {Grogin}, {Koekemoer}, {Nedkova}, {Hayes}, {Prichard}, {Siana}, {Smith}, {Windhorst}, {Ashcraft}, {Bagley}, {Baronchelli}, {Barro}, {Blanche}, {Broussard}, {Carleton}, {Chartab}, {Codoreanu}, {Cohen}, {Conselice}, {Dai}, {Darvish}, {Dav{\'e}}, {Degroot}, {de Mello}, {Dickinson}, {Emami}, {Ferguson}, {Ferreira}, {Finkelstein}, {Finkelstein}, {Gardner}, {Gawiser}, {Gburek}, {Giavalisco}, {Grazian}, {Gronwall}, {Guo}, {Arrabal Haro}, {Hemmati}, {Howell}, {Jansen}, {Ji}, {Kaviraj}, {Kim}, {Kurczynski}, {Lazar}, {Lucas}, {MacKenty}, {Mantha}, {Martin}, {Martin}, {McCabe}, {Mobasher}, {Morales}, {O'Connell}, {Olsen}, {Otteson}, {Ravindranath}, {Redshaw}, {Rutkowski}, {Robertson}, {Sattari}, {Soto}, {Sun}, {Taamoli}, {Vanzella}, {Yung}, {Zabelle}, \& {UVCANDELS Team}}]{mehta24}
{Mehta}, V., {Rafelski}, M., {Sunnquist}, B., {et~al.} 2024, \apjs, 275, 17, \dodoi{10.3847/1538-4365/ad7d8f}

\bibitem[{{Merlin} {et~al.}(2016){Merlin}, {Amor{\'\i}n}, {Castellano}, {Fontana}, {Buitrago}, {Dunlop}, {Elbaz}, {Boucaud}, {Bourne}, {Boutsia}, {Brammer}, {Bruce}, {Capak}, {Cappelluti}, {Ciesla}, {Comastri}, {Cullen}, {Derriere}, {Faber}, {Ferguson}, {Giallongo}, {Grazian}, {Lotz}, {Micha{\l}owski}, {Paris}, {Pentericci}, {Pilo}, {Santini}, {Schreiber}, {Shu}, \& {Wang}}]{merlin16}
{Merlin}, E., {Amor{\'\i}n}, R., {Castellano}, M., {et~al.} 2016, \aap, 590, A30, \dodoi{10.1051/0004-6361/201527513}

\bibitem[{{Miller} {et~al.}(2023){Miller}, {van Dokkum}, \& {Mowla}}]{miller23}
{Miller}, T.~B., {van Dokkum}, P., \& {Mowla}, L. 2023, \apj, 945, 155, \dodoi{10.3847/1538-4357/acbc74}

\bibitem[{{Miller} {et~al.}(2022){Miller}, {Whitaker}, {Nelson}, {van Dokkum}, {Bezanson}, {Brammer}, {Heintz}, {Leja}, {Suess}, \& {Weaver}}]{miller22}
{Miller}, T.~B., {Whitaker}, K.~E., {Nelson}, E.~J., {et~al.} 2022, \apjl, 941, L37, \dodoi{10.3847/2041-8213/aca675}

\bibitem[{{Mintz} {et~al.}(2025){Mintz}, {Setton}, {Greene}, {Leja}, {Wang}, {Burnham}, {Suess}, {Atek}, {Bezanson}, {Brammer}, {Cutler}, {Dayal}, {Feldmann}, {Furtak}, {Glazebrook}, {Khullar}, {Kokorev}, {Labb{\'e}}, {Maseda}, {Miller}, {Mitsuhashi}, {Nanayakkara}, {Pan}, {Price}, {Weaver}, \& {Whitaker}}]{mintz25}
{Mintz}, A., {Setton}, D.~J., {Greene}, J.~E., {et~al.} 2025, arXiv e-prints, arXiv:2506.16510, \dodoi{10.48550/arXiv.2506.16510}

\bibitem[{{Momcheva} {et~al.}(2016){Momcheva}, {Brammer}, {van Dokkum}, {Skelton}, {Whitaker}, {Nelson}, {Fumagalli}, {Maseda}, {Leja}, {Franx}, {Rix}, {Bezanson}, {Da Cunha}, {Dickey}, {F{\"o}rster Schreiber}, {Illingworth}, {Kriek}, {Labb{\'e}}, {Ulf Lange}, {Lundgren}, {Magee}, {Marchesini}, {Oesch}, {Pacifici}, {Patel}, {Price}, {Tal}, {Wake}, {van der Wel}, \& {Wuyts}}]{momcheva16}
{Momcheva}, I.~G., {Brammer}, G.~B., {van Dokkum}, P.~G., {et~al.} 2016, \apjs, 225, 27, \dodoi{10.3847/0067-0049/225/2/27}

\bibitem[{{Morales} {et~al.}(2024){Morales}, {Finkelstein}, {Leung}, {Bagley}, {Cleri}, {Dave}, {Dickinson}, {Ferguson}, {Hathi}, {Jones}, {Koekemoer}, {Papovich}, {P{\'e}rez-Gonz{\'a}lez}, {Pirzkal}, {Smith}, {Wilkins}, \& {Yung}}]{morales24}
{Morales}, A.~M., {Finkelstein}, S.~L., {Leung}, G. C.~K., {et~al.} 2024, \apjl, 964, L24, \dodoi{10.3847/2041-8213/ad2de4}

\bibitem[{{Morishita} {et~al.}(2025{\natexlab{a}}){Morishita}, {Liu}, {Stiavelli}, {Treu}, {Bergamini}, \& {Zhang}}]{morishita25_Z}
{Morishita}, T., {Liu}, Z., {Stiavelli}, M., {et~al.} 2025{\natexlab{a}}, arXiv e-prints, arXiv:2507.10521, \dodoi{10.48550/arXiv.2507.10521}

\bibitem[{{Morishita} {et~al.}(2025{\natexlab{b}}){Morishita}, {Mason}, {Kreilgaard}, {Trenti}, {Treu}, {Vulcani}, {Zhang}, {Abdurro'uf}, {Alavi}, {Atek}, {Bah{\'e}}, {Brada{\v{c}}}, {Bradley}, {Bunker}, {Coe}, {Colbert}, {Gelli}, {Hayes}, {Jones}, {Kodama}, {Leethochawalit}, {Liu}, {Malkan}, {Mehta}, {Metha}, {Newman}, {Rafelski}, {Roberts-Borsani}, {Rutkowski}, {Scarlata}, {Stiavelli}, {Sutanto}, {Takahashi}, {Teplitz}, \& {Wang}}]{morishita25}
{Morishita}, T., {Mason}, C.~A., {Kreilgaard}, K.~C., {et~al.} 2025{\natexlab{b}}, \apj, 983, 152, \dodoi{10.3847/1538-4357/adbbdc}

\bibitem[{{Mosleh} {et~al.}(2020){Mosleh}, {Hosseinnejad}, {Hosseini-ShahiSavandi}, \& {Tacchella}}]{mosleh20}
{Mosleh}, M., {Hosseinnejad}, S., {Hosseini-ShahiSavandi}, S.~Z., \& {Tacchella}, S. 2020, \apj, 905, 170, \dodoi{10.3847/1538-4357/abc7cc}

\bibitem[{{Naidu} {et~al.}(2022){Naidu}, {Oesch}, {Setton}, {Matthee}, {Conroy}, {Johnson}, {Weaver}, {Bouwens}, {Brammer}, {Dayal}, {Illingworth}, {Barrufet}, {Belli}, {Bezanson}, {Bose}, {Heintz}, {Leja}, {Leonova}, {Marques-Chaves}, {Stefanon}, {Toft}, {van der Wel}, {van Dokkum}, {Weibel}, \& {Whitaker}}]{naidu22}
{Naidu}, R.~P., {Oesch}, P.~A., {Setton}, D.~J., {et~al.} 2022, arXiv e-prints, arXiv:2208.02794, \dodoi{10.48550/arXiv.2208.02794}

\bibitem[{{Naidu} {et~al.}(2025){Naidu}, {Oesch}, {Brammer}, {Weibel}, {Li}, {Matthee}, {Chisholm}, {Pollock}, {Heintz}, {Johnson}, {Shen}, {Hviding}, {Leja}, {Tacchella}, {Ganguly}, {Witten}, {Atek}, {Belli}, {Bose}, {Bouwens}, {Dayal}, {Decarli}, {de Graaff}, {Fudamoto}, {Giovinazzo}, {Greene}, {Illingworth}, {Inoue}, {Kane}, {Labbe}, {Leonova}, {Marques-Chaves}, {Meyer}, {Nelson}, {Roberts-Borsani}, {Schaerer}, {Simcoe}, {Stefanon}, {Sugahara}, {Toft}, {van der Wel}, {van Dokkum}, {Walter}, {Watson}, {Weaver}, \& {Whitaker}}]{naidu25}
{Naidu}, R.~P., {Oesch}, P.~A., {Brammer}, G., {et~al.} 2025, arXiv e-prints, arXiv:2505.11263, \dodoi{10.48550/arXiv.2505.11263}

\bibitem[{{Narayanan} {et~al.}(2025){Narayanan}, {Stark}, {Finkelstein}, {Torrey}, {Li}, {Cullen}, {Topping}, {Marinacci}, {Sales}, {Shen}, \& {Vogelsberger}}]{narayanan25}
{Narayanan}, D., {Stark}, D.~P., {Finkelstein}, S.~L., {et~al.} 2025, \apj, 982, 7, \dodoi{10.3847/1538-4357/adb41c}

\bibitem[{{Nelson} {et~al.}(2016){Nelson}, {van Dokkum}, {F{\"o}rster Schreiber}, {Franx}, {Brammer}, {Momcheva}, {Wuyts}, {Whitaker}, {Skelton}, {Fumagalli}, {Hayward}, {Kriek}, {Labb{\'e}}, {Leja}, {Rix}, {Tacconi}, {van der Wel}, {van den Bosch}, {Oesch}, {Dickey}, \& {Ulf Lange}}]{nelson16}
{Nelson}, E.~J., {van Dokkum}, P.~G., {F{\"o}rster Schreiber}, N.~M., {et~al.} 2016, \apj, 828, 27, \dodoi{10.3847/0004-637X/828/1/27}

\bibitem[{{Nelson} {et~al.}(2023){Nelson}, {Suess}, {Bezanson}, {Price}, {van Dokkum}, {Leja}, {Wang}, {Whitaker}, {Labb{\'e}}, {Barrufet}, {Brammer}, {Eisenstein}, {Gibson}, {Hartley}, {Johnson}, {Heintz}, {Mathews}, {Miller}, {Oesch}, {Sandles}, {Setton}, {Speagle}, {Tacchella}, {Tadaki}, {{\"U}bler}, \& {Weaver}}]{nelson23}
{Nelson}, E.~J., {Suess}, K.~A., {Bezanson}, R., {et~al.} 2023, \apjl, 948, L18, \dodoi{10.3847/2041-8213/acc1e1}

\bibitem[{{Oesch} {et~al.}(2023){Oesch}, {Brammer}, {Naidu}, {Bouwens}, {Chisholm}, {Illingworth}, {Matthee}, {Nelson}, {Qin}, {Reddy}, {Shapley}, {Shivaei}, {van Dokkum}, {Weibel}, {Whitaker}, {Wuyts}, {Covelo-Paz}, {Endsley}, {Fudamoto}, {Giovinazzo}, {Herard-Demanche}, {Kerutt}, {Kramarenko}, {Labbe}, {Leonova}, {Lin}, {Magee}, {Marchesini}, {Maseda}, {Mason}, {Matharu}, {Meyer}, {Neufeld}, {Prieto Lyon}, {Schaerer}, {Sharma}, {Shuntov}, {Smit}, {Stefanon}, {Wyithe}, \& {Xiao}}]{oesch23}
{Oesch}, P.~A., {Brammer}, G., {Naidu}, R.~P., {et~al.} 2023, \mnras, 525, 2864, \dodoi{10.1093/mnras/stad2411}

\bibitem[{{{\"O}stlin} {et~al.}(2025){{\"O}stlin}, {P{\'e}rez-Gonz{\'a}lez}, {Melinder}, {Gillman}, {Iani}, {Costantin}, {Boogaard}, {Rinaldi}, {Colina}, {Ulrik N{\o}rgaard-Nielsen}, {Dicken}, {Greve}, {Wright}, {Alonso-Herrero}, {{\'A}lvarez-M{\'a}rquez}, {Annunziatella}, {Bik}, {Bosman}, {Caputi}, {Crespo Gomez}, {Eckart}, {Garcia-Marin}, {Hjorth}, {Ilbert}, {Jermann}, {Kendrew}, {Labiano}, {Langeroodi}, {Le Fevre}, {Libralato}, {Meyer}, {Moutard}, {Peissker}, {Pye}, {Tikkanen}, {Topinka}, {Walter}, {Ward}, {van der Werf}, {van Dishoeck}, {G{\"u}del}, {Henning}, {Lagage}, {Ray}, \& {Vandenbussche}}]{ostlin25}
{{\"O}stlin}, G., {P{\'e}rez-Gonz{\'a}lez}, P.~G., {Melinder}, J., {et~al.} 2025, \aap, 696, A57, \dodoi{10.1051/0004-6361/202451723}

\bibitem[{{P{\'e}rez-Gonz{\'a}lez} {et~al.}(2024){P{\'e}rez-Gonz{\'a}lez}, {Rinaldi}, {Caputi}, {{\'A}lvarez-M{\'a}rquez}, {Annunziatella}, {Langeroodi}, {Moutard}, {Boogaard}, {Iani}, {Melinder}, {Costantin}, {{\"O}stlin}, {Colina}, {Greve}, {Wright}, {Alonso-Herrero}, {Bik}, {Bosman}, {Crespo G{\'o}mez}, {Dicken}, {Eckart}, {Garc{\'\i}a-Mar{\'\i}n}, {Gillman}, {G{\"u}del}, {Henning}, {Hjorth}, {Jermann}, {Labiano}, {Meyer}, {Pei{\ensuremath{\beta}}ker}, {Pye}, {Ray}, {Tikkanen}, {Walter}, \& {van der Werf}}]{ppg24}
{P{\'e}rez-Gonz{\'a}lez}, P.~G., {Rinaldi}, P., {Caputi}, K.~I., {et~al.} 2024, \apjl, 969, L10, \dodoi{10.3847/2041-8213/ad517b}

\bibitem[{{Picouet} {et~al.}(2023){Picouet}, {Arnouts}, {Le Floc'h}, {Moutard}, {Kraljic}, {Ilbert}, {Sawicki}, {Desprez}, {Laigle}, {Schiminovich}, {de la Torre}, {Gwyn}, {McCracken}, {Dubois}, {Dav{\'e}}, {Toft}, {Weaver}, {Shuntov}, \& {Kauffmann}}]{picouet23}
{Picouet}, V., {Arnouts}, S., {Le Floc'h}, E., {et~al.} 2023, \aap, 675, A164, \dodoi{10.1051/0004-6361/202245756}

\bibitem[{{Price} {et~al.}(2025){Price}, {Suess}, {Williams}, {Bezanson}, {Khullar}, {Nelson}, {Wang}, {Weaver}, {Fujimoto}, {Kokorev}, {Greene}, {Brammer}, {Cutler}, {Dayal}, {Furtak}, {Labbe}, {Leja}, {Miller}, {Nanayakkara}, {Pan}, \& {Whitaker}}]{price25}
{Price}, S.~H., {Suess}, K.~A., {Williams}, C.~C., {et~al.} 2025, \apj, 980, 11, \dodoi{10.3847/1538-4357/ada0b1}

\bibitem[{{Rieke} {et~al.}(2024){Rieke}, {Alberts}, {Shivaei}, {Lyu}, {Willmer}, {P{\'e}rez-Gonz{\'a}lez}, \& {Williams}}]{rieke24}
{Rieke}, G.~H., {Alberts}, S., {Shivaei}, I., {et~al.} 2024, \apj, 975, 83, \dodoi{10.3847/1538-4357/ad6cd2}

\bibitem[{{Rieke} {et~al.}(2023{\natexlab{a}}){Rieke}, {Kelly}, {Misselt}, {Stansberry}, {Boyer}, {Beatty}, {Egami}, {Florian}, {Greene}, {Hainline}, {Leisenring}, {Roellig}, {Schlawin}, {Sun}, {Tinnin}, {Williams}, {Willmer}, {Wilson}, {Clark}, {Rohrbach}, {Brooks}, {Canipe}, {Correnti}, {DiFelice}, {Gennaro}, {Girard}, {Hartig}, {Hilbert}, {Koekemoer}, {Nikolov}, {Pirzkal}, {Rest}, {Robberto}, {Sunnquist}, {Telfer}, {Wu}, {Ferry}, {Lewis}, {Baum}, {Beichman}, {Doyon}, {Dressler}, {Eisenstein}, {Ferrarese}, {Hodapp}, {Horner}, {Jaffe}, {Johnstone}, {Krist}, {Martin}, {McCarthy}, {Meyer}, {Rieke}, {Trauger}, \& {Young}}]{rieke23}
{Rieke}, M.~J., {Kelly}, D.~M., {Misselt}, K., {et~al.} 2023{\natexlab{a}}, \pasp, 135, 028001, \dodoi{10.1088/1538-3873/acac53}

\bibitem[{{Rieke} {et~al.}(2023{\natexlab{b}}){Rieke}, {Robertson}, {Tacchella}, {Hainline}, {Johnson}, {Hausen}, {Ji}, {Willmer}, {Eisenstein}, {Pusk{\'a}s}, {Alberts}, {Arribas}, {Baker}, {Baum}, {Bhatawdekar}, {Bonaventura}, {Boyett}, {Bunker}, {Cameron}, {Carniani}, {Charlot}, {Chevallard}, {Chen}, {Curti}, {Curtis-Lake}, {Danhaive}, {DeCoursey}, {Dressler}, {Egami}, {Endsley}, {Helton}, {Hviding}, {Kumari}, {Looser}, {Lyu}, {Maiolino}, {Maseda}, {Nelson}, {Rieke}, {Rix}, {Sandles}, {Saxena}, {Sharpe}, {Shivaei}, {Skarbinski}, {Smit}, {Stark}, {Stone}, {Suess}, {Sun}, {Topping}, {{\"U}bler}, {Villanueva}, {Wallace}, {Williams}, {Willott}, {Whitler}, {Witstok}, \& {Woodrum}}]{rieke23_jades}
{Rieke}, M.~J., {Robertson}, B., {Tacchella}, S., {et~al.} 2023{\natexlab{b}}, \apjs, 269, 16, \dodoi{10.3847/1538-4365/acf44d}

\bibitem[{{Rinaldi} {et~al.}(2023){Rinaldi}, {Caputi}, {Costantin}, {Gillman}, {Iani}, {P{\'e}rez-Gonz{\'a}lez}, {{\"O}stlin}, {Colina}, {Greve}, {Noorgard-Nielsen}, {Wright}, {Alonso-Herrero}, {{\'A}lvarez-M{\'a}rquez}, {Eckart}, {Garc{\'\i}a-Mar{\'\i}n}, {Hjorth}, {Ilbert}, {Kendrew}, {Labiano}, {Le F{\`e}vre}, {Pye}, {Tikkanen}, {Walter}, {van der Werf}, {Ward}, {Annunziatella}, {Azzollini}, {Bik}, {Boogaard}, {Bosman}, {Crespo G{\'o}mez}, {Jermann}, {Langeroodi}, {Melinder}, {Meyer}, {Moutard}, {Peissker}, {Topinka}, {van Dishoeck}, {G{\"u}del}, {Henning}, {Lagage}, {Ray}, {Vandenbussche}, {Waelkens}, {Navarro-Carrera}, \& {Kokorev}}]{rinaldi2023}
{Rinaldi}, P., {Caputi}, K.~I., {Costantin}, L., {et~al.} 2023, \apj, 952, 143, \dodoi{10.3847/1538-4357/acdc27}

\bibitem[{{Roberts-Borsani} {et~al.}(2021){Roberts-Borsani}, {Treu}, {Mason}, {Schmidt}, {Jones}, \& {Fontana}}]{robertsborsani21}
{Roberts-Borsani}, G., {Treu}, T., {Mason}, C., {et~al.} 2021, \apj, 910, 86, \dodoi{10.3847/1538-4357/abe45b}

\bibitem[{{Rujopakarn} {et~al.}(2023){Rujopakarn}, {Williams}, {Daddi}, {Schramm}, {Sun}, {Alberts}, {Rieke}, {Tan}, {Tacchella}, {Giavalisco}, \& {Silverman}}]{rujopakarn23}
{Rujopakarn}, W., {Williams}, C.~C., {Daddi}, E., {et~al.} 2023, \apjl, 948, L8, \dodoi{10.3847/2041-8213/accc82}

\bibitem[{{Santini} {et~al.}(2022){Santini}, {Castellano}, {Fontana}, {Fortuni}, {Menci}, {Merlin}, {Pagul}, {Testa}, {Calabr{\`o}}, {Paris}, \& {Pentericci}}]{santini22}
{Santini}, P., {Castellano}, M., {Fontana}, A., {et~al.} 2022, \apj, 940, 135, \dodoi{10.3847/1538-4357/ac9a48}

\bibitem[{{Sarrouh} {et~al.}(2024){Sarrouh}, {Muzzin}, {Iyer}, {Mowla}, {Withers}, {Martis}, {Abraham}, {Asada}, {Brada{\v{c}}}, {Brammer}, {Desprez}, {Estrada-Carpenter}, {Matharu}, {Noirot}, {Sawicki}, {Strait}, {Willott}, \& {Zabl}}]{sarrouh24}
{Sarrouh}, G. T.~E., {Muzzin}, A., {Iyer}, K.~G., {et~al.} 2024, \apjl, 967, L17, \dodoi{10.3847/2041-8213/ad43e8}

\bibitem[{{Sarrouh} {et~al.}(2025){Sarrouh}, {Asada}, {Martis}, {Willott}, {Iyer}, {Noirot}, {Muzzin}, {Sawicki}, {Brammer}, {Desprez}, {Rihtar{\v{s}}i{\v{c}}}, {Zabl}, {Abraham}, {Brada{\v{c}}}, {Doyon}, {Antwi-Danso}, {Berek}, {Brown}, {Estrada-Carpenter}, {Favaro}, {Felicioni}, {Forrest}, {Gaspar}, {Gould}, {Gledhill}, {Harshan}, {Jahan}, {Jagga}, {Jude{\v{z}}}, {Marchesini}, {Markov}, {Matharu}, {MacFarland}, {Merchant}, {M{\'e}rida}, {Mowla}, {Myers}, {Omori}, {Pacifici}, {Ravindranath}, {Robbins}, {Strait}, {Sok}, {Tan}, {Tripodi}, {Wilson}, \& {Withers}}]{sarrouh25}
{Sarrouh}, G. T.~E., {Asada}, Y., {Martis}, N.~S., {et~al.} 2025, arXiv e-prints, arXiv:2506.21685.
\newblock \doarXiv{2506.21685}

\bibitem[{{Scoville} {et~al.}(2007){Scoville}, {Aussel}, {Brusa}, {Capak}, {Carollo}, {Elvis}, {Giavalisco}, {Guzzo}, {Hasinger}, {Impey}, {Kneib}, {LeFevre}, {Lilly}, {Mobasher}, {Renzini}, {Rich}, {Sanders}, {Schinnerer}, {Schminovich}, {Shopbell}, {Taniguchi}, \& {Tyson}}]{scoville07}
{Scoville}, N., {Aussel}, H., {Brusa}, M., {et~al.} 2007, \apjs, 172, 1, \dodoi{10.1086/516585}

\bibitem[{{Setton} {et~al.}(2024){Setton}, {Khullar}, {Miller}, {Bezanson}, {Greene}, {Suess}, {Whitaker}, {Antwi-Danso}, {Atek}, {Brammer}, {Cutler}, {Dayal}, {Feldmann}, {Fujimoto}, {Furtak}, {Glazebrook}, {Goulding}, {Kokorev}, {Labbe}, {Leja}, {Ma}, {Marchesini}, {Nanayakkara}, {Pan}, {Price}, {Siegel}, {Shipley}, {Weaver}, {van Dokkum}, {Wang}, \& {Williams}}]{setton24}
{Setton}, D.~J., {Khullar}, G., {Miller}, T.~B., {et~al.} 2024, \apj, 974, 145, \dodoi{10.3847/1538-4357/ad6a18}

\bibitem[{{Shapley} {et~al.}(2025){Shapley}, {Sanders}, {Topping}, {Reddy}, {Berg}, {Bouwens}, {Brammer}, {Carnall}, {Cullen}, {Dav{\'e}}, {Dunlop}, {Ellis}, {F{\"o}rster Schreiber}, {Furlanetto}, {Glazebrook}, {Illingworth}, {Jones}, {Kriek}, {McLeod}, {McLure}, {Narayanan}, {Oesch}, {Pahl}, {Pettini}, {Schaerer}, {Stark}, {Steidel}, {Tang}, {Clarke}, {Donnan}, \& {Kehoe}}]{shapley25}
{Shapley}, A.~E., {Sanders}, R.~L., {Topping}, M.~W., {et~al.} 2025, \apj, 980, 242, \dodoi{10.3847/1538-4357/adad68}

\bibitem[{{Shipley} {et~al.}(2018){Shipley}, {Lange-Vagle}, {Marchesini}, {Brammer}, {Ferrarese}, {Stefanon}, {Kado-Fong}, {Whitaker}, {Oesch}, {Feinstein}, {Labb{\'e}}, {Lundgren}, {Martis}, {Muzzin}, {Nedkova}, {Skelton}, \& {van der Wel}}]{shipley18}
{Shipley}, H.~V., {Lange-Vagle}, D., {Marchesini}, D., {et~al.} 2018, \apjs, 235, 14, \dodoi{10.3847/1538-4365/aaacce}

\bibitem[{{Simmonds} {et~al.}(2023){Simmonds}, {Tacchella}, {Maseda}, {Williams}, {Baker}, {Witten}, {Johnson}, {Robertson}, {Saxena}, {Sun}, {Witstok}, {Bhatawdekar}, {Boyett}, {Bunker}, {Charlot}, {Curtis-Lake}, {Egami}, {Eisenstein}, {Ji}, {Maiolino}, {Sandles}, {Smit}, {{\"U}bler}, \& {Willott}}]{simmonds23}
{Simmonds}, C., {Tacchella}, S., {Maseda}, M., {et~al.} 2023, \mnras, 523, 5468, \dodoi{10.1093/mnras/stad1749}

\bibitem[{{Skelton} {et~al.}(2014){Skelton}, {Whitaker}, {Momcheva}, {Brammer}, {van Dokkum}, {Labb{\'e}}, {Franx}, {van der Wel}, {Bezanson}, {Da Cunha}, {Fumagalli}, {F{\"o}rster Schreiber}, {Kriek}, {Leja}, {Lundgren}, {Magee}, {Marchesini}, {Maseda}, {Nelson}, {Oesch}, {Pacifici}, {Patel}, {Price}, {Rix}, {Tal}, {Wake}, \& {Wuyts}}]{skelton14}
{Skelton}, R.~E., {Whitaker}, K.~E., {Momcheva}, I.~G., {et~al.} 2014, \apjs, 214, 24, \dodoi{10.1088/0067-0049/214/2/24}

\bibitem[{{Smit} {et~al.}(2012){Smit}, {Bouwens}, {Franx}, {Illingworth}, {Labb{\'e}}, {Oesch}, \& {van Dokkum}}]{smit12}
{Smit}, R., {Bouwens}, R.~J., {Franx}, M., {et~al.} 2012, \apj, 756, 14, \dodoi{10.1088/0004-637X/756/1/14}

\bibitem[{{Smit} {et~al.}(2014){Smit}, {Bouwens}, {Labb{\'e}}, {Zheng}, {Bradley}, {Donahue}, {Lemze}, {Moustakas}, {Umetsu}, {Zitrin}, {Coe}, {Postman}, {Gonzalez}, {Bartelmann}, {Ben{\'\i}tez}, {Broadhurst}, {Ford}, {Grillo}, {Infante}, {Jimenez-Teja}, {Jouvel}, {Kelson}, {Lahav}, {Maoz}, {Medezinski}, {Melchior}, {Meneghetti}, {Merten}, {Molino}, {Moustakas}, {Nonino}, {Rosati}, \& {Seitz}}]{smit14}
{Smit}, R., {Bouwens}, R.~J., {Labb{\'e}}, I., {et~al.} 2014, \apj, 784, 58, \dodoi{10.1088/0004-637X/784/1/58}

\bibitem[{{Stark} {et~al.}(2013){Stark}, {Schenker}, {Ellis}, {Robertson}, {McLure}, \& {Dunlop}}]{stark13}
{Stark}, D.~P., {Schenker}, M.~A., {Ellis}, R., {et~al.} 2013, \apj, 763, 129, \dodoi{10.1088/0004-637X/763/2/129}

\bibitem[{{Stefanon} {et~al.}(2017){Stefanon}, {Yan}, {Mobasher}, {Barro}, {Donley}, {Fontana}, {Hemmati}, {Koekemoer}, {Lee}, {Lee}, {Nayyeri}, {Peth}, {Pforr}, {Salvato}, {Wiklind}, {Wuyts}, {Ashby}, {Castellano}, {Conselice}, {Cooper}, {Cooray}, {Dolch}, {Ferguson}, {Galametz}, {Giavalisco}, {Guo}, {Willner}, {Dickinson}, {Faber}, {Fazio}, {Gardner}, {Gawiser}, {Grazian}, {Grogin}, {Kocevski}, {Koo}, {Lee}, {Lucas}, {McGrath}, {Nandra}, {Newman}, \& {van der Wel}}]{stefanon17}
{Stefanon}, M., {Yan}, H., {Mobasher}, B., {et~al.} 2017, \apjs, 229, 32, \dodoi{10.3847/1538-4365/aa66cb}

\bibitem[{{Straatman} {et~al.}(2016){Straatman}, {Spitler}, {Quadri}, {Labb{\'e}}, {Glazebrook}, {Persson}, {Papovich}, {Tran}, {Brammer}, {Cowley}, {Tomczak}, {Nanayakkara}, {Alcorn}, {Allen}, {Broussard}, {van Dokkum}, {Forrest}, {van Houdt}, {Kacprzak}, {Kawinwanichakij}, {Kelson}, {Lee}, {McCarthy}, {Mehrtens}, {Monson}, {Murphy}, {Rees}, {Tilvi}, \& {Whitaker}}]{straatman16}
{Straatman}, C. M.~S., {Spitler}, L.~R., {Quadri}, R.~F., {et~al.} 2016, \apj, 830, 51, \dodoi{10.3847/0004-637X/830/1/51}

\bibitem[{{Suess} {et~al.}(2019{\natexlab{a}}){Suess}, {Kriek}, {Price}, \& {Barro}}]{suess19}
{Suess}, K.~A., {Kriek}, M., {Price}, S.~H., \& {Barro}, G. 2019{\natexlab{a}}, \apj, 877, 103, \dodoi{10.3847/1538-4357/ab1bda}

\bibitem[{{Suess} {et~al.}(2019{\natexlab{b}}){Suess}, {Kriek}, {Price}, \& {Barro}}]{suess19b}
---. 2019{\natexlab{b}}, \apjl, 885, L22, \dodoi{10.3847/2041-8213/ab4db3}

\bibitem[{{Suess} {et~al.}(2024){Suess}, {Weaver}, {Price}, {Pan}, {Wang}, {Bezanson}, {Brammer}, {Cutler}, {Labbe}, {Leja}, {Williams}, {Whitaker}, {Dayal}, {de Graaff}, {Feldmann}, {Franx}, {Fudamoto}, {Fujimoto}, {Furtak}, {Goulding}, {Greene}, {Khullar}, {Kokorev}, {Kriek}, {Lorenz}, {Marchesini}, {Maseda}, {Matthee}, {Miller}, {Mitsuhashi}, {Mowla}, {Muzzin}, {Naidu}, {Nanayakkara}, {Nelson}, {Oesch}, {Setton}, {Shipley}, {Smit}, {Spilker}, {van Dokkum}, \& {Zitrin}}]{suess24}
{Suess}, K.~A., {Weaver}, J.~R., {Price}, S.~H., {et~al.} 2024, arXiv e-prints, arXiv:2404.13132, \dodoi{10.48550/arXiv.2404.13132}

\bibitem[{{Sun} {et~al.}(2024){Sun}, {Helton}, {Egami}, {Hainline}, {Rieke}, {Willmer}, {Eisenstein}, {Johnson}, {Rieke}, {Robertson}, {Tacchella}, {Alberts}, {Baker}, {Bhatawdekar}, {Boyett}, {Bunker}, {Charlot}, {Chen}, {Chevallard}, {Curtis-Lake}, {Danhaive}, {DeCoursey}, {Ji}, {Lyu}, {Maiolino}, {Rujopakarn}, {Sandles}, {Shivaei}, {{\"U}bler}, {Willott}, \& {Witstok}}]{sun24}
{Sun}, F., {Helton}, J.~M., {Egami}, E., {et~al.} 2024, \apj, 961, 69, \dodoi{10.3847/1538-4357/ad07e3}

\bibitem[{{Sun} {et~al.}(2025){Sun}, {Fudamoto}, {Lin}, {Helton}, {Hsiao}, {Egami}, {Akhtarkavan}, {Bunker}, {Cai}, {DeCoursey}, {Eisenstein}, {Fan}, {Harikane}, {Ji}, {Jin}, {Liu}, {Liu}, {Ma}, {Maiolino}, {Ouchi}, {Tee}, {Wang}, {Willmer}, {Wu}, {Xu}, {Yang}, {Zhang}, \& {Zhu}}]{sun25}
{Sun}, F., {Fudamoto}, Y., {Lin}, X., {et~al.} 2025, arXiv e-prints, arXiv:2503.15587, \dodoi{10.48550/arXiv.2503.15587}

\bibitem[{{Tacchella} {et~al.}(2015){Tacchella}, {Carollo}, {Renzini}, {F{\"o}rster Schreiber}, {Lang}, {Wuyts}, {Cresci}, {Dekel}, {Genzel}, {Lilly}, {Mancini}, {Newman}, {Onodera}, {Shapley}, {Tacconi}, {Woo}, \& {Zamorani}}]{tacchella15}
{Tacchella}, S., {Carollo}, C.~M., {Renzini}, A., {et~al.} 2015, Science, 348, 314, \dodoi{10.1126/science.1261094}

\bibitem[{{Tacchella} {et~al.}(2025){Tacchella}, {McClymont}, {Scholtz}, {Maiolino}, {Ji}, {Villanueva}, {Charlot}, {D'Eugenio}, {Helton}, {Williams}, {Witstok}, {Bhatawdekar}, {Carniani}, {Chevallard}, {Curti}, {Hainline}, {Ji}, {Johnson}, {Leja}, {Li}, {Maseda}, {Pusk{\'a}s}, {Rieke}, {Robertson}, {Shivaei}, {Silcock}, {Simmonds}, {{\"U}bler}, {Willmer}, \& {Willott}}]{tacchella25}
{Tacchella}, S., {McClymont}, W., {Scholtz}, J., {et~al.} 2025, \mnras, 540, 851, \dodoi{10.1093/mnras/staf718}

\bibitem[{{Tan} {et~al.}(2024){Tan}, {Muzzin}, {Sarrouh}, {Antwi-Danso}, {Sok}, {Jagga}, {Abraham}, {Asada}, {Desprez}, {Iyer}, {Martis}, {M{\'e}rida}, {Mowla}, {Noirot}, {Omori}, {Sawicki}, {Tripodi}, \& {Willott}}]{tan24}
{Tan}, V. Y.~Y., {Muzzin}, A., {Sarrouh}, G. T.~E., {et~al.} 2024, arXiv e-prints, arXiv:2412.07829, \dodoi{10.48550/arXiv.2412.07829}

\bibitem[{{Tanaka} {et~al.}(2024){Tanaka}, {Silverman}, {Nakazato}, {Onoue}, {Shimasaku}, {Fudamoto}, {Fujimoto}, {Ding}, {Faisst}, {Valentino}, {Jin}, {Hayward}, {Kokorev}, {Ceverino}, {Kalita}, {Casey}, {Liu}, {Kaminsky}, {Fei}, {Andika}, {Lambrides}, {Akins}, {Kartaltepe}, {Koekemoer}, {McCracken}, {Rhodes}, {Robertson}, {Franco}, {Liu}, {Chartab}, {Gillman}, {Gozaliasl}, {Hirschmann}, {Huertas-Company}, {Massey}, {Roy}, {Sattari}, {Shuntov}, {Sterling}, {Toft}, {Trakhtenbrot}, {Yoshida}, \& {Zavala}}]{tanaka24}
{Tanaka}, T.~S., {Silverman}, J.~D., {Nakazato}, Y., {et~al.} 2024, \pasj, 76, 1323, \dodoi{10.1093/pasj/psae091}

\bibitem[{{Taniguchi} {et~al.}(2007){Taniguchi}, {Scoville}, {Murayama}, {Sanders}, {Mobasher}, {Aussel}, {Capak}, {Ajiki}, {Miyazaki}, {Komiyama}, {Shioya}, {Nagao}, {Sasaki}, {Koda}, {Carilli}, {Giavalisco}, {Guzzo}, {Hasinger}, {Impey}, {LeFevre}, {Lilly}, {Renzini}, {Rich}, {Schinnerer}, {Shopbell}, {Kaifu}, {Karoji}, {Arimoto}, {Okamura}, \& {Ohta}}]{taniguchi07}
{Taniguchi}, Y., {Scoville}, N., {Murayama}, T., {et~al.} 2007, \apjs, 172, 9, \dodoi{10.1086/516596}

\bibitem[{{Taylor} {et~al.}(2009){Taylor}, {Franx}, {van Dokkum}, {Quadri}, {Gawiser}, {Bell}, {Barrientos}, {Blanc}, {Castander}, {Damen}, {Gonzalez-Perez}, {Hall}, {Herrera}, {Hildebrandt}, {Kriek}, {Labb{\'e}}, {Lira}, {Maza}, {Rudnick}, {Treister}, {Urry}, {Willis}, \& {Wuyts}}]{taylor09}
{Taylor}, E.~N., {Franx}, M., {van Dokkum}, P.~G., {et~al.} 2009, \apjs, 183, 295, \dodoi{10.1088/0067-0049/183/2/295}

\bibitem[{{Tomczak} {et~al.}(2014){Tomczak}, {Quadri}, {Tran}, {Labb{\'e}}, {Straatman}, {Papovich}, {Glazebrook}, {Allen}, {Brammer}, {Kacprzak}, {Kawinwanichakij}, {Kelson}, {McCarthy}, {Mehrtens}, {Monson}, {Persson}, {Spitler}, {Tilvi}, \& {van Dokkum}}]{tomczack14}
{Tomczak}, A.~R., {Quadri}, R.~F., {Tran}, K.-V.~H., {et~al.} 2014, \apj, 783, 85, \dodoi{10.1088/0004-637X/783/2/85}

\bibitem[{{Tomczak} {et~al.}(2016){Tomczak}, {Quadri}, {Tran}, {Labb{\'e}}, {Straatman}, {Papovich}, {Glazebrook}, {Allen}, {Brammer}, {Cowley}, {Dickinson}, {Elbaz}, {Inami}, {Kacprzak}, {Morrison}, {Nanayakkara}, {Persson}, {Rees}, {Salmon}, {Schreiber}, {Spitler}, \& {Whitaker}}]{tomczak16}
---. 2016, \apj, 817, 118, \dodoi{10.3847/0004-637X/817/2/118}

\bibitem[{{Trussler} {et~al.}(2024){Trussler}, {Conselice}, {Adams}, {Austin}, {Caruana}, {Harvey}, {Li}, {Lovell}, {Seeyave}, {Vijayan}, \& {Wilkins}}]{trussler24}
{Trussler}, J., {Conselice}, C., {Adams}, N., {et~al.} 2024, arXiv e-prints, arXiv:2404.07163, \dodoi{10.48550/arXiv.2404.07163}

\bibitem[{{Trussler} {et~al.}(2025){Trussler}, {Conselice}, {Adams}, {Austin}, {Caruana}, {Harvey}, {Li}, {Lovell}, {Seeyave}, {Vijayan}, \& {Wilkins}}]{trussler25}
{Trussler}, J. A.~A., {Conselice}, C.~J., {Adams}, N., {et~al.} 2025, \mnras, 537, 3662, \dodoi{10.1093/mnras/staf213}

\bibitem[{{van der Wel} {et~al.}(2011){van der Wel}, {Straughn}, {Rix}, {Finkelstein}, {Koekemoer}, {Weiner}, {Wuyts}, {Bell}, {Faber}, {Trump}, {Koo}, {Ferguson}, {Scarlata}, {Hathi}, {Dunlop}, {Newman}, {Dickinson}, {Jahnke}, {Salmon}, {de Mello}, {Kocevski}, {Lai}, {Grogin}, {Rodney}, {Guo}, {McGrath}, {Lee}, {Barro}, {Huang}, {Riess}, {Ashby}, \& {Willner}}]{vanderwel11}
{van der Wel}, A., {Straughn}, A.~N., {Rix}, H.~W., {et~al.} 2011, \apj, 742, 111, \dodoi{10.1088/0004-637X/742/2/111}

\bibitem[{{Wang} {et~al.}(2023){Wang}, {Fujimoto}, {Labb{\'e}}, {Furtak}, {Miller}, {Setton}, {Zitrin}, {Atek}, {Bezanson}, {Brammer}, {Leja}, {Oesch}, {Price}, {Chemerynska}, {Cutler}, {Dayal}, {van Dokkum}, {Goulding}, {Greene}, {Fudamoto}, {Khullar}, {Kokorev}, {Marchesini}, {Pan}, {Weaver}, {Whitaker}, \& {Williams}}]{wang23}
{Wang}, B., {Fujimoto}, S., {Labb{\'e}}, I., {et~al.} 2023, \apjl, 957, L34, \dodoi{10.3847/2041-8213/acfe07}

\bibitem[{{Wang} {et~al.}(2025){Wang}, {Leja}, {Atek}, {Bezanson}, {Burnham}, {Dayal}, {Feldmann}, {Greene}, {Johnson}, {Labbe}, {Maseda}, {Nanayakkara}, {Price}, {Suess}, {Weaver}, \& {Whitaker}}]{wang25}
{Wang}, B., {Leja}, J., {Atek}, H., {et~al.} 2025, arXiv e-prints, arXiv:2504.15255, \dodoi{10.48550/arXiv.2504.15255}

\bibitem[{{Wang} {et~al.}(2024{\natexlab{a}}){Wang}, {Sun}, {Zhou}, {Xu}, {Cheng}, {Li}, {Chen}, {Mo}, {Dekel}, {Yang}, {Wang}, {Zheng}, {Cai}, {Elbaz}, {Dai}, \& {Huang}}]{wang24}
{Wang}, T., {Sun}, H., {Zhou}, L., {et~al.} 2024{\natexlab{a}}, arXiv e-prints, arXiv:2403.02399, \dodoi{10.48550/arXiv.2403.02399}

\bibitem[{{Wang} {et~al.}(2024{\natexlab{b}}){Wang}, {Teplitz}, {Sun}, {Rafelski}, {Grogin}, {Prichard}, {Sunnquist}, {Alavi}, {Windhorst}, {Koekemoer}, {Ashcraft}, {Bagley}, {Baronchelli}, {Barro}, {Blanche}, {Brammer}, {Broussard}, {Carleton}, {Chartab}, {Cheng}, {Codoreanu}, {Cohen}, {Colbert}, {Conselice}, {Dai}, {Darvish}, {Dav{\'e}}, {DeGroot}, {De Mello}, {Dickinson}, {Emami}, {Ferguson}, {Ferreira}, {Finkelstein}, {Finkelstein}, {Gardner}, {Gawiser}, {Gburek}, {Giavalisco}, {Grazian}, {Gronwall}, {Guo}, {Arrabal Haro}, {Hathi}, {Hayes}, {Hemmati}, {Howell}, {Iyer}, {Jansen}, {Ji}, {Kaviraj}, {Kurczynski}, {Lazar}, {Lucas}, {MacKenty}, {Mehta}, {Mantha}, {Martin}, {Martin}, {McCabe}, {Mobasher}, {Nedkova}, {O'Connell}, {Olsen}, {Otteson}, {Ravindranath}, {Redshaw}, {Robertson}, {Rutkowski}, {Sattari}, {Scarlata}, {Siana}, {Smith}, {Soto}, {Vanzella}, {Yung}, \& {Zabelle}}]{wang24b}
{Wang}, X., {Teplitz}, H.~I., {Sun}, L., {et~al.} 2024{\natexlab{b}}, Research Notes of the American Astronomical Society, 8, 26, \dodoi{10.3847/2515-5172/ad1f6f}

\bibitem[{{Waskom} {et~al.}(2023){Waskom}, {Gelbart}, {Botvinnik}, {Ostblom}, {Hobson}, {Lukauskas}, {Gemperline}, {Augspurger}, {Halchenko}, {Warmenhoven}, {Cole}, {Ter Hoeven}, {De Ruiter}, {Vanderplas}, {Hoyer}, {Pye}, {Miles}, {Swain}, {Meyer}, {Martin}, {Bachant}, {Molin}, {Quintero}, {Kunter}, {Villalba}, {Brian}, {Fitzgerald}, {Evans}, \& {Williams}}]{waskom23}
{Waskom}, M., {Gelbart}, M., {Botvinnik}, O., {et~al.} 2023, {mwaskom/seaborn: v0.13.0 (September 2023)}, v0.13.0,  Zenodo, \dodoi{10.5281/zenodo.592845}

\bibitem[{{Weaver} {et~al.}(2023){Weaver}, {Davidzon}, {Toft}, {Ilbert}, {McCracken}, {Gould}, {Jespersen}, {Steinhardt}, {Lagos}, {Capak}, {Casey}, {Chartab}, {Faisst}, {Hayward}, {Kartaltepe}, {Kauffmann}, {Koekemoer}, {Kokorev}, {Laigle}, {Liu}, {Long}, {Magdis}, {McPartland}, {Milvang-Jensen}, {Mobasher}, {Moneti}, {Peng}, {Sanders}, {Shuntov}, {Sneppen}, {Valentino}, {Zalesky}, \& {Zamorani}}]{weaver23}
{Weaver}, J.~R., {Davidzon}, I., {Toft}, S., {et~al.} 2023, \aap, 677, A184, \dodoi{10.1051/0004-6361/202245581}

\bibitem[{{Weibel} {et~al.}(2024){Weibel}, {Oesch}, {Barrufet}, {Gottumukkala}, {Ellis}, {Santini}, {Weaver}, {Allen}, {Bouwens}, {Bowler}, {Brammer}, {Carnall}, {Cullen}, {Dayal}, {Dickinson}, {Donnan}, {Dunlop}, {Giavalisco}, {Grogin}, {Illingworth}, {Koekemoer}, {Labbe}, {Marchesini}, {McLeod}, {McLure}, {Naidu}, {P{\'e}rez-Gonz{\'a}lez}, {Shuntov}, {Stefanon}, {Toft}, \& {Xiao}}]{weibel24}
{Weibel}, A., {Oesch}, P.~A., {Barrufet}, L., {et~al.} 2024, \mnras, 533, 1808, \dodoi{10.1093/mnras/stae1891}

\bibitem[{{Weibel} {et~al.}(2025{\natexlab{a}}){Weibel}, {Oesch}, {Williams}, {Jespersen}, {Shuntov}, {Whitaker}, {Atek}, {Bezanson}, {Brammer}, {Chemerynska}, {Cloonan}, {Dayal}, {Furtak}, {Hutter}, {Ji}, {Maseda}, \& {Xiao}}]{weibel25_pano}
{Weibel}, A., {Oesch}, P.~A., {Williams}, C.~C., {et~al.} 2025{\natexlab{a}}, arXiv e-prints, arXiv:2507.06292, \dodoi{10.48550/arXiv.2507.06292}

\bibitem[{{Weibel} {et~al.}(2025{\natexlab{b}}){Weibel}, {de Graaff}, {Setton}, {Miller}, {Oesch}, {Brammer}, {Lagos}, {Whitaker}, {Williams}, {Baggen}, {Bezanson}, {Boogaard}, {Cleri}, {Greene}, {Hirschmann}, {Hviding}, {Kuruvanthodi}, {Labb{\'e}}, {Leja}, {Maseda}, {Matthee}, {McConachie}, {Naidu}, {Roberts-Borsani}, {Schaerer}, {Suess}, {Valentino}, {van Dokkum}, \& {Wang}}]{weibel25}
{Weibel}, A., {de Graaff}, A., {Setton}, D.~J., {et~al.} 2025{\natexlab{b}}, \apj, 983, 11, \dodoi{10.3847/1538-4357/adab7a}

\bibitem[{{Whitaker} {et~al.}(2012){Whitaker}, {van Dokkum}, {Brammer}, \& {Franx}}]{whitaker12}
{Whitaker}, K.~E., {van Dokkum}, P.~G., {Brammer}, G., \& {Franx}, M. 2012, \apjl, 754, L29, \dodoi{10.1088/2041-8205/754/2/L29}

\bibitem[{{Whitaker} {et~al.}(2011){Whitaker}, {Labb{\'e}}, {van Dokkum}, {Brammer}, {Kriek}, {Marchesini}, {Quadri}, {Franx}, {Muzzin}, {Williams}, {Bezanson}, {Illingworth}, {Lee}, {Lundgren}, {Nelson}, {Rudnick}, {Tal}, \& {Wake}}]{whitaker11}
{Whitaker}, K.~E., {Labb{\'e}}, I., {van Dokkum}, P.~G., {et~al.} 2011, \apj, 735, 86, \dodoi{10.1088/0004-637X/735/2/86}

\bibitem[{{Williams} {et~al.}(2023){Williams}, {Tacchella}, {Maseda}, {Robertson}, {Johnson}, {Willott}, {Eisenstein}, {Willmer}, {Ji}, {Hainline}, {Helton}, {Alberts}, {Baum}, {Bhatawdekar}, {Boyett}, {Bunker}, {Carniani}, {Charlot}, {Chevallard}, {Curtis-Lake}, {de Graaff}, {Egami}, {Franx}, {Kumari}, {Maiolino}, {Nelson}, {Rieke}, {Sandles}, {Shivaei}, {Simmonds}, {Smit}, {Suess}, {Sun}, {{\"U}bler}, \& {Witstok}}]{williams23}
{Williams}, C.~C., {Tacchella}, S., {Maseda}, M.~V., {et~al.} 2023, \apjs, 268, 64, \dodoi{10.3847/1538-4365/acf130}

\bibitem[{{Williams} {et~al.}(2024){Williams}, {Alberts}, {Ji}, {Hainline}, {Lyu}, {Rieke}, {Endsley}, {Suess}, {Sun}, {Johnson}, {Florian}, {Shivaei}, {Rujopakarn}, {Baker}, {Bhatawdekar}, {Boyett}, {Bunker}, {Cameron}, {Carniani}, {Charlot}, {Curtis-Lake}, {DeCoursey}, {de Graaff}, {Egami}, {Eisenstein}, {Gibson}, {Hausen}, {Helton}, {Maiolino}, {Maseda}, {Nelson}, {P{\'e}rez-Gonz{\'a}lez}, {Rieke}, {Robertson}, {Saxena}, {Tacchella}, {Willmer}, \& {Willott}}]{williams24}
{Williams}, C.~C., {Alberts}, S., {Ji}, Z., {et~al.} 2024, \apj, 968, 34, \dodoi{10.3847/1538-4357/ad3f17}

\bibitem[{{Williams} {et~al.}(2025){Williams}, {Oesch}, {Weibel}, {Brammer}, {Cloonan}, {Whitaker}, {Barrufet}, {Bezanson}, {Bowler}, {Dayal}, {Franx}, {Greene}, {Hutter}, {Ji}, {Labb{\'e}}, {Manning}, {Maseda}, \& {Xiao}}]{williams25}
{Williams}, C.~C., {Oesch}, P.~A., {Weibel}, A., {et~al.} 2025, \apj, 979, 140, \dodoi{10.3847/1538-4357/ad97bc}

\bibitem[{{Williams} {et~al.}(1996){Williams}, {Blacker}, {Dickinson}, {Dixon}, {Ferguson}, {Fruchter}, {Giavalisco}, {Gilliland}, {Heyer}, {Katsanis}, {Levay}, {Lucas}, {McElroy}, {Petro}, {Postman}, {Adorf}, \& {Hook}}]{williams96}
{Williams}, R.~E., {Blacker}, B., {Dickinson}, M., {et~al.} 1996, \aj, 112, 1335, \dodoi{10.1086/118105}

\bibitem[{{Willott} {et~al.}(2024){Willott}, {Desprez}, {Asada}, {Sarrouh}, {Abraham}, {Brada{\v{c}}}, {Brammer}, {Estrada-Carpenter}, {Iyer}, {Martis}, {Matharu}, {Mowla}, {Muzzin}, {Noirot}, {Sawicki}, {Strait}, {Rihtar{\v{s}}i{\v{c}}}, \& {Withers}}]{willott24}
{Willott}, C.~J., {Desprez}, G., {Asada}, Y., {et~al.} 2024, \apj, 966, 74, \dodoi{10.3847/1538-4357/ad35bc}

\bibitem[{{Wisnioski} {et~al.}(2015){Wisnioski}, {F{\"o}rster Schreiber}, {Wuyts}, {Wuyts}, {Bandara}, {Wilman}, {Genzel}, {Bender}, {Davies}, {Fossati}, {Lang}, {Mendel}, {Beifiori}, {Brammer}, {Chan}, {Fabricius}, {Fudamoto}, {Kulkarni}, {Kurk}, {Lutz}, {Nelson}, {Momcheva}, {Rosario}, {Saglia}, {Seitz}, {Tacconi}, \& {van Dokkum}}]{wisnioski15}
{Wisnioski}, E., {F{\"o}rster Schreiber}, N.~M., {Wuyts}, S., {et~al.} 2015, \apj, 799, 209, \dodoi{10.1088/0004-637X/799/2/209}

\bibitem[{{Wisnioski} {et~al.}(2019){Wisnioski}, {F{\"o}rster Schreiber}, {Fossati}, {Mendel}, {Wilman}, {Genzel}, {Bender}, {Wuyts}, {Davies}, {{\"U}bler}, {Bandara}, {Beifiori}, {Belli}, {Brammer}, {Chan}, {Davies}, {Fabricius}, {Galametz}, {Lang}, {Lutz}, {Nelson}, {Momcheva}, {Price}, {Rosario}, {Saglia}, {Seitz}, {Shimizu}, {Tacconi}, {Tadaki}, {van Dokkum}, \& {Wuyts}}]{wisnioski19}
{Wisnioski}, E., {F{\"o}rster Schreiber}, N.~M., {Fossati}, M., {et~al.} 2019, \apj, 886, 124, \dodoi{10.3847/1538-4357/ab4db8}

\bibitem[{{Withers} {et~al.}(2023){Withers}, {Muzzin}, {Ravindranath}, {Sarrouh}, {Abraham}, {Asada}, {Brada{\v{c}}}, {Brammer}, {Desprez}, {Iyer}, {Martis}, {Mowla}, {Noirot}, {Sawicki}, {Strait}, \& {Willott}}]{withers23}
{Withers}, S., {Muzzin}, A., {Ravindranath}, S., {et~al.} 2023, \apjl, 958, L14, \dodoi{10.3847/2041-8213/ad01c0}

\bibitem[{{Wold} {et~al.}(2024){Wold}, {Malhotra}, {Rhoads}, {Weaver}, \& {Wang}}]{wold2024}
{Wold}, I. G.~B., {Malhotra}, S., {Rhoads}, J.~E., {Weaver}, J.~R., \& {Wang}, B. 2024, arXiv e-prints, arXiv:2407.19023, \dodoi{10.48550/arXiv.2407.19023}

\bibitem[{{Wold} {et~al.}(2025){Wold}, {Malhotra}, {Rhoads}, {Weaver}, \& {Wang}}]{wold25}
---. 2025, \apj, 980, 200, \dodoi{10.3847/1538-4357/ada8a6}

\bibitem[{{Wolf} {et~al.}(2003){Wolf}, {Meisenheimer}, {Rix}, {Borch}, {Dye}, \& {Kleinheinrich}}]{wolf03}
{Wolf}, C., {Meisenheimer}, K., {Rix}, H.~W., {et~al.} 2003, \aap, 401, 73, \dodoi{10.1051/0004-6361:20021513}

\bibitem[{{Wuyts} {et~al.}(2008){Wuyts}, {Labb{\'e}}, {F{\"o}rster Schreiber}, {Franx}, {Rudnick}, {Brammer}, \& {van Dokkum}}]{wuyts08}
{Wuyts}, S., {Labb{\'e}}, I., {F{\"o}rster Schreiber}, N.~M., {et~al.} 2008, \apj, 682, 985, \dodoi{10.1086/588749}

\bibitem[{{Zaidi} {et~al.}(2024){Zaidi}, {Marchesini}, {Papovich}, {Antwi-Danso}, {Nonino}, {Annunziatella}, {Brammer}, {Esdaile}, {Glazebrook}, {Iyer}, {Labb{\'e}}, {Marsan}, {Muzzin}, \& {Wake}}]{zaidi24}
{Zaidi}, K., {Marchesini}, D., {Papovich}, C., {et~al.} 2024, \apj, 969, 84, \dodoi{10.3847/1538-4357/ad45fa}

\bibitem[{{Zavala} {et~al.}(2023){Zavala}, {Buat}, {Casey}, {Finkelstein}, {Burgarella}, {Bagley}, {Ciesla}, {Daddi}, {Dickinson}, {Ferguson}, {Franco}, {Jim{\'e}nez-Andrade}, {Kartaltepe}, {Koekemoer}, {Le Bail}, {Murphy}, {Papovich}, {Tacchella}, {Wilkins}, {Aretxaga}, {Behroozi}, {Champagne}, {Fontana}, {Giavalisco}, {Grazian}, {Grogin}, {Kewley}, {Kocevski}, {Kirkpatrick}, {Lotz}, {Pentericci}, {P{\'e}rez-Gonz{\'a}lez}, {Pirzkal}, {Ravindranath}, {Somerville}, {Trump}, {Yang}, {Yung}, {Almaini}, {Amor{\'\i}n}, {Annunziatella}, {Arrabal Haro}, {Backhaus}, {Barro}, {Bell}, {Bhatawdekar}, {Bisigello}, {Buitrago}, {Calabr{\`o}}, {Castellano}, {Ch{\'a}vez Ortiz}, {Chworowsky}, {Cleri}, {Cohen}, {Cole}, {Cooke}, {Cooper}, {Cooray}, {Costantin}, {Cox}, {Croton}, {Dav{\'e}}, {de La Vega}, {Dekel}, {Elbaz}, {Estrada-Carpenter}, {Fern{\'a}ndez}, {Finkelstein}, {Freundlich}, {Fujimoto}, {Garc{\'\i}a-Argum{\'a}nez}, {Gardner}, {Gawiser}, {G{\'o}mez-Guijarro}, {Guo}, {Hamilton}, {Hathi}, {Holwerda}, {Hirschmann},
  {Huertas-Company}, {Hutchison}, {Iyer}, {Jaskot}, {Jha}, {Jogee}, {Juneau}, {Jung}, {Kassin}, {Kurczynski}, {Larson}, {Leung}, {Long}, {Lucas}, {Magnelli}, {Mantha}, {Matharu}, {McGrath}, {McIntosh}, {Medrano}, {Merlin}, {Mobasher}, {Morales}, {Newman}, {Nicholls}, {Pandya}, {Rafelski}, {Ronayne}, {Rose}, {Ryan}, {Santini}, {Seill{\'e}}, {Shah}, {Shen}, {Simons}, {Snyder}, {Stanway}, {Straughn}, {Teplitz}, {Vanderhoof}, {Vega-Ferrero}, {Wang}, {Weiner}, {Willmer}, {Wuyts}, \& {Ceers Team}}]{zavala23}
{Zavala}, J.~A., {Buat}, V., {Casey}, C.~M., {et~al.} 2023, \apjl, 943, L9, \dodoi{10.3847/2041-8213/acacfe}

\bibitem[{{Zhu} {et~al.}(2024){Zhu}, {Rieke}, {Ji}, {Simmonds}, {Sun}, {Sun}, {Alberts}, {Bhatawdekar}, {Bunker}, {Cargile}, {Carniani}, {de Graaff}, {Hainline}, {Helton}, {Jones}, {Lyu}, {Rieke}, {Rinaldi}, {Robertson}, {Scholtz}, {{\"U}bler}, {Williams}, \& {Willmer}}]{zhu24}
{Zhu}, Y., {Rieke}, M.~J., {Ji}, Z., {et~al.} 2024, arXiv e-prints, arXiv:2409.11464, \dodoi{10.48550/arXiv.2409.11464}

\end{thebibliography}

\end{document}